\documentclass[twocolumn,tighten]{aastex701}

\usepackage{amsmath}
\usepackage{multirow}

\newcommand{\rOC}{\ensuremath{\rho}\,Oph\,C}
\defcitealias{Gunderson2025b}{Paper 1} 

\newcommand{\mki}{
  Kavli Institute for Astrophysics and Space Research, 
  Massachusetts Institute of Technology , 77 Massachusetts  Ave., 
  Cambridge, MA 02139, USA
}
\newcommand{\macro}{The MACRO Consortium; \href{http://www.macroconsortium.org}{http://www.macroconsortium.org}}

\newcounter{ion} \newcommand{\eli}[2]{\setcounter{ion}{#2}#1{~\sc\roman{ion}}}

\newcommand{\Change}[1]{{\color{red}\bf #1}}
\renewcommand{\Change}[1]{#1}
\newcommand{\ChangeTwo}[1]{{\color{violet}\bf #1}}
\renewcommand{\ChangeTwo}[1]{#1}

\begin{document}

\title{A Multiwavelength View of $\rho$\,Oph II: Disentangling the Variability of the Multi-Star Component C}

\author[orcid=0000-0003-2602-6703, sname=Gunderson, gname=Sean]{Sean J.\ Gunderson}
\affiliation{\mki}
\affiliation{\macro}
\email[show]{seang97@mit.edu}

\author[orcid=0000-0001-7946-1034, sname=Golay, gname=Walter]{Walter W.\ Golay}
\email{wgolay@cfa.harvard.edu}
\affiliation{Center for Astrophysics $|$ Harvard \& Smithsonian, 60 Garden St., Cambridge, MA 02138, USA}
\affiliation{\macro}

\author[orcid=0000-0002-3534-7691, sname=Codd, gname=Jackson]{Jackson Codd}
\affiliation{Department of Physics and Astronomy, Macalester College, 1600 Grand Avenue, Saint Paul, MN 55105, USA}
\affiliation{\macro}
\email{jackson@coddsquad.com}

\author[orcid=0000-0002-3860-6230,sname=Huenemoerder,gname=David]{David P.\ Huenemoerder}
\affiliation{\mki}
\email{dph@mit.edu}

\author[orcid=0000-0002-0284-0578, gname=Felipe, sname=Navarete]{Felipe Navarete}
\affiliation{Laborat\'orio Nacional de Astrof\'isica (LNA/MCTI), Rua dos Estados Unidos, 154, 37504-364, Itajubá, MG, Brazil}
\email{fnavarete@lna.br}

\author[orcid=0000-0003-1299-8878, sname=Erba,gname=Christiana]{Christiana Erba}
\affiliation{Space Telescope Science Institute, 3700 San Martin Drive, Baltimore, MD 21218, USA}
\affiliation{Department of Physics, California State University, Fresno, Fresno, CA 93740, USA}
\email{christi.erba@gmail.com}

\author[orcid=0009-0005-9304-0742, sname=Alexandrea, gname=Moreno]{Alexandrea Moreno}
\affiliation{Department of Physics and Astronomy, University of Iowa, 30 N. Dubuque Street, Iowa City, IA 52242, USA}
\email{alexandrea-moreno@uiowa.edu}

\author[orcid=0000-0002-1821-7019, sname=Cannon, gname=John]{John M.\ Cannon}
\affiliation{Department of Physics and Astronomy, Macalester College, 1600 Grand Avenue, Saint Paul, MN 55105, USA}
\affiliation{\macro}
\email{jcannon@macalester.edu}

\author[orcid=0000-0002-7204-5502, sname=Ignace, gname=Richard]{Richard Ignace} \affiliation{Department
 of Physics \& Astronomy, East Tennessee State University, Johnson City, TN 37614 USA}
\email{ignace@etsu.edu}

\author[orcid=0000-0002-1131-3059,sname=Pradhan,gname=Pragati]{Pragati Pradhan}
\affiliation{Embry Riddle Aeronautical University, Department of
 Physics \& Astronomy, 3700 Willow Creek Road Prescott, AZ 86301, USA}
 \email{pradhanp@erau.edu}

\author{the MACRO consortium}
\altaffiliation{A full list of MACRO Consortium authors for the current year is available at \url{https://macroconsortium.org/about-us/members}}
\affiliation{\macro}
\email{macro@macroconsortium.org}

\begin{abstract}

We present a multiwavelength analysis of the star $\rho$\,Oph\,C, covering X-ray, optical, near-infrared, and radio observations to test for the presence of interferometrically-detected cool star companions. The X-ray observations from Chandra, XMM-Newton, and NuSTAR show flare-like events lasting minutes or days in time and spectral properties consistent with those of cooler stars instead of the primary magnetic B star. The near-infrared data is also consistent with the proposed trinary system properties due to the presence of CO-band heads. After removing the primary B star's rotation, the optical data from TESS shows a residual signal at the level of 0.8\,mmags, matching the variations expected from cool stars in orbit around a bright B star. Finally, multi-epoch radio data reveals that $\rho$\,Oph\,C exhibits significant large scale flux variations that are atypical for a magnetic massive star. Taken as a whole, the multi-epoch and -wavelength data presents a consistent picture of $\rho$\,Oph\,C as a rare trinary system composed of a hot star and two cool stars.
\end{abstract}


\section{Introduction} 

The ``star'' $\rho$\,Oph is an archetypal example of how seemingly single stars can in reality be composed of multiple objects. $\rho$\,Oph is a collection of several young stars, spanning both age (from early formation to the main sequence) and size (as there are both massive and low-mass stars). The primary components are the main-sequence massive stars AB, C, and D, all of which are B-types.

$\rho$\,Oph is also the namesake of the $\rho$\,Oph cloud complex, which is centered approximately 1 degree south of the $\rho$\,Oph star system. The star forming regions of the cloud complex are not as young as the Orion Nebula \citep{Shulz2024} but have the unique feature of the pre-main sequence (PMS) stars having a higher than average multiplicity rate \citep{Barsony2003}.

Having been formed from this cloud complex, the $\rho$\,Oph star system should presumably exhibit the statistically high rate of multiplicity. \citet{Shultz2025} discovered that $\rho$\,Oph\,A is a binary of two B-type stars from high resolution optical data. Then from X-ray and radio data, \citet[hereafter Paper 1]{Gunderson2025b} found evidence that $\rho$\,Oph\,B is a binary. For component D, \citet{Novakovic2007} suggests that it is a visual binary with a 680\,day orbit.

\citetalias{Gunderson2025b} specifically found that $\rho$\,Oph\,B appears to be a rare hot+cool star binary pair; here ``hot'' refers to OB-type stars and ``cool'' refers to GKM-type. Such combinations tend to occur as a consequence of stellar evolution, such as Algol, a B8V+K0IV+A5V triple system \citep{Lestrade1993}. \citetalias{Gunderson2025b} used this to suggest $\rho$\,Oph\,B is also of the ``Demon Star'' type, in reference to Algol's moniker, but this failed to account for an important factor: the age of the stars.

Even though the B-type stars in $\rho$\,Oph are well into their main sequence phase, cool stars would not yet have reached the Zero-Age Main Sequence (ZAMS). This would mean $\rho$\,Oph\,B consists of a hot star and a PMS star, an even more important configuration to study due to their extreme rarity \citep[see statistics for finding companions in B star surveys by][]{Gullikson2016}.

While most stars in the Galaxy are in multiple star systems, hot, massive stars in particular show high rates of forming in binaries \citep{Moe2025}. There is growing evidence that current unanswered questions can be explained by OB stars evolving in binary pairs (e.g., $\gamma$\,Cas stars \citep{Tsujimoto2018,Tsujimoto2023,Gunderson2025c,Naze2026}, hot subdwarfs \citep{Pelisoli2020}, and runaway massive stars \citep{Renzo2019}). It should be noted that OB binaries tend to form in binaries of similar spectral type instead of with a variety of spectral types \citep{Gullikson2016}.

It is surprising then that a second hot+PMS system might be in $\rho$\,Oph. Using interferometric imaging in near-infrared wavelengths, \citet{Frost2025} found evidence that \rOC\ is a \textit{trinary} system with two low-mass companions with mass ratios of $\mu_1 = 0.216^{+0.006}_{-0.004}$ and $\mu_2=0.083\pm0.006$. The most recent determination of the primary B-type star's mass, from extensive ESPaDOnS data, by \citet{Shultz2026} is $M_\mathrm{B}=4.8\pm0.4\,M_\odot$, meaning that the companions are only $M_1=1.03\pm0.09$ and $M_2=0.40\pm0.11$. At the age of the the system \citep[7\,Myr;][]{Shultz2026}, neither of these stars would be ZAMS yet.

Such a system combination suggests significant ramifications for the evolution of the PMSs. The accretion disks of PMS stars are known to be chemically altered by UV light from external sources, but most cases of this process are assumed to be from distant sources within the same star forming cluster \citep{Berne2024,Ndugu2024}. The projected separations of the two PMS stars \Change{from the B star} are only $r_1 = 0.93\pm0.01$\,AU and $r_2 = 11.54\pm0.04$\,AU. The primary B star could photoevaporate the accretion disks since the incident flux will be substantially higher than other intracluster sources. At the same time, the intense wind from the B star \citep[for typical wind mass-loss rates and terminal velocities, see][]{Pradhan2023} provides an additional mechanical means of stripping the accretion disk.

This makes \rOC\ an important source for understanding the chemical evolution of PMS disks in the presence of UV flux along with the evolution of hot+cool star binaries under different conditions. However, such investigations would not be easy to conduct due to the variability in the system. Low-mass stars, whether PMS or ZAMS are known to be variable across the electromagnetic spectra. The primary, ubiquitous feature is their coronal flaring \citep{Feinstein2024}, which can be solely an X-ray phenomenon or extend to all wavelengths through so-called white light flares. Low-mass stars can also show more coherent modulation due to star spots changing the local flux as the star rotates. This is to say nothing of potential interactions between magnetospheres (more on this in \S~\ref{sec:Radio}).

This is in addition to well-known coherent variability of the primary B star itself.  The primary B star is an archetypical example of a magnetic massive star. Magnetic early-type stars are known to have periodic and repeatable light curves across a wide range of wavelengths that are modulated on the stellar rotation period. For about 25\% of cases, this photometric variability is seen at radio wavelengths \citep{Drake:1987, Linsky:1992, Leone:1994, Babcock:1949, Stibbs:1950}. Three dimensional modeling of incoherent gyro-synchrotron radiation from non-thermal electrons \citep{Trigilio:2004, Leto:2006} interacting with a magnetosphere can reproduce the periodic light curves and flat spectral energy distributions (SEDs) that are observed over a wide range of frequencies \citep[$0.1-200\,$GHz;][]{Townsend2005,Leto:2021, Leto:2026}, including for \rOC\ \citep{Leto2020}. 


\rOC\ has so far been well-described by the centrifugal breakout (CBO) model, in which the forces from a centrifugally constrained magnetosphere exceed magnetic pressure, causing the field lines to temporarily break open in a reconnection event \citep{Owocki:2022}. The resulting current sheet accelerates electrons \citep{Usov:1992} that then produce radio emission. CBO events have been observed in other rapidly-rotating ($P_{\rm rot}\lesssim2$\,days) BA-type systems characterized by weaker winds, similar to \rOC. Based on the extensive modeling from \citet{Leto2020} and \citet{Shultz2026}, \rOC's optical and radio emission show periodicity that matches well with CBO theory. However, these same works have shown  additional unexplained variability beyond the primary rotation period. 



Confirmation of the presence of the trinary companions proposed by \citet{Frost2025} may be able to explain these additional signals. We address this hypothesis using a multiwavelength approach. It is also our second paper in a broader effort by the MACRO consortium to conduct a comprehensive, multiwavelength survey of the young stellar population of $\rho$\,Oph.

The paper is organized as follows. In \S~\ref{sec:X-rayObs}, we examine the X-ray data. In \S~\ref{sec:Infrared}, we present the near infrared data. In \S~\ref{sec:Optical}, we analyze recent TESS and serendipitous Chandra Aspect Camera Assembly data. In \S~\ref{sec:Radio}, we detail the radio data of the source. Finally, in \S~\ref{sec:Conclusions}, we discuss implications for the \rOC\ system and summarize our conclusions.

\section{Data Reduction} \label{sec:DataReduction}

A summary of the observations from public observatories used in our analysis is given in Table~\ref{tab:ObsIds}. The Chandra data were reprocessed with the standard pipeline in \textsc{ciao} version 4.17 \citep{Fruscione2006}. NuSTAR data were processed with the standard pipeline in the \textsc{heasoft} version 6.34. XMM data were processed using \textsc{sas} version 20.0.0 \citep{Gabriel2004}.

\begin{deluxetable}{llllcl}
    \tablecaption{$\rho$\,Oph C Data Information.\label{tab:ObsIds}}
    \tablehead{
        \colhead{Wavelengths} & \colhead{Observatory} & \colhead{Obs Id} & \colhead{Start Date} & \colhead{Exposure time (ks)} & \colhead{Emission State}
    }
    \startdata
        Radio & VLA & NVSS (AC496) & 1997-09-27 & 0.17 & ... \\
        & & VLASS1.2 & 2019-06-29 & 0.03 & ... \\
        & & 19A-446 & 2019-08-26 & 0.187 & ... \\
        & & VLASS2.2 & 2022-02-08 & 0.03 & ... \\
        & & VLASS3.2 & 2024-09-24 & 0.03 & ... \\
        & ASKAP & RACS (AS110) & 2019-04-25 & 0.9 & ... \\
        & & RACS (AS110) & 2021-01-18 & 0.9 & ... \\
        & & RACS (AS110) & 2022-01-20 & 0.9 & ... \\
        & & RACS (AS110) & 2022-04-11 & 0.9 & ... \\
        & & WALLABY (AS202) & 2023-09-07 & 28.8 & ... \\
        & & RACS (AS110) & 2024-01-18 & 0.9 & ... \\
        & & RACS (AS110) & 2024-11-16 & 0.9 & ... \\
        & & WALLABY (AS202) & 2025-04-24 & 28.8 & ... \\
        & & WALLABY (AS202) & 2025-05-11 & 28.8 & ... \\
        & & FLASH (AS209) & 2025-10-08 & 7.2 & ... \\
        NIR & SOAR/TripleSpec & SO2025B-001 & 2025-08-13 & 0.005 & ... \\
        Optical & TESS & 203822414 & 2025-04-09 & ... & ... \\ 
        X-ray & XMM & 0720690101 & 2013-08-29 & 53 & Quiescent and Flares\\
        & & 0760900101 & 2016-02-22 & 141.9 & Low\\
        & & 0870920101 & 2020-09-13 14:26:50& 79 & Quiescent\\
        & NuSTAR & 30601025002 & 2020-09-13 03:06:12 & 136.33 & Flare\\
        & Chandra & 24759 & 2022-03-21 & 18.18 & Quiescent\\
        & & 24760 & 2022-03-22 & 28.58 & Quiescent\\
        & & 26367 & 2022-03-26 & 9.93 & Quiescent\\
        & & 24763 & 2022-05-03 12:50:14 & 12.90 & Flare\\
        & & 26406 & 2022-05-03 22:09:19 & 16.84 & Flare\\
        & & 24672 & 2022-05-08 & 24.73 & Flare\\
        & & 26416 & 2022-05-12 & 24.53 & Flare\\
        & & 24761 & 2022-07-21 & 27.20 & Flare\\
        & & 24762 & 2022-07-24 & 10.92 & Quiescent\\
        & & 26481 & 2022-07-30 & 17.82 & Quiescent\\
    \enddata
\end{deluxetable}

Near-infrared spectroscopic observations were carried out with the TripleSpec\,4.1 NIR Imaging Spectrograph \citep[TripleSpec,][]{Schlawinl14} at the Southern Astrophysical Research Telescope (SOAR, Cerro Pach\'on, Chile). TripleSpec is a cross-dispersed spectrograph with a fixed {1.1$\times$28\arcsec} slit,  providing full spectral coverage from 0.94 to 2.47\,{\micron} (R$\sim$3,500), encompassing the entire YJHK photometric range.

The TripleSpec/SOAR data were processed using a modified version of the IDL-based \texttt{Spextool} pipeline \citep{Cushing2004} for use at SOAR. The data reduction steps include flat-field correction, wavelength calibration using a CuHeAr arc lamp, removal of emission sky features by subtracting A and B exposures, and extraction of the one-dimensional spectra. We used observations of the telluric standard star HIP\,79229 (Ks = 6.5 mag), observed at similar airmass values, to perform the telluric correction and flux calibration of the spectrum.

Finally, we analyzed radio observations of the $\rho$\,Oph field from the Karl G. Jansky Very Large Array (VLA). The nearby star $\rho$\,Oph\,A was targeted in L-band (1-2\,GHz) with the VLA in the most extended A-configuration, with a 186\,s on-source integration time under program VLA/19A-446 (PI: J. Ling); however, the compact synthesized beam ($\theta_{\rm beam}\approx1.3\arcsec$ for A-configuration) and the primary beam's field-of-view ($\theta_{\rm pb}\approx28\arcmin$ for L-band) also afford a serendipitous measurement of \rOC's flux. We retrieved the calibrated visibilities from the NRAO Data Archive and imaged \Change{them} using CASA (v6.7.2.42, \citealt{Casa2022}) with standard {\tt tclean} parameters for wideband imaging to \Change{produce} Stokes I and V images. \Change{Since \rOC\ was not located at the phase tracking center of these observations, we corrected the images for primary beam attenuation using the CASA task {\tt pbcor}.} Due to the relatively compact synthesized beam in A-configuration and to minimize compute time, it was necessary to place ``outlier fields" centered on several nearby bright sources to approach the theoretical sensitivity limits across the image. These sources are NVSS~J162331-231334 ($F_{1.4 {\rm GHz}}=0.1613\pm0.0049$\,Jy; $\Delta\theta_{\rm \rOC}=30.4\arcmin$), PKS~1623-22 ($F_{1.4 {\rm GHz}}=1.2458\pm0.0374$\,Jy; $\Delta\theta_{\rm \rOC}=31.5\arcmin$), and PKS~1617-235 ($F_{1.4 {\rm GHz}}=1.2201\pm0.039$\,Jy; $\Delta\theta_{\rm \rOC}=65.5\arcmin$; \citealt{Condon:1998}). The brightness of $\rho$\,Oph\,A and \rOC\ allowed for self calibration of the field to improve the overall image fidelity at the expense of absolute astrometric calibration, so we applied a single round of phase-only self calibration using a 16-second solution interval to the Stokes I image. When using this approach, the mean RMS across the full-bandwidth Stokes I and V images was $42\,\mu$Jy. 

In addition to the targeted L-band (1-2\,GHz) pointing from VLA/19A-446, we also investigated various radio survey data products. We extracted fluxes from the Stokes I Quicklook images of the VLA Sky Survey (VLASS, \citealt{Lacy:2020}), which observed the field of $\rho$\,Oph on three occasions in S-band (2-4\,GHz) from 2019 to 2024. We also checked the NRAO VLA Sky Survey (NVSS, \citealt{Condon:1998}), which observed in a narrower L-band bandwidth centered at 1.4\,GHz in 1997. Finally, we investigated the public data archive of Australian Square Kilometer Array Pathfinder (ASKAP) and found that \rOC\ was observed ten times between 2019 and 2025 at frequencies ranging from 855.5\,MHz to 1655.5\,MHz via three separate surveys including RACS (887.5\,MHz, 943.5\,MHz, 1367.5\,MHz, 1655.5\,MHz with 15-minute integration time; \citealt{McConnell:2020}), FLASH (855.5\,MHz with 2-hour integration time; \citealt{Allison:2022}), and WALLABY (1367.5\,MHz with 8-hour integration time; \citealt{Koribalski:2020}), all of which had Stokes I and V continuum images available, except for the first epoch of RACS in 2019 observing at 887.5\,MHz. In all cases, fluxes were extracted as the amplitude of a 2D elliptical gaussian fit to the image at the position of \rOC. Fit fidelity was evaluated by comparing morphological parameters (size and orientation) to the synthesized beam and by inspecting the residual image. \Change{All derived fluxes are reported in \autoref{table:radio-fluxes}.}

\section{X-ray Observations}\label{sec:X-rayObs}
\subsection{Light curves} \label{sec:X-raylc}

\begin{figure*}
    \centering
    \includegraphics[width=\linewidth]{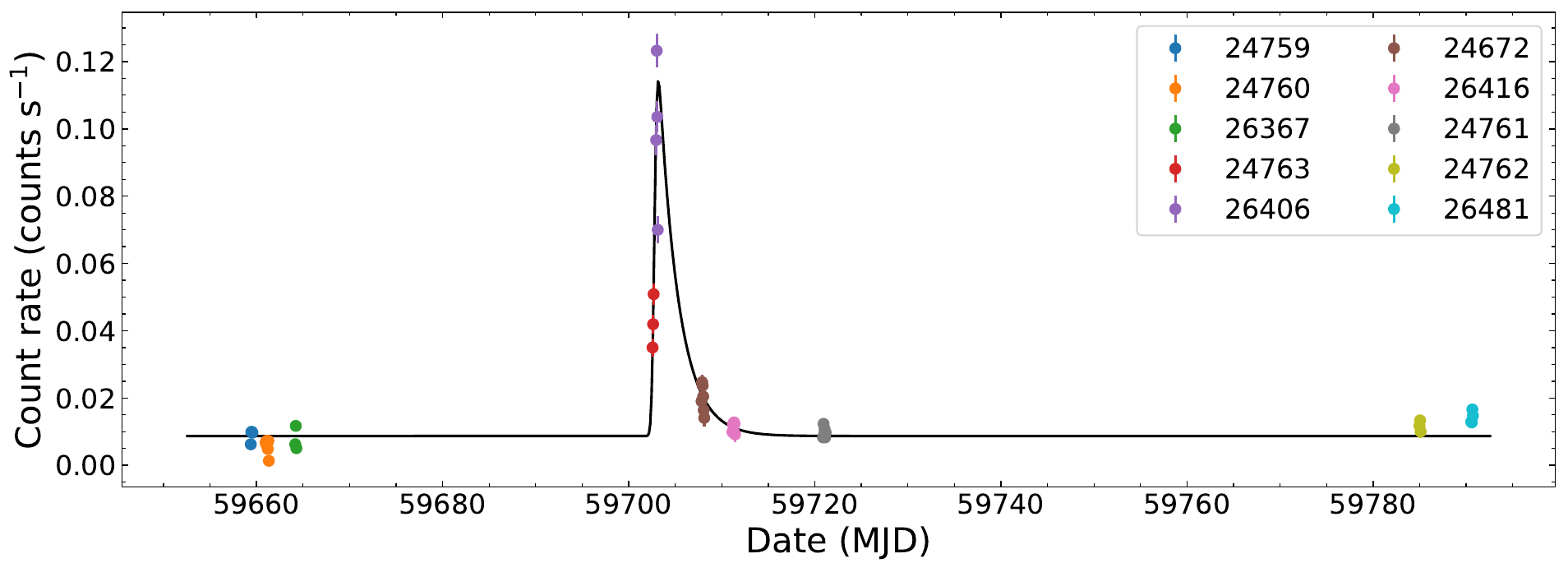}\\
    \includegraphics[width=\linewidth]{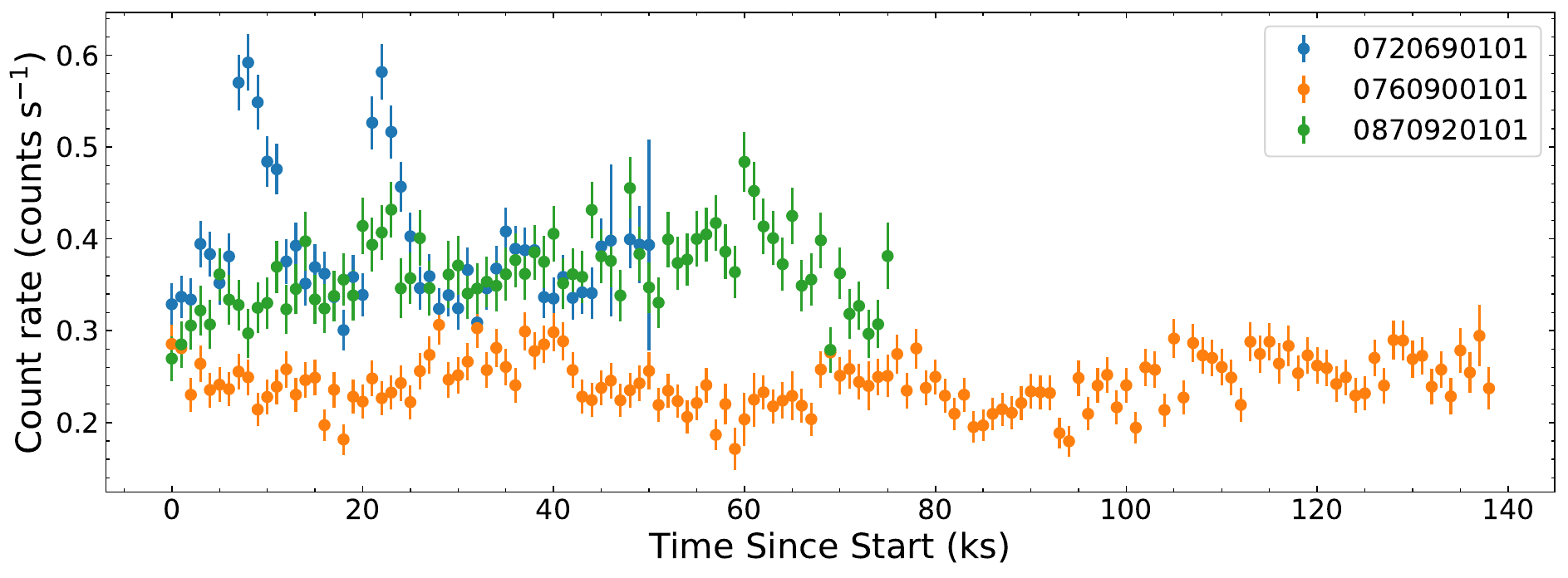}\\
    \includegraphics[width=\linewidth]{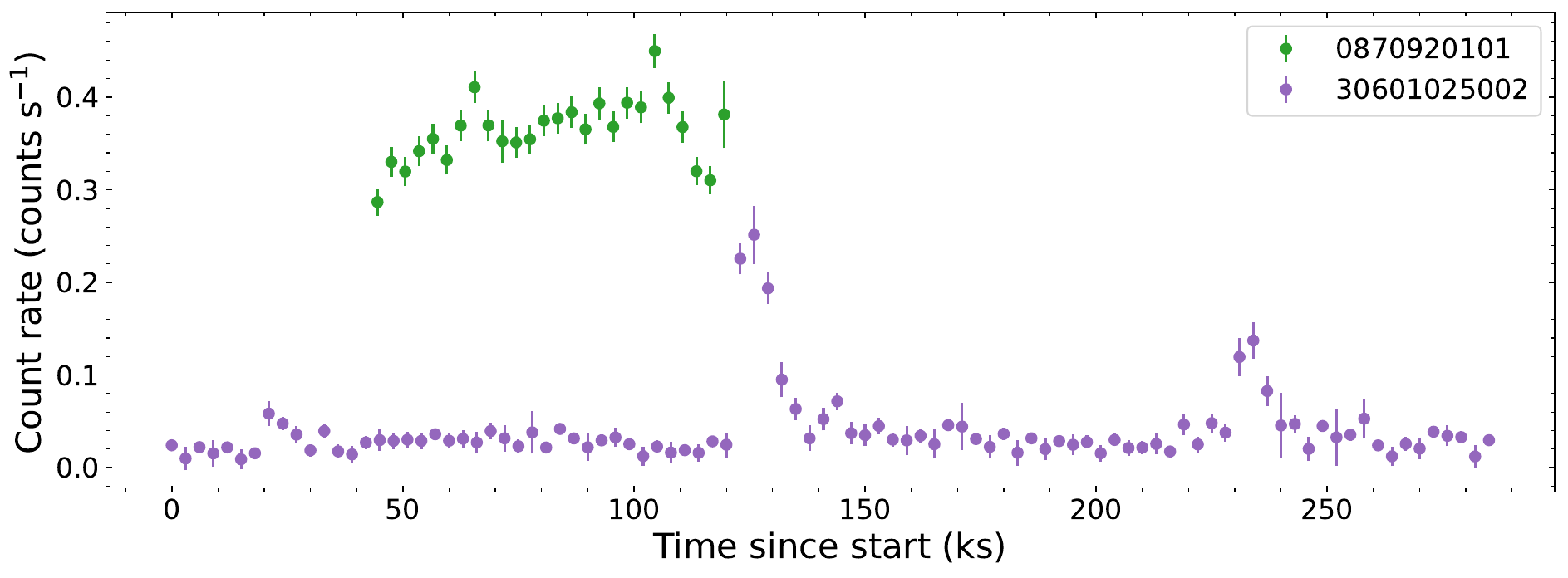}
    \caption{\textbf{Top panel:} Chandra zeroth order light curves with 5\,ks bins. \textbf{Middle panel:} XMM PN light curves with 1\,ks bins plotted as time since the start of each observation. \textbf{Bottom panel:} Simultaneous XMM PN (green) and NuSTAR FPMA (purple) light curves with 3\,ks bins plotted as time since start of the NuSTAR observation.}
    \label{fig:lightcurves}
\end{figure*}

There is a surprising level of X-ray variability within and between observations. Figure~\ref{fig:lightcurves} shows the X-ray light curves from Chandra (top), XMM (middle), and NuSTAR (bottom, with the simultaneous XMM data). As a presumed single magnetic B star system, \rOC\ displays an unusually high level of variability. In classical magnetic massive stars, X-rays are produced through magnetically channeled winds shocking at the magnetic equator in a largely stable process \citep{Townsend2005,Wade2006}. Most X-ray emission variability arises from cyclical changes in the observer's line-of-sight owing to rotational or orbital motion \citep[e.g., $\theta^1$\,Ori\,C,][]{Gagne1997}. \Change{The origin of X-rays in a CBO system is potentially similar, though there are differences in X-ray luminosity that suggest a separate and/or additional mechanism is operating \citep{Owocki:2022}; this will be discussed in more detail in \S~\ref{sec:CBOXrays}.}

In contrast, the light curves in Figure~\ref{fig:lightcurves} show features that are morphologically similar to the coronal activity of cool stars \citep[e.g., the varied flares in EV\,Lac;][]{Huenemoerder2010}. This includes classical flares, small-scale outbursts, and epochal changes in X-ray flux. The Chandra light curve highlights this the most with small scale fluctuations in the quiescent times along with a substantial flare-like event that lasted for $\sim$20 days.

To characterize the properties of the flare event, we employ a simplified model from \citet{Tovar2022}. The flare is taken to have a brightness given by
\begin{equation}
    f(t) = \int_{-\infty}^t g(t')h(t-t')dt' + C,
\end{equation}
where the impulsive heating of the corona is described using a Gaussian 
\begin{equation}
    g(t') = K \exp\left[{-\frac{(t'-t_\mathrm{peak})^2}{2\tau_\mathrm{heat}^2}}\right]
\end{equation}
and the cooling is assumed to be exponential
\begin{equation}
    h(t') = \exp\left(-\frac{t'}{\tau_\mathrm{cool}}\right).
\end{equation}
As described by \citet{Tovar2022}, the convolution of the heating and cooling terms accounts for both processes occurring simultaneously. It also allows for an analytic solution with
\begin{eqnarray}
    f(t) = &&\sqrt{\frac{\pi}{2}}K\tau_\mathrm{heat}\exp\left[\frac{t_\mathrm{peak}-t}{\tau_\mathrm{cool}}+\frac{1}{2}\left(\frac{\tau_\mathrm{heat}}{\tau_\mathrm{cool}}\right)^2\right]\times\nonumber\\
    &&\left[1-\mathrm{erf}\left(\frac{1}{\sqrt{2}}\frac{t_\mathrm{peak}-t}{\tau_\mathrm{heat}}+\frac{\tau_\mathrm{heat}}{\sqrt{2}\tau_\mathrm{cool}}\right)\right] + C.\label{eq:FlareModel}
\end{eqnarray}
The free parameters of this model are the flare amplitude \Change{$K$}, time of peak flare $t_\mathrm{peak}$, heating time scale $\tau_\mathrm{heat}$, cooling time scale $\tau_\mathrm{cool}$, and the quiescent X-ray emission $C$. It is also possible to include a second cooling term in the convolution to describe both fast and slow cooling; however, the Chandra data sampling is too sparse to constrain two cooling models.

The best fit of Equation~\eqref{eq:FlareModel} is also shown in Figure~\ref{fig:lightcurves} as the black line. The best fit parameters are given in Table~\ref{tab:FlareMod}. One important caveat of our best fit is the assumption that the flare is a single event. The sparsity of the data may mean that we have captured different portions of multiple flares that simply appear to be one by chance. As a sanity check on the time scales, we can compare against the flare decay times measured from the COUP survey of the Orion Nebula \citep{Favata2005}, which \Change{have a minimum to maximum measured range of $0.02 - 4.59$\,days}. Our measured timescale is well within the this range, so it is reasonable to interpret the Chandra data using a single-flare model.

\begin{deluxetable}{ccc}
    \tablecaption{Chandra Flare Model.\label{tab:FlareMod}}
    \tablehead{
        \colhead{Parameter} & \colhead{Value} & \colhead{Units}
    }
    \startdata
        $K$ & $0.2768 \pm 0.0395$ & Counts\,s$^{-1}$\,day$^{-1}$ \\
        $t_\mathrm{peak}$ & $59702.3273 \pm 0.0690$ & MJD \\
        $\tau_\mathrm{heat}$ & $0.2388 \pm 0.0853$ & day \\
        $\tau_\mathrm{cool}$ & $0.4701 \pm 0.0583$ & day \\
        $C$ & $0.0087 \pm 0.0005$ & Counts\,s$^{-1}$ \\
    \enddata
\end{deluxetable}

\subsection{Spectral Fits}\label{sec:SpecFits}

Operating under the assumption that the X-ray emission from \rOC\ is coming from at least one cool-star component, we employed a physically motivated differential emission distribution (DEM) model.
Following the work by \citet{Gudel2004}, the DEM from flare heated corona can be described using a double power-law distribution
\begin{equation}
    \mathrm{DEM}(T)\propto\begin{cases}
        (T/T_\mathrm{peak})^\alpha & T \leq T_\mathrm{peak}\\
        (T/T_\mathrm{peak})^\beta & T > T_\mathrm{peak},
    \end{cases}\label{eq:DEM}
\end{equation}
where $T_\mathrm{peak}$ is the peak temperature of the corona, $\alpha > -1$ describes the coronal heating, and $\beta < -1$ describes the coronal cooling. This is a simplification of the many-parameter version described in \citet{Gudel2004}, but with fewer parameters we can reduce model degeneracies. Reconstructed DEM models produce a similar shaped distribution that can be described with only these three parameters \citep[e.g.,][]{Huenemoerder2013}.

\begin{figure*}
    \centering
    \includegraphics[width=\linewidth]{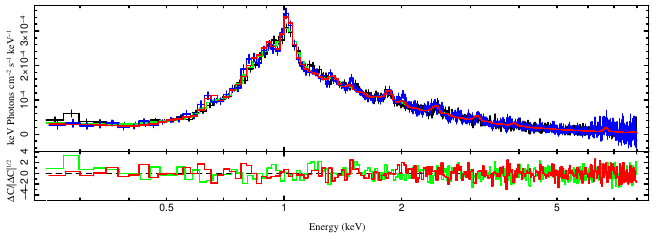}
    \includegraphics[width=\linewidth]{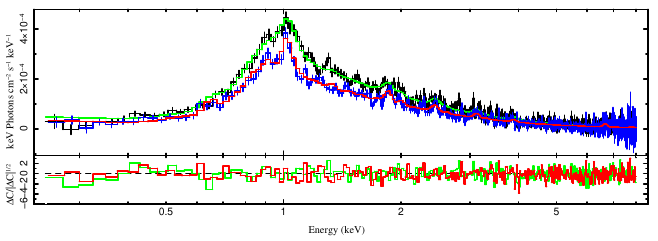}
    \caption{Top Panel: XMM PN (black) and MOS1+2 (blue) low state with best fitting DEM model (green and red). Both data and model are binned by a constant factor of 5 (bin size of 25\,eV). Bottom Panel: XMM PN (black) and MOS1+2 (blue) quiescent state with best fitting DEM model (green and red). Both data and model are binned by a constant factor of 5 (bin size of 25\,eV).}
    \label{fig:XraySpec-XMM}
\end{figure*}

\begin{figure*}
    \centering
    \includegraphics[width=\linewidth]{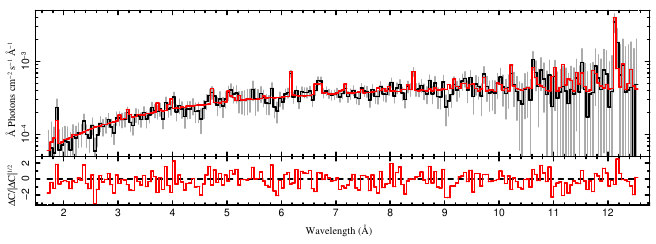}
    \includegraphics[width=\linewidth]{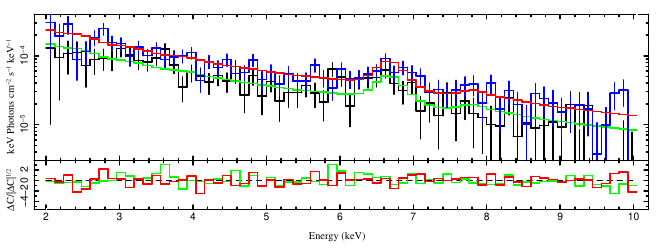}
    \caption{Top Panel: Total Chandra data (black) with best fitting model (red), binned by a constant factor of 10 (bin size of 0.05\,\AA). Bottom Panel: NuSTAR FPMA (black) and FPMB (blue) with best fitting model (green and red). Both spectra are binned by a constant factor of 3 (bin size of 0.12\,keV).}
    \label{fig:XraySpec-ChandraNuSTAR}
\end{figure*}

Our best fitting models are shown in Figures~\ref{fig:XraySpec-XMM} and \ref{fig:XraySpec-ChandraNuSTAR} with corresponding parameter values in Table~\ref{tab:XrayModelFits}. \Change{Additionally, we computed unabsorbed X-ray fluxes $\mathcal{F}_X$ and luminosities $L_X$ for each model along with a predicted radio flux $\mathcal{F}_R$.} The data were split into four categories based on the light curves discussed above: XMM Low (ObsID 0760900101),  XMM Quiescence (ObsId 0870920101), Chandra Flare, and NuSTAR Flare. The latter two use all observations made by those two telescopes. \Change{While we could fit the Chandra quiescent state data separately, the quiescent Chandra observations provide no additional information over the XMM data. We instead combined all the Chandra data to fit the median spectrum, which is dominated by the flare.} These four cases cover the breadth of emission states of \rOC\ for detailed comparison. \Change{The energy ranges fit in these data sets were $0.2-8$\,keV for XMM, $1-7.3$\,keV ($1.7-12.5$\,\AA) for Chandra, and $2-10$\,keV for NuSTAR.}

\begin{deluxetable}{c|c|c|c|c|c}
    \tablecaption{Plasma model best fit parameters with 68 percent errors\label{tab:XrayModelFits}}
    \tablehead{
        \colhead{Parameters} & \multicolumn{4}{c}{Value} & \colhead{Unit}\\
        \hline
        & XMM & XMM & Chandra & NuSTAR \\
        & Low & Quiescence & Flare & Flare
    }
    \startdata
        $T_\mathrm{peak}$ & $9.85_{-0.67}^{+0.85}$ & $10.81_{-1.05}^{+1.91}$ & $59.39_{-11.05}^{+10.72}$ & $54.60_{-8.79}^{+8.52}$ & MK \\
        $\alpha$ & $-0.31_{-0.13}^{+0.15}$ & $-0.32_{-0.15}^{+0.80}$ & $-0.28_{-0.24}^{+0.31}$ & $-0.30$ &  \\
        $\beta$ & $-2.50_{-0.08}^{+0.07}$ & $-2.61_{-0.12}^{+0.11}$ & $-5.63_{-2.67}^{+1.93}$ & $-6.13_{-2.51}^{+2.17}$ & \\
        Norm$\times10^{-3}$ & $2.22_{-0.26}^{+0.30}$ & $3.41_{-0.05}^{+0.08}$ & $3.02_{-0.48}^{+0.75}$ & $1.14_{-0.14}^{+0.16}$ & $10^{-14}\;\mathrm{VEM}/{4\pi d^2}$ \\
        $A(\mathrm{Ne})$ & $1.10_{-0.08}^{+0.08}$ & $1.05_{-0.15}^{+0.16}$ & $1.15_{-0.25}^{+0.29}$ & ... &   \\
        $A(\mathrm{Mg})$ & $0.26_{-0.04}^{+0.05}$ & $0.18_{-0.05}^{+0.06}$ & $0.25_{-0.06}^{+0.07}$ & ... &  \\
        $A(\mathrm{Si})$ & $0.22_{-0.03}^{+0.03}$ & $0.26_{-0.04}^{+0.04}$ & $0.35_{-0.06}^{+0.07}$ & ... &  \\
        $A(\mathrm{S})$ & $0.29_{-0.06}^{+0.05}$ & $0.29_{-0.06}^{+0.07}$ & $0.39_{-0.06}^{+0.08}$ & ... & \\
        $A(\mathrm{Fe})$ & $0.24_{-0.02}^{+0.02}$ & $0.23_{-0.03}^{+0.03}$ & $0.19_{-0.04}^{+0.04}$ & $0.44_{-0.08}^{+0.09}$ &  \\
        $N_\mathrm{H}$ & $0.590_{-0.018}^{+0.017}$ & $0.599_{-0.006}^{+0.011}$ & $0.305_{-0.074}^{+0.072}$ & ... & $10^{22}\,\mathrm{cm}^{-2}$\\
        \hline
        $\mathcal{F}_X$ & $3.03$ & $4.55$ & $4.79$ & $1.95$ & $10^{-12}$\,ergs\,cm$^{-2}$\,s$^{-1}$\\
        $L_X$ & $7.00$ & $10.52$ & $11.07$ & $4.51$ & $10^{30}$\,ergs\,s$^{-1}$\\
        $\mathcal{F}_R$ & $0.58$ & $0.87$ & $0.92$ & $0.37$ &  mJy\\
        \hline
    \enddata
    \tablecomments{Solar Abundances are from \citet{Anders1989}. Reported X-ray fluxes are unabsorbed in the 0.2--12\,keV band. X-ray luminosities assume a distance of $d=139$\,pc \citep{Shultz2026}. Radio fluxes calculated assuming a G\"{u}del-Benz relation at 5\,GHz \citep{Gudel1993,Benz1994}.}
\end{deluxetable}

Model fits were performed using the Interactive Spectral Interpretation System \citep[\textsc{isis;}][]{Houck2000}. Poisson statistics were adopted and a Cash fit statistic \citep{Cash1979} used. We used Markov chain Monte Carlo routines for minimizing the Cash statistic. Best fit parameters were extracted from the resulting chains after the fit statistic stabilized. The total model used for the fit was the broken power law DEM from Equation~\eqref{eq:DEM} multiplied by a \texttt{phabs} absorption model\footnote{\url{https://heasarc.gsfc.nasa.gov/docs/software/xspec/manual/node274.html}}.

Focusing first on the XMM Low state, this is shown in the top panel of Figure~\ref{fig:XraySpec-XMM}. Here we used the PN and Metal Oxide Semiconductor (MOS) detectors. Due to \rOC\ being an off axis source from the observation targets ($\rho$\,Oph\,A and B), it was not detected in the Reflection Grating Spectrometer. This limited our ability to constrain low energy elemental abundances due to the limited spectral resolution of the PN and MOS. Only the abundances of Ne, Mg, Si, S, and Fe were elected to be constrained.

The best fit abundances show further evidence of at least one cool star being present in the system. The over abundance of Ne compared to other metals is consistent with a first-ionization potential effect observed in cool star coronal emission \citep{Huenemoerder2013}. A similar determination was made in \citetalias{Gunderson2025b} for $\rho$\,Oph\,B. By contrast, magnetically confined wind shocks in OB stars show strong emission lines from all metals \citep{Pradhan2023}, which is not seen in our best-fit modeling. Finally, it is worth noting that OB stars can be described with a much simpler DEM model consisting of only a single power law slope \citep{Huenemoerder2020}.\footnote{An exception to this are the so called ``weak-wind'' stars like $\mu$\,Col \citep{Huenemoerder2012}, but these are unique cases that are still poorly understood.}

These are solidified in the XMM Quiescence fit, shown in the bottom panel in Figure~\ref{fig:XraySpec-XMM}. The best fit parameters are found to be statistically the same as in the XMM Low case, with larger error bars due to the difference in observation length. It should also be noted that due to differences in effective exposure time for the PN and MOS, this observation encountered a difference in observed flux between the PN and MOS. We accounted for this through an overall cross calibration constant applied to the MOS instantiation of the model to ensure the applied models gave the same flux for the MOS and PN.

The XMM Quiescence parameters indicate that the source of the higher X-ray counts in Figure~\ref{fig:lightcurves} derived from an increase in the emission measure (EM) of the emitting plasma. Such effects can be explained through a prior flare that has cooled to nominal coronal temperatures but the amount or volume of hot gas has not settled to pre-flare levels again. Similar events have been observed in Algol where the post-flare plasma stays at a high EM but is no longer as hot \citep{Nordon2007}.

In comparison, the Chandra Flare in the top panel of Figure~\ref{fig:XraySpec-ChandraNuSTAR} shows significant changes in the plasma peak temperature, $\beta$, and H column density $N_\mathrm{H}$. The latter can be explained by the bandpass used in the Chandra data. Due to the loss of the soft portion of Chandra's ACIS detector from contamination \citep{Marshall2004,Odell2017}, we were limited to $\lambda<12.5$\,\AA\ in the Chandra spectrum. This is soft enough that absorption was required when conducting model fits but limits the full diagnostic potential of the parameter. The resulting measure of  $N_H\approx 0.305\times10^{22}$\,cm$^{-2}$ is thus likely underestimated.

\begin{figure}
    \centering
    \includegraphics[width=\linewidth]{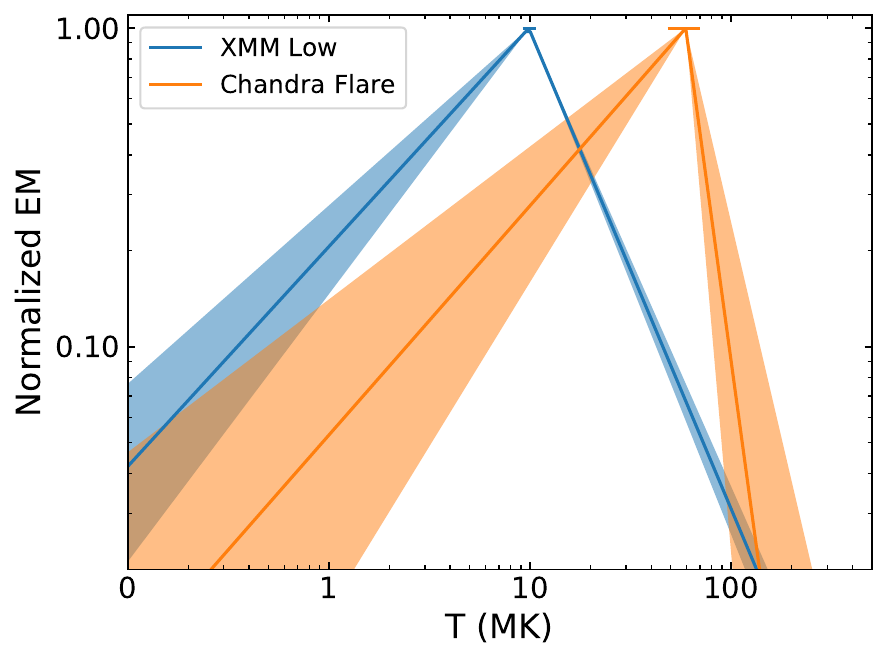}
    \caption{Comparison of the normalized EM distributions for the XMM Low and Chandra Flare states. Shaded regions correspond to the 68\% uncertainties quoted in Table~\ref{tab:XrayModelFits}. The horizontal error bars represent the uncertainty in the maximum flare temperature.}
    \label{fig:EMComparison}
\end{figure}

The peak temperature, however, is a major change in the system as expected for a flare. As noted before, our treatment of the Chandra data as corresponding to a single flare event means that this temperature is an average peak flare temperature for however many events actually occurred within the 20 days. Of the parameter changes, the most intriguing is that of $\beta$. Even during the flare, the slope $\alpha$ stays statistically constant yet $\beta$ goes through a significant change. This appears concerning, as if we are within a regime of parameter space that is unphysical but gives good model fits, until this change is considered alongside the maximum temperature. Consider the EM distribution,
\begin{eqnarray}
    \mathrm{EM}(T)&&=\int_0^T \mathrm{DEM}(T')dT'\nonumber\\
    &&\propto \begin{cases}
        (T/T_\mathrm{peak})^{\alpha+1} & T\leq T_\mathrm{peak}\\
        (T/T_\mathrm{peak})^{\beta+1} & T > T_\mathrm{peak}
    \end{cases},
\end{eqnarray}
where integration constants are ignored to focus on the slopes of the distribution. In Figure~\ref{fig:EMComparison}, we show $\mathrm{EM}(T)$ for the XMM Low and Chandra Flare model fits. Here we can see why $\beta$ changes to such a steep distribution. Since the peak temperature of the corona increased, the $\beta$ parameter had to become steeper to reduce the amount of $T>100$\,MK plasma.

Our final model fit is to the NuSTAR data. While the flares caught in the NuSTAR data are a small portion of the overall observation, the significant change in count rate and limited band pass of the instrument results in the flares dominating the spectrum. We thus fit the NuSTAR spectrum alone to get a secondary flare model under the presumption that we will see a different characteristic set of parameters from the Chandra flare. This fit required a significant change in routine due to the aforementioned band pass. Since NuSTAR is sensitive only down to ${\sim}2$\,keV, $N_\mathrm{H}$ is not constrainable, so we did not include any absorption. This also meant that abundances of Ne and Mg were not constrainable, plus the limited spectral resolution makes Si and S difficult to fit. We thus only allowed Fe to be a free abundance. Finally, we fixed $\alpha=-0.30$ since the lack of low energies makes it difficult to constrain the slope before the peak temperature. Since $\alpha$ does not change between non-flare and flare states, it is reasonable to assume that even during the NuSTAR flares we would find the same parameter.

Some of these difficulties could be alleviated by fitting the simultaneous XMM observation with the NuSTAR spectra. However, since the XMM data does not occur at the same time as the NuSTAR flare, they represent two different spectra. Fitting the NuSTAR data alone has the added benefit of comparing pre-flare and post-flare plasmas.


The NuSTAR model fits are given in the bottom panel of Figure~\ref{fig:XraySpec-ChandraNuSTAR}. The overall model fit is identical to the Chandra case, showing that the flares are characteristically similar. However, the normalization is significantly different than the Chandra flare. These changes are physically interesting given the overall timescale difference of the flares in question. The NuSTAR flares, of which there appear to be three along with reheating events, are much shorter in duration, so while they reach a similar temperature as the Chandra flare, the overall amount of hot plasma is less. It is concerning, though, that the normalization is half of the XMM Low state despite covering a flare. There is a systematic uncertainty caused by the limited NuSTAR band pass, but a similar effect was not seen in \citetalias{Gunderson2025b} when fitting the available NuSTAR data of $\rho$\,Oph\,B.

The change in the Fe abundance might also suggest further variability in the system. A flare from $\rho$\,Oph\,B modeled in \citetalias{Gunderson2025b} was proposed to have dredged up chromospheric material, changing the metal abundances in the corona. A similar effect may be signified by the Fe abundance here, but without high resolution line spectra after the flare, we cannot say for certain if this is the case. Alternatively, this might be a sign that we are looking at flares from two different stars. We noted in the introduction that our analysis is attempting to find spectral evidence of the two interferometric identified companion stars. While we cannot determine which of the two cool stars are causing the X-ray emission, this change could signify the flares observed by NuSTAR are from a different companion than the one observed by Chandra.

\subsection{X-ray Emission Line Fitting}

We considered the possibility of measuring line shifts and broadening due to velocity flows caused 
by a large flare.  Plasma bulk motions were detected in a large X-ray flare on HR~9024 by 
\citet{Argiroffi2019}.  However, \rOC\ was observed at about $2.5\,\mathrm{arcmin}$ off-
axis.  The image size (and hence dispersed spectral line widths) grows roughly quadratically with off-axis angle and has about doubled at the observed position of \rOC, meaning resolving power is about half the on-axis performance. 

We fit the positions and widths of the stronger lines (Si, Mg, and Ne), but did not detect Doppler shifts significantly different from zero (to about $100\,\mathrm{km\,s^{-1}}$ accuracy) or widths significantly broader than instrumental (with a threshold of about $1000\,\mathrm{km\,s^{-1}}$).  Line fluxes can be robustly measured and we give the results of fits in Table~\ref{tab:LineFlux}.  Lines were fit by folding Gaussian profiles through the instrument response. Line groups had relative positions constrained and a common width parameter.  Unresolved H-like doublets were modeled with two Gaussians with offsets and flux ratios constrained (though with the off-axis blurring, a single Gaussian would have been sufficient to obtain a flux). The continuum under the lines was derived from the plasma model fit to the  spectrum; this fit also included the line-of-sight column density so that the line fluxes are as-emitted.
\begin{deluxetable}{lcc}
  \tablecaption{$\rho$\,Oph C Line Fluxes.\label{tab:LineFlux}}
   \tablehead{
    \colhead{Feature} & 
    \colhead{$\lambda_0$} & 
    \colhead{$f_x$} \\
    &
    \colhead{[$\mathrm{\AA}$]}&
    \colhead{[$\mathrm{10^{-6}\,phot\,cm^2\,s^{-1}}$]}
  }
  \startdata
\eli{Fe}{25}& $1.86$& $3.88 \pm 1.22$\\
\eli{S}{16}& $4.73$& $1.62 \pm 0.64$\\
\eli{Si}{14}& $6.18$& $3.37 \pm 0.68$\\
\eli{Si}{13}& $6.65$& $2.75 \pm 0.65$\\
\eli{Si}{13}& $6.74$& $1.83 \pm 0.87$\\
\eli{Mg}{12}& $8.42$& $3.71 \pm 0.99$\\
\eli{Ne}{10}& $9.71$& $2.01 \pm 0.92$\\
\eli{Ne}{10}& $10.24$& $2.52 \pm 1.09$\\
\eli{Fe}{24}& $10.62$& $4.56 \pm 1.87$\\
\eli{Ne}{10}& $12.14$& $42.21 \pm 14.60$\\
  \enddata
  \tablecomments{Emission line fluxes for features with a
  signal-to-noise ratio $\ge 2$. 
}
\end{deluxetable}

\subsection{The CBO--X-ray Connection}\label{sec:CBOXrays}

Before continuing, we will now consider the alternate case that our measured X-rays (despite their spectroscopic similarities to cool stars) are coming from the magnetosphere of the B star. \citet{Owocki:2022} derived that the luminosity from a CBO event scales with the wind kinetic energy
\begin{equation}
    \frac{L_\mathrm{CBO}(p)}{L_\mathrm{wind}} = 2\eta_c^{1/p}\left(\frac{v_\mathrm{rot}}{v_\mathrm{\infty}}\right)^2,\label{eq:Lcbo}
\end{equation}
where $p$ is the multipole index, $v_\mathrm{rot}$ is the surface rotational velocity, \ChangeTwo{$v_\mathrm{\infty}$ is the wind terminal velocity}, $L_\mathrm{wind} = \dot{M}v_\infty^2/2$ is the wind kinetic energy, and
\begin{equation}
    \eta_c = \frac{B_\mathrm{eq}^2 R_*^2}{\dot{M}v_\mathrm{orb}}
\end{equation}
is the centrifugal magnetic confinement parameter, where \Change{$B_\mathrm{eq}$ is the equatorial magnetic field strength} and \ChangeTwo{$v_\mathrm{orb}$ is the Keplerian orbital velocity.} The dependence of the luminosity on $p$ makes it difficult for a precise measurement since we do not know the exact magnetic field topology. We can calculate a lower-bound, though, by assuming a simple dipole case $p=2.$ Using the stellar parameter values measured by \citet{Shultz2026}, \ChangeTwo{the X-ray luminosity from the B star would be $L_\mathrm{CBO}=47.2L_\mathrm{wind}\approx 4.5\times10^{32}$\,ergs\,s$^{-1}$.}

Compared to the unabsorbed luminosities given in Table~\ref{tab:XrayModelFits}, the CBO X-ray flux is an order of magnitude larger than both quiescence and flaring states. The discrepancy in the measured versus theoretical X-ray luminosity could mean that the magnetic topology is much more complex than a simple dipole. If we instead use our measured luminosity $L_\mathrm{X}\sim10^{31}$\,ergs\,s$^{-1}\approx L_\mathrm{CBO}$ to solve for the multipole index, we find that the index would need to be $p\approx3.86$ for the B star to be X-ray source.

\ChangeTwo{\citet{Shultz2026} used a moderately complex field of two harmonic components for modeling the magnetic field over the B star. Although a value of $p\approx3.86$ suggests the presence of a quadrapolar component to the field, the contribution from such a component to the overall magnetic energy would be comparatively minimal. It is thus reasonable to describe the star's magnetic topology using the $p=2$ dipolar approximation.\footnote{On average, O and early-B type magnetic stars have approximately dipolar fields \citep{Wade:2016,Petite2013}.}}

\Change{Under the assumption that the X-ray emission is from the CBO mechanism, our measured X-ray luminosity suggests an inefficient conversion of CBO energy to X-rays: $\varepsilon = L_X/L_\mathrm{CBO}\sim 0.1$. \citet{Owocki:2022} suggested a similar, albeit significantly smaller efficiency factor $\varepsilon\sim 10^{-8}$ for the radio regime. The physical explanation for this efficiency factor is difficult to pin down without knowing the process that is generating the X-rays in a CBO system. Three likely possibilities are (1) the gas localized around the breakout event being heated by the magnetic field reconnection; (2) particles energized by the reconnection event streaming up the magnetic field and shocking with the stellar atmosphere/sphere; or (3) gas ejected by breakout event(s) shocking with prior ejected gas. Options (1) and (2) could possibly be connected to the loss of energy in electrons due to collisions found by \citet{Das2025}.}

\Change{However, the above assumes that all of the observed X-rays are from the magnetic B-type component. Magnetic B star X-ray fluxes follow a distribution that is centered around a median value of $\log(L_X/L_\mathrm{Bol})\sim-7$, but there is considerable width to the distribution \citep[see Figure 2 of][]{Naze2014}. Based on the unabsorbed X-ray luminosity in Table~\ref{tab:XrayModelFits} and $\log(L_\mathrm{Bol}/L_\odot)=2.4$ from \citet{Shultz2026}, the scaling for the B star in \rOC\ would be $\log(L_X/L_\mathrm{Bol})=-5.14$. Based on \citet{Naze2014}'s survey, this would make the B-type component the brightest in X-rays (by luminosity ratio) magnetic massive star. More likely, the X-ray emission from the B-type star is only a small a component of the measured quiescent luminosity.}

\Change{One can, in theory, have multiple X-ray spectral components for each star in an unresolved system, but this is liable to create significant degeneracies. Both hot and cool star spectra are well described by multi-temperature thermal plasma models. Without priors to restrict the relative normalization of each component, appropriate parameter transformation could not be applied to reduce the degeneracies between normalizations and temperatures. We encountered this same modeling degeneracy for the three stars, hence why we did not have separate models for the three stars in \rOC.}

\Change{There is evidence for the B star providing some flux to system. Assuming the X-ray emission is from the closer companion, which we predict to have $L_\mathrm{Bol}\approx 0.57 L_\odot$ (see \S~\ref{sec:Infrared} for predicted stellar parameters of the cool stars), then the luminosity ratio is $\log(L_X/L_\mathrm{Bol})\approx-2.5$. This is within measured luminosity ratios for cool stars, but is at the high end of the distribution \citep{He2019}. The B star providing a measurable but spectrally ambiguous flux in the system would reduce the X-ray luminosity ratio to statistically more like values.}

\Change{Taken all together, the X-ray data are not only consistent with the presence of at least one cool star companion but require the presence of such cool stars to explain the observed temporal and spectral properties. B-type stars, even when magnetic, are not known to exhibit flare-like events in addition to the differences in EM distributions and luminosity ratio trends between hot and cool stars.}

\section{Infrared}\label{sec:Infrared}

\begin{figure*}
    \centering
    \includegraphics[width=\linewidth]{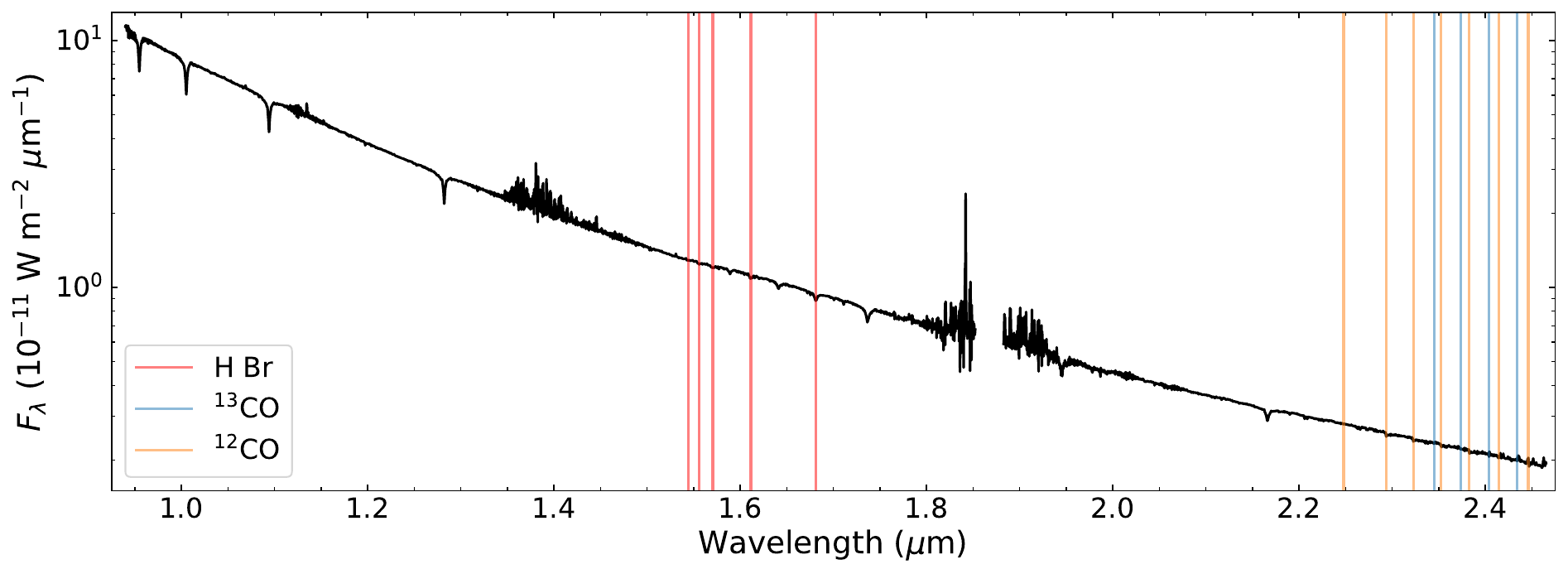}
    \includegraphics[width=\linewidth]{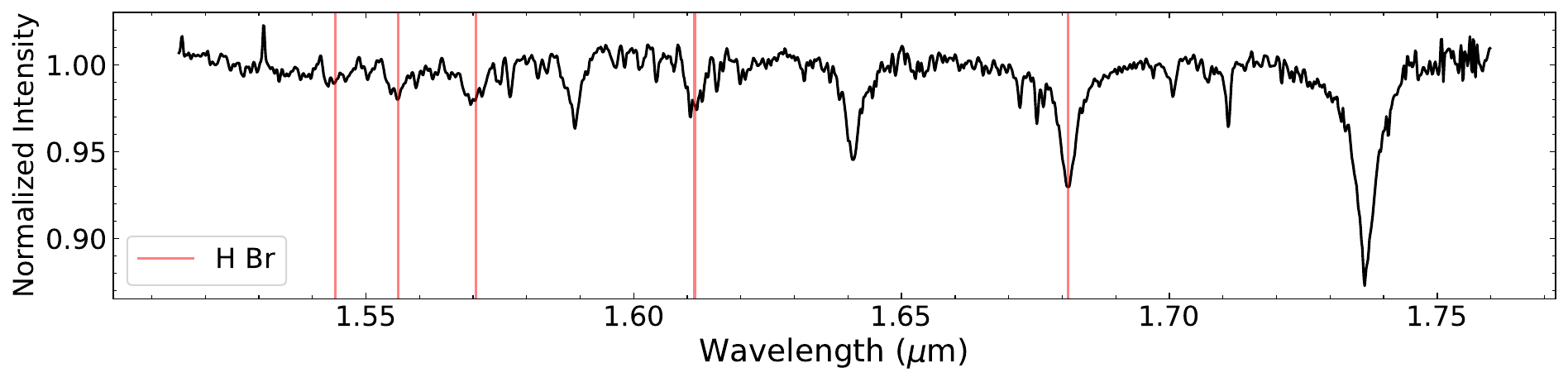}
    \includegraphics[width=\linewidth]{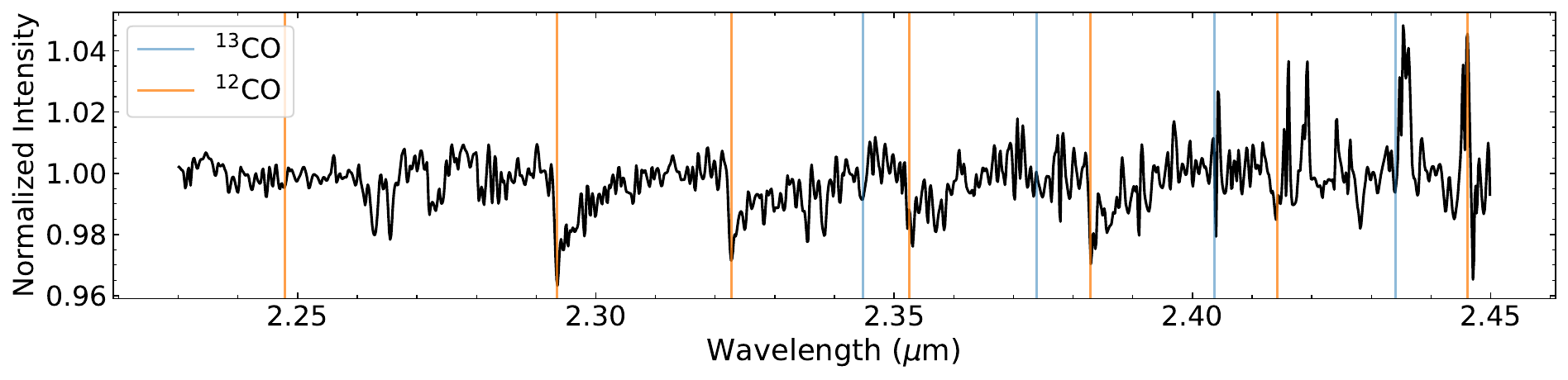}
    \caption{\textbf{Top Panel:} Extinction ($R_\mathrm{V}\approx5.2$ \citep{Ducati2003}, $E(B-V) = 0.59$ \citep{Shultz2026}) corrected SOAR/TripleSpec spectrum of $\rho$\,Oph\,C with Br and CO bandhead absorption line regions labeled. \textbf{Middle Panel:} Zoom in of the H Br absorption regions. \textbf{Bottom Panel:} Zoom in of the CO bandhead absorption regions.}
    \label{fig:TripleSpec}
\end{figure*}

Due to the brightness of B stars, it is not possible to measure stellar properties of nearby cool stars in optical wavelengths. Instead, longer wavelengths where the B star is sufficiently dim for spectral features from the cool star to appear are required. For this, we collected near infrared (NIR) data using the TripleSpec instrument at SOAR \citep{Schlawinl14}, shown in the top panel of Figure~\ref{fig:TripleSpec}. 

The first step in analyzing the NIR data was cross checking some of the stellar parameters found by \citet{Shultz2026}; namely, the spectral type. For this, we used the B star spectral type formulas from \citet{RomanLopes2018} that use the equivalent width (EW) of H Brackett (H Br; not to be confused with the element Bromine) lines. The H Br lines are marked in red in Figure\,\ref{fig:TripleSpec} in the top panel and focused on in the second panel. We specifically computed the EW of Br13 $\lambda$16113.714 and Br11 $\lambda$16811.111. For these, we used the built-in functions in the SpecUtils \citep{specutils2025} Python package for computing EWs. This resulted in values of $\mathrm{EW}(\mathrm{Br11}) = 3.03$\,\AA\ and $\mathrm{EW}(\mathrm{Br11}) = 0.78$\,\AA.

\begin{figure*}
    \centering
    \includegraphics[width=\linewidth]{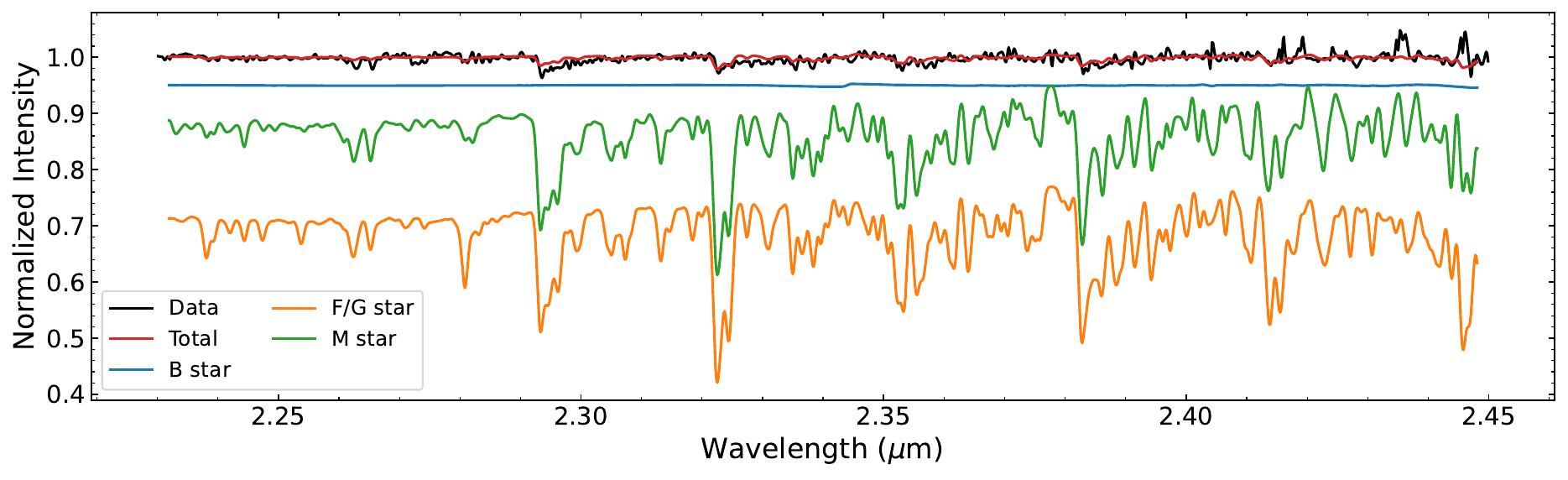}
    \caption{\Change{Illustration of the reduction in equivalent width of lines (in this case CO bandheads) by the addition of the B star continuum on the M and G star lines. Model spectra have arbitrary vertical offsets applied for visibility.}}
    \label{fig:TripleSpecFit}
\end{figure*}

The total EW of these lines is $\mathrm{EW}(\mathrm{Br11}+\mathrm{Br13}) = 3.81$\,\AA, corresponding to a B1 V star with a ${\sim}25$\,kK surface temperature based on the spectral type classifications from \citet{RomanLopes2018}. In comparison, the work by \citet{Shultz2026} measured that the effective surface temperature is $T_\mathrm{eff}=15.3-18.3$\,kK, covering a range of spectral types from B3--5. \Change{Our Br line measurement is a significant deviation from the optically determined temperature. One possible explanation is that the Br absorption lines are being filled in by the continuum flux of the two companions, reducing their equivalent width (for an example of this, see Figure~\ref{fig:TripleSpecFit}).}

\Change{The chemical peculiarity of the primary magnetic B star \citep{Shultz2026} can cause the same filling of the absorption lines. Chemically peculiar stars are known to have excess NIR -- IR emission due to the line blanketing effect from metals being levitated by the magnetic field \citep{Kochukhov2009}. The line blanketing causes UV and more energetic emission to scatter, heating up the layers below the levitated metals, thereby causing longer wavelengths to appear from a hotter star \citep{Strom1969,Kochukhov2005}. The line blanketing is not easy to disentangle from the effects of multiplicity, making it difficult to extract quantitative information from the NIR spectrum; this will be discussed in more detail below.}


\Change{Further evidence of the cool companions can be seen} in the 2.25-2.5\,$\mu$m region, which contain prominent CO bandhead absorption lines, shown in the bottom panel of Figure~\ref{fig:TripleSpec}. CO bandheads are not expected in a B-type main sequence star but are known to occur in cool stars ubiquitously. These lines can thus serve as an additional benchmark for the presence of companions in \rOC. \citet{Frost2025} found that the interfometric companions in the system have (inclination angle projected) masses of $1.03\,M_\odot$ and $0.4\,M_\odot$, corresponding to a \Change{late-F/early-G} and M-type star. Based on the age of the system \citep[${\sim}7$\,Myr;][]{Shultz2026}, these stars will be PMS stars, so we can estimate their stellar properties using evolutionary tracks from the Mesa Isochrones and Stellar Tracks \citep[MIST;][]{Choi2016}. The predicted values of the MIST tracks are given in Table~\ref{tab:NIRAtmoFits}.

\begin{deluxetable}{lcccc}
    \centering
    \tablecaption{Stellar Properties of the Cool Companions from MIST Isochrones.\label{tab:NIRAtmoFits}}
    \tablehead{
        \colhead{Star} & \colhead{$M$ ($M_\odot$)\tablenotemark{a}} & \colhead{$T_\mathrm{MIST}$ (kK)} & \colhead{$\log g_\mathrm{MIST}$} & \colhead{$R_\mathrm{MIST}$ ($R_\odot$)}
    }
    \startdata
        F/G & 1.03 & 4.38 & 4.21 & 1.31 \\
        M & 0.40 & 3.54 & 4.14 & 0.89 \\
    \enddata
    \tablenotetext{a}{Masses of the companion stars based on the mass ratios from \citet{Frost2025}.}
\end{deluxetable}

\Change{We next examined whether these spectra could reproduce the observed features using tabulated BT-NextGen atmosphere models \citep{Allard2011,Allard2012}. However, the aforementioned ambiguity of the line blanketing created significant difficulties in fitting the spectrum with a self-consistent model. Many parameters were either unconstrained or converged to values that were not consistent with the age and multi-wavelength constraints. Typically, chemically peculiar stars require specialized atmospheric models that account for their enhanced metal abundances and the line blanketing.}

\Change{The modeling was further complicated by the nature of the stars in the system because two potentially have disks. The primary B star is classified as a Centrifugal Magnetosphere star, meaning a disk-like structure of gas is present within the magnetosphere\citep{Leto:2020B}. This disk is dynamically supported by the balance between the magnetic field below the star's Alfv\'{e}n radius, the centrifugal force due to the star's rotation, and the gas pressure of the disk. For the M star, at its current age as a PMS, it possibly still has an accretion disk. Circumstellar disks are which are known to have excess NIR emission that typical stellar atmospheric modeling do not account for. All together, we found that we were unable to successfully find a combination of simple BT-NextGen models that could adequately fit the system.}

\Change{Instead, we focus on our hypothesis that the CO bandheads are from the cool stars. While we have highlighted that longer wavelengths can have a better contrast for seeing features of these stars, the B star is still \textit{much} brighter. To test whether this is possible, we show in Figure~\ref{fig:TripleSpecFit} the normalized model spectra of the three stars using the stellar parameters from \citet{Shultz2026} for the B star and the MIST isochrone predicted parameters in Table~\ref{tab:NIRAtmoFits}. The observed spectrum is given in black and while the total of the three spectra is in red. Despite the overall brightness of the B star, the bandheads of the two cool stars are deep enough that they can qualitatively match the observed spectrum.}

The spectroscopic NIR data adds further weight to the interferometric determination of two companion stars in \rOC. The continuum and spectral features in the TripleSpec require that two cool, PMS stars are present in the system. Not including both companions causes results that are not consistent with prior determined system age and stellar properties. We can thus conclude robustly that \rOC\ is trinary system consisting of a magnetic B-type star, a close to zero age main sequence G star, and an M-type PMS star.

\section{Optical} \label{sec:Optical}

As noted previously, the optical data is unlikely to reveal instantaneous spectral features in the system due to the B star dominating over the cool stars. Additionally, much of the specific optical information has been thoroughly analyzed \citet{Shultz2026} from the exceptional resolution offered by ESPaDOnS data. However, it is possible for differentials to be identified if the cool stars cause additional variabilities.

The orbital motion of the system can of course cause variabilities, but the orbital periods calculated from the separations found by \citet{Frost2025} are on the order of months at the shortest. Such timescales will require detailed spectroscopic campaigns of exceptional resolution to observe. Instead, if the background flux from one or both of the cool stars is inherently variable, either from white light flares or rotation, then a sufficiently sensitive observation may be able to detect such changes.

\begin{figure*}
    \centering
    \includegraphics[width=\linewidth]{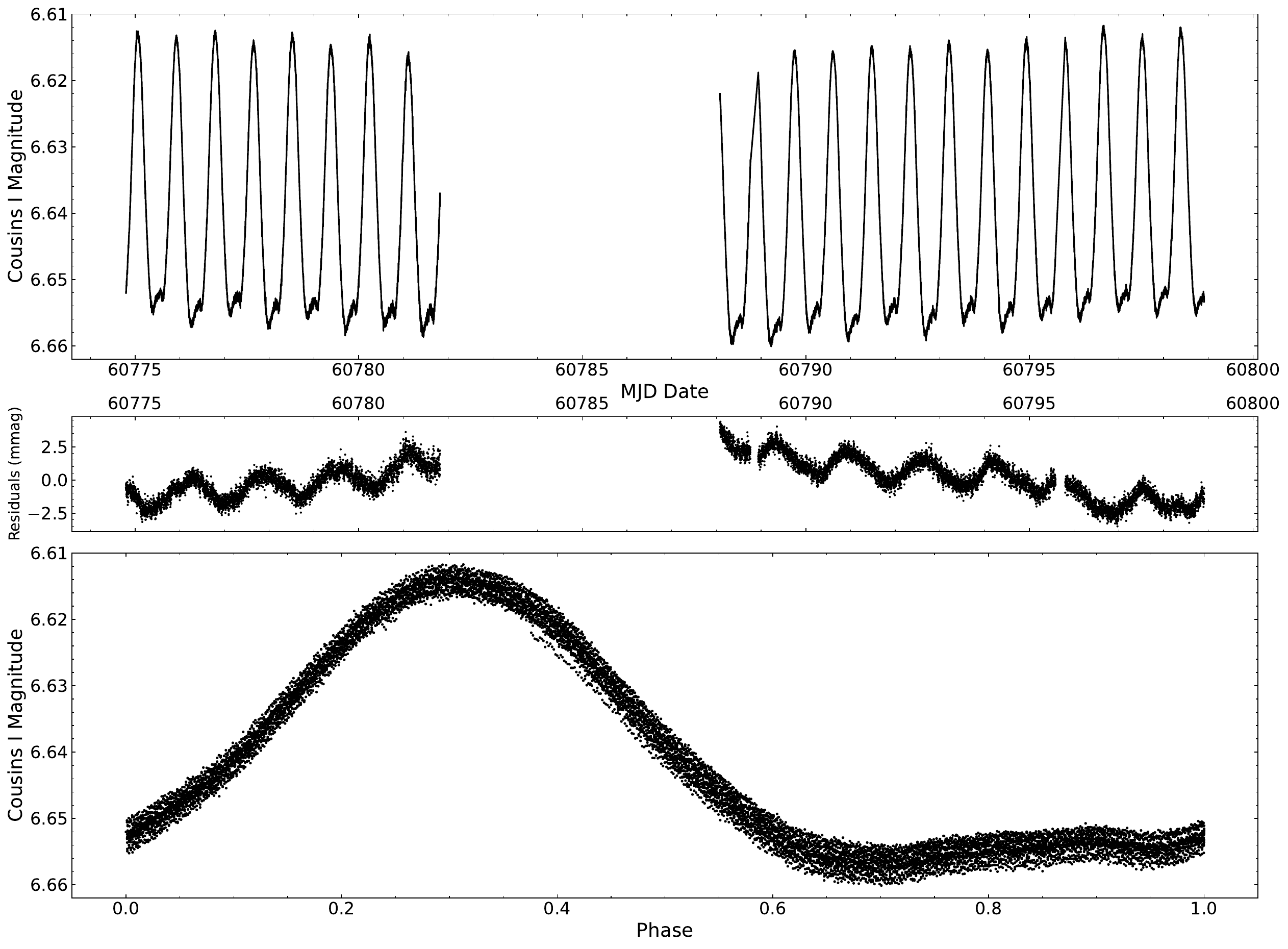}
    \includegraphics[width=\linewidth]{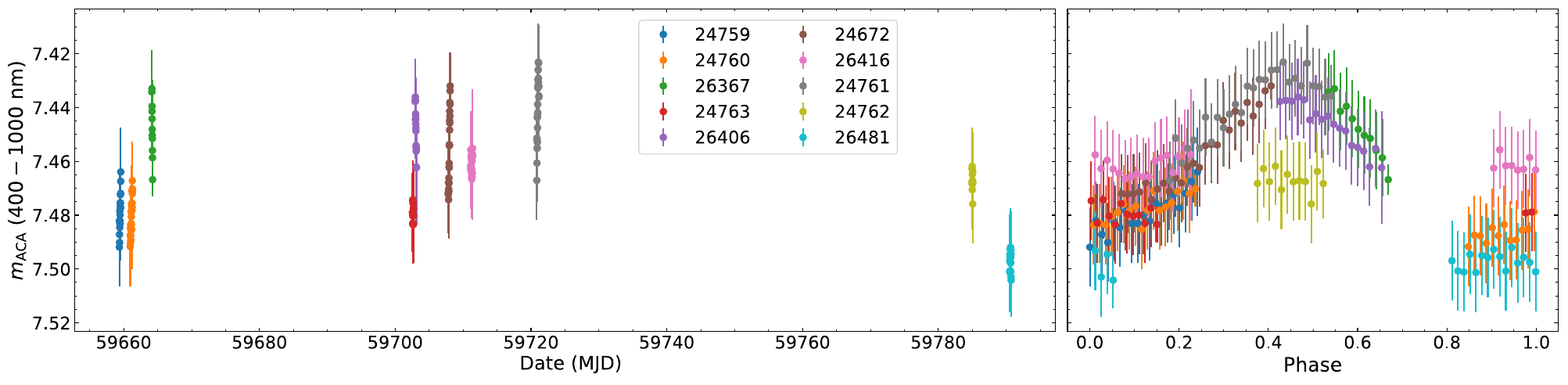}
    \caption{\textbf{Top Panel:} \textit{TESS} light curve of \rOC\ converted to magnitudes. \textbf{Second Panel:} Residual magnitude after subtracting a Fourier series fit to the TESS light curve. \textbf{Third Panel:}  Same light curve but phased to the $P=0.8639$\,d period using the start of the TESS observation as $t_0$. \textbf{Bottom Left Panel:} Chandra ACA light curve of \rOC\ with 1\,ks bins. \textbf{Bottom Right Panel:} Chandra ACA light curves now phased to the primary period determined from the TESS data}
    \label{fig:TESSlightcurve}
\end{figure*}

\rOC\ has a well measured rotational variability from both K2 \citep{Jilinski2006} and radio \citep{Leto2020}, which complicates the problem of detecting additional variabilities. However, as can be seen in the recent TESS data shown in the top panel of Figure~\ref{fig:TESSlightcurve}, the optical variability of the system is more complex than a single stellar rotation.

This is further highlighted in the third panel where we show the light curve phased to the dominant spin period. For phasing, we measured the dominant period using a Python implementation of the Bretthorst period searching algorithm \citep{Bretthorst1988}, finding $P=0.86396 \pm 0.00018$\,d; a perfect match to the prior determined K2 period and those of \citet{Shultz2026} using the same TESS data. The light curve was then phased to the start of the TESS observation. This phasing reveals there are additional periodicities in the system, causing both systematic flux changes at any given phase of the main period and additional structure in the light curve.

\citet{Shultz2026} made a similar conclusion due to the detection of about a dozen harmonic frequencies of the main period. In their work, the authors used the first and second harmonics to reveal further residual variations in the light curve that are not associated with the harmonics of the rotation period. These periods may be associated with further dynamics in the system, but using only up-to the second harmonic limits the ability to ascertain the coherence of the residuals.

To better reveal the residual variations, we refit the TESS light curve with a Fourier series ranging from 2-20 harmonic frequencies. Using such a large number of terms has a risk of over fitting the data, so we computed a $\chi^2$ statistic to evaluate the goodness of fit for each series. As expected the $\chi^2$ decreased with increasing number of included frequencies. However, the value asymptoted when including at least 6 harmonic frequencies. We thus chose the 6-term Fourier series for our best-fitting model of the TESS light curve.

In the second panel of Figure~\ref{fig:TESSlightcurve}, we show the residuals after subtracting the 6-term Fourier series. The error bars are the preserved TESS uncertainties. The resulting residuals show a well-ordered periodicity to them, which is in contrast to the residuals from \citet{Shultz2026}. There is also an overall longer term trend in the data, suggesting the potential for a further long period. Using the Bretthorst algorithm again, these two periods are $P_\mathrm{short}=1.64317 \pm 0.00227$\,d and $P_\mathrm{long}=30.04808 \pm 0.20647$\,d.

The longer period $P_\mathrm{long}$ could be related to orbital modulation. While some orbital parameters can be estimated from the interferometric observations done by \citet{Frost2025}, they are only a single instantaneous measurement. Based on the projected separation of the stars, the orbital period of the B-G pair is on the order of months. $P_\mathrm{long}$ is of the right scale for this to occur, although it is still rather short compared to the inferred periods from \citet{Frost2025}. At the same time, there is also the proverbial problem of TESS's momentum maneuvers that can cause long time scale variations.

The short period on the other hand is interesting given how close it is to being a harmonic of the main period: $P_\mathrm{short}/P = 1.9019\pm0.0027$. In order to investigate if the observed \Change{${\sim}0.8$\,mmag} residual photometric variations could be attributed to variability of the companion, we computed the combined system flux and solved for the fractional flux change in the fainter star required to produce the measured change in the total system magnitude. Adopting apparent magnitudes based on the V magnitude of 11.6 and 6.5 for the fainter and brighter components \citep{Mason2025}, respectively, we find that a flux increase of approximately \Change{7\%} in the fainter component is sufficient to account for the observed variations. 


On board Chandra is also a suite of optical guidance detectors referred to as the Aspect Camera Assembly (ACA) that monitor 5 stars for observational alignment. \rOC\ was used as a guide star in all of the Chandra observations, so we have simultaneous optical data taken during the huge flare shown in the top panel of Figure~\ref{fig:lightcurves}. What this data can tell us is whether the flare was strong enough to be a ``white light'' flare that increases the optical flux from the cool star.

In the bottom left panel of Figure~\ref{fig:TESSlightcurve}, we show the light curve for the ACA camera using 1\,ks bins. We specifically plot the magnitude as measured by the ACA, calculated using
\begin{equation}
    m_\mathrm{ACA}=10.32 - 2.5\log\left(\frac{R_\mathrm{ACA}}{5263.0}\right),
\end{equation}
where $R_\mathrm{ACA}$ is the count rate measured in the 400-1000\,nm wavelength band. As expected, there is a significant level of variability detected, with apparent jumps in magnitude between some observations. Such jumps do correspond to the observations that contain the X-ray flare (red-grey points; same color scheme as in Figure~\ref{fig:lightcurves}).  However, if we phase these light curves to the $P=0.86396$\,d period from TESS, shown in the right panel of Figure~\ref{fig:TESSlightcurve}, we can see that the variability matches the primary B star's rotation exactly.

This does not explicitly rule out the X-ray flare being a white light flare. It only shows that it was not brighter than $m_\mathrm{ACA}\sim7.56$ magnitude. The wide wavelength range of the ACA extending all the way to bluer wavelengths made this particular experiment difficult given that the primary star is a main sequence B star.

There are additional features in the phased light curve that are of note, though. While the $P=0.86396$\,d period does produce a high quality phase plot, we can see the effects of the secondary and tertiary periods found in the TESS data. Observations 24760, 26416, and 24681 show this for 0.8-1.0 phase. Additionally, observation 24762 does not follow the overall rotational period at all. This shows that the low amplitude TESS variabilities extend all the way to 400\,nm, suggesting further effects at shorter wavelengths to study. Some of the scatter may be due to hot pixels from the ACA that do not get corrected for during the dithering process, so additional photometric or spectroscopic observations should be taken around 400\,nm to determine if there are any color variations in the system.

\begin{figure*}[ht!]
\centering
\includegraphics[width=\linewidth]{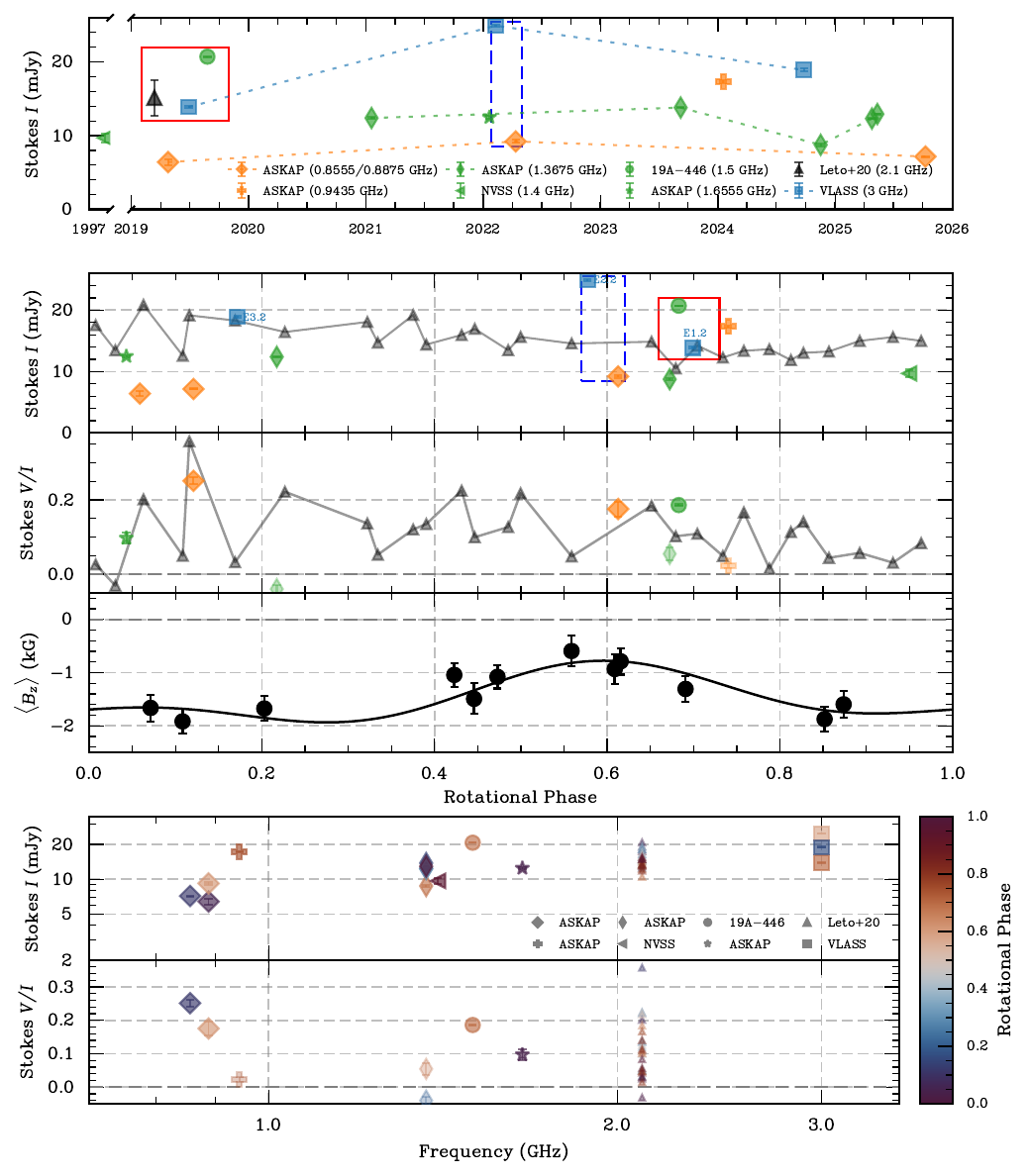}
\caption{\textbf{Top:} The long-term radio light curve of \rOC. Closely-spaced frequencies are plotted in the same color, and related data is connected with a dashed line. Individual datasets are indicated with different marker shapes. \Change{\textbf{Middle:} The same data} with integration times of $t_{\rm int}\lesssim2\,$hours with Stokes~$I$ (upper) and $V/I$ (fractional circular polarization, middle), longitudinal magnetic field $\langle B_z\rangle$, and best-fitting dipole harmonic model corresponding from \citet[][bottom]{Shultz2026} vs the rotational phase of \rOC. \Change{\textbf{Bottom:} The same Stokes~$I$ and $V/I$ shown vs observed frequency, where the color is used to indicate rotational phase.}}
\label{fig:radio-lcs}
\end{figure*}

\section{Radio}\label{sec:Radio}
We have compiled a long-term radio light curve of \rOC\ with stochastic sampling over the 2019-2026 period, spanning frequencies from 0.8555 -- 3\,GHz, and including a measurement from 1997 \citep[NRAO VLA Sky Survey (NVSS);][]{Condon:1998} and several ASKAP fluxes reported by \citet{Das:2025}. We compare our results with Australia Telescope Compact Array (ATCA) observations from a six-day campaign that covered the full rotational phase at several frequencies \citep{Leto2020}.

As shown in Figure~\ref{fig:radio-lcs}, \rOC\ shows significant unexpected variability in the radio flux. Even when \Change{computing the rotational phase $\phi$} from the period, this variability is inconsistent with the star's rotation cycle (see \Change{middle} panels of \autoref{fig:radio-lcs}). The most extreme examples of a change in flux are from VLA/19A-446 (PI: J. Ling) at 1.5\,GHz in early 2019 (green circle, \Change{$\phi=0.68$)} at $20.72\pm0.04$\,mJy and from ASKAP at 1.3675\,GHz in late 2024 (green diamond, \Change{$\phi=0.67$}) to $8.75\pm0.16$\,mJy, demonstrating a $2.4\times$ change over a several-year-long baseline. Variations of the flux with the stellar rotation cycle do not contribute to the observed variability, since these two highlighted data points \Change{both occur near the same phase}. 

The lowest-frequency data spanning $855.5-943.5$\,MHz (orange diamonds in \autoref{fig:radio-lcs}) also display interesting evolution. The \Change{flux} variability within this band already does not demonstrate a monotonic evolution of the flux with frequency over this small bandwidth \Change{as demonstrated in the bottom panels of \autoref{fig:radio-lcs}. The 855.5\,MHz point at $\phi=0.12$ compared to the 887.5\,MHz point with an available Stokes~$V$ at $\phi=0.61$} is statistically consistent with a difference only in the unpolarized flux. \Change{Taking the unpolarized (Stokes $I-V$) component of the same measurements} as static, the spectral index would be extraordinarily steep at $\approx9.\Change{4}$.

At 943.5\,MHz, a significant enhancement is observed in early 2024 \Change{($\phi=0.74$, orange plus)} that remarkably is consistent with no detectable circular polarization for a $3\sigma$ upper limit of $V<0.39$\,mJy. This enhancement corresponds to a fractional polarization of $V/I\lesssim2\%$. The 887.5\,MHz observation at \Change{$\phi=0.61$ is separated from the 943.5\,MHz at $\phi=0.74$ by $\Delta\phi=0.13$}, and comparing the two indicates a complete loss of the polarized component and therefore a $\sim2.3\times$ increase in the unpolarized flux. This is notably the same phase-dependent Stokes~$I$ evolution (increasing with phase) observed at $\approx1.4$\,GHz and nearly the same phase \Change{between the ASKAP observation at $\phi=0.67$ and VLA/19A-446 at $\phi=0.68$. In contrast, the Stokes~$V$ falls with increasing phase at the 887.5 vs 943.5\,MHz comparison, whereas it rises with phase in the $\approx1.4$\,GHz}. However, since \Change{the lower frequency} data span a wider range of phase \Change{than the higher frequency ($\Delta\phi=0.13$ vs $0.01$, respectively)}, we can't rule out rotational modulation \textit{a priori}.

The variability in the lowest frequencies seems to be two examples of several apparent sign flips in the broadband spectral index of data closely spaced in time. \Change{One} instance is demonstrated by \Change{the 2019 VLASS epoch ($\phi=0.70$)}, which has a flux below the mean value of the 2.1\,GHz light curve \citep[][black triangle]{Leto2020} but \Change{the VLASS fluxes} remain above lower-frequency observations in the following \Change{two epochs in 2022 ($\phi=0.58$ and 2024 ($\phi=0.17$)}. Another example is the VLA/19A-446 1.5\,GHz flux \Change{($\phi=0.68$)}, which falls above both the 2.1\,GHz light curve \Change{mean value \citep[][black triangle]{Leto2020} and the 2019 VLASS observation ($\phi=0.70$)}, but subsequent observations at \Change{$\approx1.5\,$GHz} fall below VLASS and above lower-frequency observations, except for the 943.5\,MHz observation in 2024 \Change{($\phi=0.74$, orange plus)}. 

Finally, we consider data that are serendipitously closely spaced in both time and phase. There are two such groups of data, the first of which is highlighted by a red\Change{, solid-line} box in \autoref{fig:radio-lcs}, falling at a central phase of $0.67$ and spanning five months. The first ASKAP 887.5\,MHz observation (orange diamond) also falls within this time range, but is half a rotational phase separated. The second group consists of VLASS Epoch~2.2 and the second 887.5\,MHz measurement from ASKAP in 2022 \Change{highlighted with a blue, dashed-line box}, which bracket the central value of \Change{$0.58$} and are separated by two months. The ASKAP 1655.5\,MHz observation (green star) is just 19\,days before the VLASS observation, but is also quite separated in rotational phase. Since all these observations are at different frequencies, the significance of any variability requires an understanding of the underlying frequency dependence from a particular model prediction, so we defer detailed discussion to the following subsections.

If originating from the B star's magnetosphere, this represents a marked deviation from the expectations for early-type rapid rotators \Change{given that the $\sim$\,kG nearly-dipole fields \citep{Kochukhov:2019} of early-type stars are known to be extremely stable \citep{Braithwaite:2004, Wade:2016} over thousands of rotational cycles \citep{Shultz:2018} and only decrease on evolutionary timescales \citep{Landstreet:2007, Landstreet:2008, Sikora:2019, Shultz:2019}}. In the following subsections, we consider several potential origins of the variability and the resulting interpretations. 

\subsection{Incoherent Emission from the B Star Magnetosphere} \label{subsec:incoherent-radio}

\citet{Leto2020} observed \rOC\ with ATCA (1-18\,GHz) and modeled the broadband spectrum and phase-folded multi-frequency light curves using the 3D radio emission from an obliquely rotating magnetosphere, \Change{which models the presence of incoherent, gyro-synchrotron radiation \citep{Leto:2016} powered by non-thermal electrons generated by CBOs in the magnetic equatorial region}. \rOC's inferred stellar parameters easily place it in the regime of supporting centrifugal magnetospheres \Change{\citep{Shultz2026}, and \citet{Leto:2026} recently included \rOC\ as a member of the sample used to derive a scaling relation applicable between the incoherent gyro-synchrotron emission and expected CBO energy given the stellar parameters}.

\Change{Critically, the connection between CBOs and the incoherent radio emission predicts that the radio luminosity should not evolve in time beyond obliquity-driven rotational modulation since the magnetosphere should remain stable}. Additionally, multiple radio measurements of early-type stars indicate flux measurement repeatability on timescales of several years \citep[e.g.,][]{Leto:2012}. The scaling relation originally observed by \citet{Leto:2021} and confirmed by \citet{Shultz:2022, Das:2025, Leto:2026} states that the radio spectral luminosity is correlated with the polar magnetic field strength ($B_{\rm p}$), the stellar radius ($R_*$), and the rotation period ($P_{\rm rot}$) as 
\begin{equation} \label{eq:cbo-radio}
L_{\nu,{\rm rad}}\propto \frac{B_{\rm p}^2R_*^4}{P_{\rm rot}^2}.
\end{equation}
This precisely corresponds to the ``split monopole" ($p=1$) case of \citet{Owocki:2022} up to a scaling factor between $L_{\rm CBO}$ and $L_{\nu,{\rm rad}}$, where $L_\mathrm{CBO}$ is the rate of energy release from the reconnection events that populate the electrons responsible for observed radio emission. As noted therein, the mass-loss rate is not expected to influence the observed radio spectral luminosity and is only dependent on what should be static parameters. If we instead take the ``dipole" ($p=2$) case, then $L_{\nu,\mathrm{rad}}\sim L_{\rm CBO}\sim\sqrt{\dot{M}}$, suggesting a connection between the wind mass-flux and the radio emission. 

If the observed variability in \rOC\ (\autoref{fig:radio-lcs}) is due to the incoherent gyro-synchrotron radio halo surrounding the centrifugal magnetosphere, it implies a change in one of these intrinsic parameters for which there is no theoretical expectation or empirical evidence for on \Change{these timescales from an early-type star}. Although the modeling of \citet{Leto:2026} indicates \rOC\ is consistent with the monopole scaling relation (\autoref{eq:cbo-radio}), we can relax that requirement and consider the dipole scenario. Comparing observations at similar phases allows us to \Change{marginalize over} any modulation due to stellar rotation, which is already predicted to be $\lesssim$mJy at $\lesssim3$\,GHz \citep{Leto2020}. The $2.4\times$ variation in radio luminosity that we observe around rotational phase $\phi=0.67$ at $\approx1.4$\,GHz would then imply a $5-6\times$ variation in the mass-loss rate on timescales of only years. The unpolarized flux is more likely to trace the incoherent gyro-synchrotron radiation \citep{Leto2020}, which still yields a $2\times$ variation \Change{in the unpolarized radio emission}.

The extrema in a single frequency are not the largest {\it evolution} of flux: we \Change{highlight the observations} in the red rectangle in \autoref{fig:radio-lcs}. If we treat the evolution as a rescaling of the SED derived by \citet[][their Fig.~2]{Leto2020}, we can \Change{better} compare these closely-spaced observations. This already leads to deviations, as we noted several examples of a change in sign in the apparent spectral index of observations closely spaced in time, whereas the spectral index observed by \citet{Leto2020} is always positive over this frequency range.

\Change{Further investigating this group of data that has been marginalized over rotational modulation}, assuming a power-law rise between $2.5-5$\,GHz for the SED results in $\alpha\approx0.32$ and a 3\,GHz flux of $\approx17.2$\,mJy at the \citet{Leto2020} epoch, which implies a 20\% drop just three months later \Change{in the 2019 VLASS epoch ($\phi=0.70$)}. Applying the same approach between the VLASS observation and the VLA/19A-446 observation \Change{($\phi=0.68$}, the $1.5-3$\,GHz spectral index from the \citet{Leto2020} SED is $\approx0.66$, resulting in a predicted 1.5\,GHz flux of 8.8\,mJy during VLASS E1.2, which implies a factor of $\approx2.4\times$ rise in an even shorter two months to 20.7\,mJy in VLA/19A-446. Accounting for the Stokes~$V$ contribution still requires $>2\times$ rise, assuming the best-case scenario of the VLASS data being completely unpolarized.

Investigating the second grouping highlighted by a blue, dashed-line box, the \Change{2022} 887.5\,MHz ASKAP observation \Change{($\phi=0.61$)} and VLASS E2.2 (3\,GHz, \Change{$\phi=0.58$) are similarly marginalized over rotational modulation with $\Delta\phi=0.03$ and alos only} separated by about two months. Using the lowest measurements at $1.5-2.5$\,GHz from the \citet{Leto2020} SED, we estimate $\alpha\approx0.8$ that implies a 3\,GHz flux of 24.4\,mJy, just 0.5\,mJy below the measurement. The lack of circular polarization data \Change{currently available from} VLASS \Change{data products} or a $\leq1$\,GHz measurement from the SED makes further interpretation difficult.

The enhancement at 943.5\,MHz in early 2024 (orange plus in \autoref{fig:radio-lcs}, \Change{$\phi=0.74$}) is interesting to discuss in the context of the incoherent emission since this measurement is entirely unpolarized. Fig.~1 of \citet{Leto2020} demonstrates that for the best-fitting model, the amplitude of rotational modulation decreases at lower frequencies, so treat\Change{ing} the average of the other low-frequency data as static \Change{would} imply the largest unpolarized variation of a $\gtrsim3\times$ rise. 

From a qualitative analysis of the multi-frequency radio data, we find that the flux variations have a significant amplitude that would imply an unphysical evolution of the star's mass-loss rate based on current models. Therefore, we believe the variability is extremely unlikely to originate in the incoherent gyro-synchrotron radio emission from the magnetosphere. 

\subsection{Coherent Emission from the B Star Magnetosphere}

\rOC\ is the seventh known member of the ``Main-sequence Radio Pulse emitter'' class \citep{Das:2021} -- early-type magnetic stars exhibiting highly circularly polarized emission at $\lesssim6$\,GHz -- as identified by \citet{Leto2020}. These stars all show incoherent emission, high circular polarization, and narrow width in rotational phase \citep{Das:2025}. These properties suggest that the pulses arise from a coherent and beamed mechanism, such the electron cyclotron maser instability \citep[ECMI,][]{Melrose:1982}.

\citet{Leto2020} used the nearly constant appearance of circularly polarized flux (\Change{middle panel} of \autoref{fig:radio-lcs}) in excess of the predicted contribution from the incoherent gyro-synchrotron emission to validate the measurements of the inclination $i_{\rm rot}\approx74^\circ$ and the obliquity of the dipole $\beta\approx5^\circ$. These measurements were \Change{revisited} by \citet{Shultz2026}, finding $i_{\rm rot}=64^{\circ+6}_{\ -4}$ and $\beta=7^\circ\pm2^\circ$ and in the bottom right panel of \autoref{fig:radio-lcs}, we have plotted the $\langle B_z\rangle$ values measured therein as a function of rotational phase. \Change{Note that the convention used here for $\beta$ is the opposite of, e.g., \citet{Munoz:2020}, where $\beta$ is explicitly defined as the angle between the rotation and \textit{positive} (in $\langle B_z\rangle$) pole; here, we report $\beta$ relative to the \textit{negative} (southern) dipole axis, following convention established in the existing literature on \rOC\ given the consistently negative measurement of $\langle B_z\rangle$ \citep{Alecian:2014,Shultz2026}.}


No ``magnetic nulls" are present due to the system geometry (see the \Change{bottom-middle} panel of \autoref{fig:radio-lcs}), but \citet{Leto2020} shows circular polarization across all phases, and our new measurements show a wide range of variability in fractional circular polarization at similar phases, in particular at $\phi=0.60-0.75$, which is highly unexpected \Change{given the stable, dipole-like $\sim$kG fields of early-type stars}.

With the magnetic field strength and geometry accounted for, this leaves the other parameters of the ECMI growth rate to potentially vary\Change{: for example, the ambient electron density and therefore the plasma frequency $\nu_{\rm pe}$, the overall electron energy/velocity distribution $f(v)$, the distribution's anisotropy $\delta f(v)/\delta v_\perp$, among other possibilities}. CBOs could result in variations in the emission region's parameters; \Change{however,} detailed modeling of this has not \Change{yet} been performed. \Change{The role of CBOs may be inferred since the only region causally shared between each magnetic pole is the CBO-hosting current sheet above the magnetic equator; see Fig.~4 of \citet{Leto2020} for the \rOC\ geometry specifically, and Fig.~3 of \citet{Leto:2021} or Fig.~1 of \citet{Leto:2026}, which highlights the acceleration region relative to the auroral rings.} For example, \citet{Das:2021} notes that a large CBO could explain the pair of oppositely polarized pulses (tracing opposite magnetic poles) observed in the similarly chemically peculiar, magnetic early-type star CU\,Vir, likely emission associated with a particular magnetic azimuth.



\rOC's 2024 enhancement at 943.5\,MHz (\Change{$\phi=0.74$}) is more likely to trace ECMI at low frequencies and is only slightly offset from the lowest value of $\langle B_z\rangle$. Although not an order-of-magnitude enhancement like CU~Vir's ``giant pulse," \citep{Das:2021} it represents a factor-of-several change and a complete loss of fractional circular polarization. If an injection of energetic electrons occurred at a particular magnetic azimuth, the low-obliquity ($\beta\approx7^\circ$, \citealt{Shultz2026}) naturally explains the lack of polarization observed due to a minimal delay in the arrival time of the right-hand and left-hand circular polarization pulses from the opposite poles \Change{\citep[e.g., Fig.~1 of][]{Das:2024}}. Additionally, low obliquity implies that the ray paths do not pass through the magnetosphere's densest region, which is known to cause delays in pulse arrival times \Change{in idealized scenarios} \citep{Das:2024}. However, this mechanism alone cannot explain all the observed variability: for example, the enhancement observed in the 1.5\,GHz VLA/19A-446 observation \Change{($\phi=0.68$) relative to the 2022 ASKAP 1.3675\,GHz measurement ($\phi=0.67$)--just $\Delta\phi=0.01$ prior--} shows increases in both polarized and unpolarized flux.

If attributable to CBOs, ECMI could precisely constrain the locations, timescales, and amplitudes of the events. If the variability is from the auroral emission, \rOC\ would be the second known example of \Change{significant} ECMI variability in an MRP, after CU~Vir \citep{Das:2021}.

\subsubsection{Radiation Beaming Pattern}

Our new attribution of the X-ray flares to \rOC's cool star companion(s) has significant implications for ECMI modeling. \citet{Leto2020} used a temperature of $\approx60\,$MK to estimate the initial hollow cone's radiation pattern opening angle of $\theta_{\rm B}\approx82^\circ$ and width $\Delta\theta\approx8^\circ$, sourced from the X-ray observations of \citet{Pillitteri:2016}. However, as noted therein, the existence of companions in the system may also be the origin of the X-ray emission, which we have confidently identified as the more likely scenario (see \autoref{sec:X-rayObs}). The modeling of the auroral emission \citep{Leto:2016} using the X-ray-inferred parameters implied a very wide beaming angle of $\delta\gtrsim45^\circ$ \citep[see Fig.~3 of][for $\beta=7^\circ$]{Leto2020}, in contrast to the expectations of a tangent-plane beaming model, such as the auroral kilometric radiation scenario \Change{observed at the Earth that has a loss-cone electron energy distribution from converging field lines forming a magnetic mirror} \citep{Mutel:2008}. If the X-ray emission \Change{does not} originat\Change{e} from the same population responsible for ECMI, $\theta_{\rm B}$ and $\Delta\theta$ \Change{can} take on other values, perhaps allowing for a more physically consistent inferred beaming angle $\delta$. Varying the emission parameters based on the growth rates \citep[e.g.,][]{Lee:2013} could also provide an independent test on the magneto-ionic mode and harmonic inferred by \citet{Leto2020}.

Another highly inclined and low obliquity system HR\,5907 has shown similar ``non-stop" auroral emission \citep{Biswas:2025}. When applying the same geometric model to HR\,5907 \citep{Leto:2018}, it suffers the same consequence as \rOC\ of a wide azimuthal beaming angle $\delta\gtrsim45^\circ$. \citet{Das:2020} built a model with a simplifying assumption of exactly perpendicular launch $\vec{k}\cdot \vec{B}=0$ of a ray from the ECMI source, but accounts for the ray's refraction in the ambient density field, which was used by \citet{Biswas:2025} to infer a narrower opening angle of $15^\circ$ for HR\,5907. Therefore, refraction may also help reconcile the wide beam angle of \rOC. 

It also remains a possibility that the assumed beaming pattern is incompatible with the emission observed at high inclination and low obliquity. \citet{Trigilio:2011} originally proposed the tangent-plane beaming model to partially reconcile narrow pulse widths; however, these systems show near-constant auroral emission across rotational phase. Saturn kilometric radiation (SKR)--also powered by ECMI--does not form in low-density cavities, and \Change{weak refractive effects mean it propagates with only small deviations from the instrinsic (growth-rate-dependent) launch angle} \citep{Lamy:2011}. Detailed modeling of growth rates \citep[e.g.,][]{Mutel:2010} could reveal that the Kronian scenario can reconcile the wide beaming angles. \Change{SKR is driven by ring-type or shell electron energy distributions that extend far into the high-altitude auroral plasma \citep{Mutel:2010,Lamy:2011}, differing from the typically assumed loss-cone distribution that gives rise to the tangent plane beaming model. An analogous scenario would be consistent with the quite large} inferred radial distance of the ECMI sources \Change{and may indicate a preference for an SKR-like electron energy distribution visible at high inclinations and low obliquities. However, unlike the typical X-mode of SKR, which can be detected at frequencies nearly corresponding to the surface magnetic field strength \citep{Lamy:2011}, MRPs (including \rOC) seem to show a premature cutoff at high frequency. Although the growth rates are expected to decline as ambient density increases, \citet{Das:2022c} showed that high density alone is insufficient to account for these premature cutoffs. This model would require an explanation for why, unlike SKR, the ECMI sources do not appear to extend nearly to the surface. One possibility is that the cut-off is apparent: under a different beaming pattern arising from a distinct electron energy distribution, the highest-frequency sources could extend to the surface yet be beamed away from the observer at \rOC's high inclination and low obliquity, requiring no physical truncation of the emitting region.}

Conclusive attribution of the variability observed in \autoref{fig:radio-lcs} to ECMI requires a more detailed analysis of the available radio data, as well as an investigation into the modeling of both ECMI growth rates (and hence intrinsic radiation pattern) combined with ray-tracing of the emergent emission using the relevant physical conditions of the centrifugal magnetosphere scenario. This manuscript aims only to highlight the discovery of unexpected radio variability in \rOC\ in the context of the companions, so we defer a detailed analysis of historical \Change{and forthcoming} radio observations to a follow-up publication. We highlight the need to infer the effects of CBOs on parameters related to the ECMI growth rates\Change{--most importantly, the exact electron energy distribution properties--}as a potential direct probe \citep{Das:2021} to confidently identify the origin of the variability in \rOC.

\subsection{The Role of Orbiting Companion(s)}

The final scenario we consider is that the orbital system of \rOC\ contains two (or more) radio-emitting components. As modeled by \citet{Leto2020}, a near-instantaneous measurement (six-day span) of the broadband radio SED and rotationally-phased light curve is broadly consistent with the expectations for non-thermal gyro-synchrotron emission with highly circularly-polarized pulses from auroral radio emission superimposed on its light curve. However, measurement over this short period could miss a second, underlying source in the system with a relatively flat or weakly negative spectral index over the observed frequency range in \autoref{fig:radio-lcs} that flares on longer timescales. Both pre-main-sequence (PMS) and low-mass stars--such as the G and M companions of \rOC--are known to be radio emitters in solitary systems \citep[e.g.,][]{Callingham:2021,Pritchard:2021,Golay:2023,Cesaroni:2023}. The vast majority of non-thermal radio stars follow the G\"udel-Benz relation with a standard deviation of approximately a single dex \citep[][see their Fig.~7]{Driessen:2024}. This relates the X-ray luminosity to the radio luminosity, originally at 5\,GHz \citep{Gudel1993, Benz1994}. Applying this relation to the X-ray observations reported on here (see \autoref{tab:XrayModelFits}), we find a predicted 5\,GHz radio flux varying between $0.37-0.92$\,mJy. With a moderately conservative assumption of optically thin gyro-synchrotron radiation (i.e., negative spectral index) and up to an order of magnitude due to the intrinsic scatter of the relation, it is very possible that a magnetically active companion in the \rOC\ system could manifest as the several-mJy variability we observe in the multi-frequency radio light curves in \autoref{fig:radio-lcs}.

Another possibility lies in interactions among the system's orbital components. \citet{Frost2025} found that the inner, more massive G star is $0.93\pm0.01$\,AU away from the B star (projected), which is $\lesssim1.8\,R_{\rm A}$ given $R_*=2.6\pm0.1\,R_\odot$ where the Alfv\'en radius is $R_{\rm A}=43^{+20}_{-4}\,R_*=0.52\,$AU \citep{Shultz2026}. The less massive, PMS M star is far more distant at $11.54\pm0.04$\,AU (projected), so it is unlikely to have any significant effect on the B star's magnetosphere. Interaction could go both ways: either the G star could have existing chromospheric activity on its surface excited by the B star's wind, or the presence of the G star could excite activity in the B star's magnetosphere similar to the Jupiter-Io system, which is known to excite ECMI radiation \citep{Treumann:2006}.

One possible straightforward test for the role of the companions could be achieved via very-long-baseline interferometry (VLBI) observations. At $d=139\,$pc, the synthesized beam of the NRAO Very Long Baseline Array (VLBA) at 5\,GHz is $\approx0.42\,$AU, more than sufficient to independently resolve the B star from the G star, and certainly also the much more distant PMS M star. If one of these systems is also radio-loud, it would be easily spatially separated from the B star, and the predicted flux of $\sim\,$mJy from the G\"udel-Benz relation \citep[][see \autoref{tab:XrayModelFits}]{Gudel1993, Benz1994} is well within reach of the VLBA. Additionally, such an observation would enable studies of the B star's magnetosphere, which may be partially resolved at this scale and would only improve with higher-frequency observations (although the emission region's length scale may also shrink at higher frequency, e.g., Fig.~4 of \citet{Leto:2021}). This observation would enable confirmation of 3D modeling of the radio emission distribution \citep{Trigilio:2004, Leto:2006} and searches for potential variation in surface-brightness density over time that may be signatures of CBOs at different magnetic azimuths.  

\section{Conclusions}\label{sec:Conclusions}

We reported on a multi-wavelength analysis of the C component of $\rho$\,Oph. Recent work has found this star to be a complex system, consisting of a strongly magnetic B-type star with potentially two low mass, cool star companions \citep{Frost2025}. From X-ray data taken by Chandra, XMM, and NuSTAR, we found further evidence of at least one cool star in the system based on the flare-like events' temporal and spectral properties.

The multiplicity of \rOC\ was bolstered by NIR data collected with TripleSpec at SOAR. The NIR data shows weak but visible CO-band heads from 2.28-2.4\,$\mu$m that cannot be explained as from a main sequence B type star. \Change{Further information on the properties of the system were not able to be drawn from the NIR data due to the complexity of the system. Multiplicity alone can create degeneracies when fitting spectra, especially for those with large luminosity differences, but this system also contains a chemically peculiar B-type star and potentially a PMS star with an accretion disk. These both can cause excess infrared emission that requires specialized models that are beyond the scope of this work. However, we did show that the CO band heads are explainable as being from the cool companions as the absorption line depth of the normalized NIR spectrum can be matched by the sum of BT-NextGen atmospheric models.}



Lastly, we reported on the long-term radio lightcurve of \rOC\ in the low frequency radio regime. We found a surprising result that \rOC\ is not a stable radio emission source, as is typical of magnetic massive stars. The radio flux varies by factors of $2-3\times$ at similar rotation phases separated by months to years. In the context of magnetic massive stars, this would require the mass-loss rate of \rOC\ to be varying by factors of up to 10$\times$ over the same time scales. Such a variation in mass-loss rate is unheard of in a young massive star, necessitating further explanation. We propose three possible explanation: (1) Radio flares from one or both of the PMS stars, (2) interactions between the B star's magnetic field and the closer PMS star, or (3) updates to the theory and models of magnetic massive star radio emission\Change{--possibly both the incoherent and coherent processes}. \Change{Detailed investigation} of these \Change{possibilities} are beyond the scope of this work, but \Change{further investigation of archival} radio \Change{data, additional radio} observations, \Change{and theoretical work} should be undertaken to explore them.

Taken all together, the spectroscopic and photometric properties of \rOC\ can only be explained self-consistently by the presence of the two companions, as detected by \citet{Frost2025}. \Change{This makes \rOC\ an important laboratory for studying the evolution of PMS stars in the presence of hot stars, and as a potential prototype of a rare configuration in stellar population models. Followup observations should be undertaken with high resolution infrared telescopes to better constrain the properties of the system. Furthermore, more detailed atmospheric calculations will need to be made on these higher quality datasets given the chemically peculiarity of the system.}


\begin{acknowledgments}

Support for DPH was provided by NASA through the Smithsonian Astrophysical Observatory (SAO) contract SV3-73016 to MIT for Support of the Chandra X-Ray Center (CXC) and Science Instruments. CXC is operated by SAO for and on behalf of NASA under contract NAS8-03060.

FN thanks to Conselho Nacional de Desenvolvimento Científico e Tecnológico (CNPq) for support through process 303093/2025-0.

CE gratefully acknowledges financial support for travel related to this work provided by the STScI Director's Research Fund. SJG thanks the JWST Discretionary Fund at STScI for travel related to this work.

This paper employs a list of Chandra datasets, obtained by the Chandra X-ray Observatory, contained in the Chandra Data Collection (CDC) ~\dataset[DOI: 10.25574/cdc.441]{https://doi.org/10.25574/cdc.441}.

All TESS data used in this paper can be found in MAST: \dataset[10.17909/t9-wpz1-8s54]{https://doi.org/10.17909/t9-wpz1-8s54} \citep{Caldwell2020}

The National Radio Astronomy Observatory is a facility of the National Science Foundation operated under cooperative agreement by Associated Universities, Inc.

This scientific work uses data obtained from Inyarrimanha Ilgari Bundara, the CSIRO Murchison Radio-astronomy Observatory. We acknowledge the Wajarri Yamaji People as the Traditional Owners and native title holders of the Observatory site. CSIRO’s ASKAP radio telescope is part of the Australia Telescope National Facility (\url{https://ror.org/05qajvd42}). Operation of ASKAP is funded by the Australian Government with support from the National Collaborative Research Infrastructure Strategy. ASKAP uses the resources of the Pawsey Supercomputing Research Centre. Establishment of ASKAP, Inyarrimanha Ilgari Bundara, the CSIRO Murchison Radio-astronomy Observatory and the Pawsey Supercomputing Research Centre are initiatives of the Australian Government, with support from the Government of Western Australia and the Science and Industry Endowment Fund.

This study is partially based on observations obtained at the SOAR telescope, which is a joint project of the Minist\'{e}rio da Ci\^{e}ncia, Tecnologia e Inova\c{c}\~{o}es (MCTI/LNA) do Brasil, the US National Science Foundation’s NOIRLab, the University of North Carolina at Chapel Hill (UNC), and Michigan State University (MSU).

This research has made use of NASA's \href{https://ui.adsabs.harvard.edu/}{Astrophysics Data System}.

The MACRO Consortium is an educational and research collaboration between Augustana College (Rock Island, IL), Coe College (Cedar Rapids, IA), Knox College (Galesburg, IL), Macalester College (Saint Paul, MN), and the University of Iowa (Iowa City, IA).  MACRO operates the robotic 0.5m Robert L. Mutel Telescope (RLMT) at the Winer Observatory (Sonoita, AZ, USA) to foster immersive research and classroom experiences for undergraduate students. Data taken by MACRO is available upon request.

This research has made use of ISIS functions (ISISscripts) provided by ECAP/Remeis observatory and MIT (\url{http://www.sternwarte.uni-erlangen.de/isis/}).

We thank Dr. David Principe and Dr. Hans Moritz Guenther for their help in interpreting the pre-main sequence status of the cool stars in the system.

We also thank Dr. Paul Barrett in advising the use and implementation of the Bretthorst period searching algorithm.

We lastly thank Dr. Claude Canizares for approving the GTO time for the Chandra observations of $\rho$\,Oph.

\end{acknowledgments}

\begin{contribution}

SJG conducted the analysis of the X-ray, NIR, and optical data and was responsible for writing and submitting the manuscript.

DPH submitted the original Chandra observation proposal, validated the X-ray analysis, and edited the manuscript.

FN collected the SOAR/TripleSpec data and validated the associated analysis.

CE aided in the NIR analysis, independently verified the optical and NIR results, and edited the manuscript.

JC, WWG, and JMC conducted the radio analysis and edited the manuscript. WWG drafted the radio analysis sections of the manuscript.

All other authors provided theoretical and interpretative support along with reviewing the manuscript.

Current list of contributing MACRO Consortium members in alphabetical order: Suhrid Abrar, Brian Adams, Hayden Arko, Paul Barrett, Dick Bernstein, Emma Biskie, John M. Cannon, Cristian Cano, Jackson Codd, Theo Darci-Maher, Ross Ferguson, Nico Flota Sanchez, Alex Fluegel, Walter W, Golay, Philip Griffin, Sean J. Gunderson, Nathalie Haurberg, Prashanna Khatiwada, Olivia Laske, Walker Law, Elliott Lewis, Joshua Marine, Marine McKnight, Emma McNellis, John Momberg, Alexandrea Morena Robert Mutel, William Peterson, Margaret Ridder, Lila Schisgal, Will St John, James Wetzel, Anna Williams, Lauren Wittry

\end{contribution}

\facilities{CXO (HETG), XMM-Newton, NRAO (VLA), ATNF (ASKAP), SOAR (TripleSpec), TESS}

\software{\textsc{ciao} \citep{Fruscione2006}, \textsc{isis} \citep{Houck2000}, \textsc{sas} \citep{Gabriel2004}, \textsc{casa} \citep{Casa2022}, \textsc{specutils}}

\bibliography{bib}{}

@INPROCEEDINGS{Allard2011,
       author = {{Allard}, F. and {Homeier}, D. and {Freytag}, B.},
        title = "{Model Atmospheres From Very Low Mass Stars to Brown Dwarfs}",
    booktitle = {16th Cambridge Workshop on Cool Stars, Stellar Systems, and the Sun},
         year = 2011,
       editor = {{Johns-Krull}, Christopher and {Browning}, Matthew K. and {West}, Andrew A.},
       series = {Astronomical Society of the Pacific Conference Series},
       volume = {448},
        month = dec,
        pages = {91},
          doi = {10.48550/arXiv.1011.5405},
archivePrefix = {arXiv},
       eprint = {1011.5405},
 primaryClass = {astro-ph.SR},
       adsurl = {https://ui.adsabs.harvard.edu/abs/2011ASPC..448...91A}
}

@ARTICLE{Allard2012,
       author = {{Allard}, F. and {Homeier}, D. and {Freytag}, B.},
        title = "{Models of very-low-mass stars, brown dwarfs and exoplanets}",
      journal = {Philosophical Transactions of the Royal Society of London Series A},
         year = 2012,
        month = jun,
       volume = {370},
       number = {1968},
        pages = {2765-2777},
          doi = {10.1098/rsta.2011.0269},
archivePrefix = {arXiv},
       eprint = {1112.3591},
 primaryClass = {astro-ph.SR},
       adsurl = {https://ui.adsabs.harvard.edu/abs/2012RSPTA.370.2765A}
}

@ARTICLE{Anders1989,
       author = {{Anders}, E. and {Grevesse}, N.},
        title = "{Abundances of the elements: Meteoritic and solar}",
      journal = {\gca},
         year = 1989,
        month = jan,
       volume = {53},
       number = {1},
        pages = {197-214},
          doi = {10.1016/0016-7037(89)90286-X},
       adsurl = {https://ui.adsabs.harvard.edu/abs/1989GeCoA..53..197A}
}

@ARTICLE{Argiroffi2019,
       author = {{Argiroffi}, C. and {Reale}, F. and {Drake}, J.~J. and {Ciaravella}, A. and {Testa}, P. and {Bonito}, R. and {Miceli}, M. and {Orlando}, S. and {Peres}, G.},
        title = "{A stellar flare-coronal mass ejection event revealed by X-ray plasma motions}",
      journal = {Nature Astronomy},
         year = 2019,
        month = may,
       volume = {3},
        pages = {742-748},
          doi = {10.1038/s41550-019-0781-4},
archivePrefix = {arXiv},
       eprint = {1905.11325},
 primaryClass = {astro-ph.SR},
       adsurl = {https://ui.adsabs.harvard.edu/abs/2019NatAs...3..742A}
}

@ARTICLE{Barsony2003,
       author = {{Barsony}, M. and {Koresko}, C. and {Matthews}, K.},
        title = "{A Search for Close Binaries in the {\ensuremath{\rho}} Ophiuchi Star-forming Region}",
      journal = {\apj},
         year = 2003,
        month = jul,
       volume = {591},
       number = {2},
        pages = {1064-1074},
          doi = {10.1086/375532},
archivePrefix = {arXiv},
       eprint = {astro-ph/0303595},
 primaryClass = {astro-ph},
       adsurl = {https://ui.adsabs.harvard.edu/abs/2003ApJ...591.1064B}
}

@ARTICLE{Benz1994,
       author = {{Benz}, A.~O. and {Guedel}, M.},
        title = "{X-ray/microwave ratio of flares and coronae}",
      journal = {\aap},
         year = 1994,
        month = may,
       volume = {285},
        pages = {621-630},
       adsurl = {https://ui.adsabs.harvard.edu/abs/1994A&A...285..621B}
}

@ARTICLE{Berne2024,
       author = {{Bern{\'e}}, Olivier and {Habart}, Emilie and {Peeters}, Els and {Schroetter}, Ilane and {Canin}, Am{\'e}lie and {Sidhu}, Ameek and {Chown}, Ryan and {Bron}, Emeric and {Haworth}, Thomas J. and {Klaassen}, Pamela and {Trahin}, Boris and {Van De Putte}, Dries and {Alarc{\'o}n}, Felipe and {Zannese}, Marion and {Abergel}, Alain and {Bergin}, Edwin A. and {Bernard-Salas}, Jeronimo and {Boersma}, Christiaan and {Cami}, Jan and {Cuadrado}, Sara and {Dartois}, Emmanuel and {Dicken}, Daniel and {Elyajouri}, Meriem and {Fuente}, Asunci{\'o}n and {Goicoechea}, Javier R. and {Gordon}, Karl D. and {Issa}, Lina and {Joblin}, Christine and {Kannavou}, Olga and {Khan}, Baria and {Lacinbala}, Ozan and {Languignon}, David and {Le Gal}, Romane and {Maragkoudakis}, Alexandros and {Meshaka}, Raphael and {Okada}, Yoko and {Onaka}, Takashi and {Pasquini}, Sofia and {Pound}, Marc W. and {Robberto}, Massimo and {R{\"o}llig}, Markus and {Schefter}, Bethany and {Schirmer}, Thi{\'e}baut and {Simmer}, Thomas and {Tabone}, Benoit and {Tielens}, Alexander G.~G.~M. and {Vicente}, S{\'\i}lvia and {Wolfire}, Mark G. and {PDRs4All Team} and {Aleman}, Isabel and {Allamandola}, Louis and {Auchettl}, Rebecca and {Baratta}, Giuseppe Antonio and {Baruteau}, Cl{\'e}ment and {Bejaoui}, Salma and {Bera}, Partha P. and {Black}, John H. and {Boulanger}, Francois and {Bouwman}, Jordy and {Brandl}, Bernhard and {Brechignac}, Philippe and {Br{\"u}nken}, Sandra and {Buragohain}, Mridusmita and {Burkhardt}, Andrew and {Candian}, Alessandra and {Cazaux}, St{\'e}phanie and {Cernicharo}, Jose and {Chabot}, Marin and {Chakraborty}, Shubhadip and {Champion}, Jason and {Colgan}, Sean W.~J. and {Cooke}, Ilsa R. and {Coutens}, Audrey and {Cox}, Nick L.~J. and {Demyk}, Karine and {Meyer}, Jennifer Donovan and {Engrand}, C{\'e}cile and {Foschino}, Sacha and {Garc{\'\i}a-Lario}, Pedro and {Gavilan}, Lisseth and {Gerin}, Maryvonne and {Godard}, Marie and {Gottlieb}, Carl A. and {Guillard}, Pierre and {Gusdorf}, Antoine and {Hartigan}, Patrick and {He}, Jinhua and {Herbst}, Eric and {Hornekaer}, Liv and {J{\"a}ger}, Cornelia and {Janot-Pacheco}, Eduardo and {Kaufman}, Michael and {Kemper}, Francisca and {Kendrew}, Sarah and {Kirsanova}, Maria S. and {Knight}, Collin and {Kwok}, Sun and {Labiano}, {\'A}lvaro and {Lai}, Thomas S.-Y. and {Lee}, Timothy J. and {Lefloch}, Bertrand and {Le Petit}, Franck and {Li}, Aigen and {Linz}, Hendrik and {Mackie}, Cameron J. and {Madden}, Suzanne C. and {Mascetti}, Jo{\"e}lle and {McGuire}, Brett A. and {Merino}, Pablo and {Micelotta}, Elisabetta R. and {Morse}, Jon A. and {Mulas}, Giacomo and {Neelamkodan}, Naslim and {Ohsawa}, Ryou and {Paladini}, Roberta and {Palumbo}, Maria Elisabetta and {Pathak}, Amit and {Pendleton}, Yvonne J. and {Petrignani}, Annemieke and {Pino}, Thomas and {Puga}, Elena and {Rangwala}, Naseem and {Rapacioli}, Mathias and {Ricca}, Alessandra and {Roman-Duval}, Julia and {Roueff}, Evelyne and {Rouill{\'e}}, Ga{\"e}l and {Salama}, Farid and {Sales}, Dinalva A. and {Sandstrom}, Karin and {Sarre}, Peter and {Sciamma-O'Brien}, Ella and {Sellgren}, Kris and {Shannon}, Matthew J. and {Simonnin}, Adrien and {Shenoy}, Sachindev S. and {Teyssier}, David and {Thomas}, Richard D. and {Togi}, Aditya and {Verstraete}, Laurent and {Witt}, Adolf N. and {Wootten}, Alwyn and {Ysard}, Nathalie and {Zettergren}, Henning and {Zhang}, Yong and {Zhang}, Ziwei E. and {Zhen}, Junfeng},
        title = "{A far-ultraviolet{\textendash}driven photoevaporation flow observed in a protoplanetary disk}",
      journal = {Science},
         year = 2024,
        month = mar,
       volume = {383},
       number = {6686},
        pages = {988-992},
          doi = {10.1126/science.adh2861},
archivePrefix = {arXiv},
       eprint = {2403.00160},
 primaryClass = {astro-ph.GA},
       adsurl = {https://ui.adsabs.harvard.edu/abs/2024Sci...383..988B}
}

@ARTICLE{Bretthorst1988,
       author = {{Bretthorst}, G. Larry},
        title = "{Bayesian Spectrum Analysis and Parameter Estimation}",
      journal = {Lecture Notes in Statistics},
         year = 1988,
        month = jan,
       volume = {48},
          doi = {10.1007/978-1-4684-9399-3},
       adsurl = {https://ui.adsabs.harvard.edu/abs/1988LNS....48399.3B}
}

@ARTICLE{Casa2022,
       author = {{CASA Team} and {Bean}, Ben and {Bhatnagar}, Sanjay and {Castro}, Sandra and {Donovan Meyer}, Jennifer and {Emonts}, Bjorn and {Garcia}, Enrique and {Garwood}, Robert and {Golap}, Kumar and {Gonzalez Villalba}, Justo and {Harris}, Pamela and {Hayashi}, Yohei and {Hoskins}, Josh and {Hsieh}, Mingyu and {Jagannathan}, Preshanth and {Kawasaki}, Wataru and {Keimpema}, Aard and {Kettenis}, Mark and {Lopez}, Jorge and {Marvil}, Joshua and {Masters}, Joseph and {McNichols}, Andrew and {Mehringer}, David and {Miel}, Renaud and {Moellenbrock}, George and {Montesino}, Federico and {Nakazato}, Takeshi and {Ott}, Juergen and {Petry}, Dirk and {Pokorny}, Martin and {Raba}, Ryan and {Rau}, Urvashi and {Schiebel}, Darrell and {Schweighart}, Neal and {Sekhar}, Srikrishna and {Shimada}, Kazuhiko and {Small}, Des and {Steeb}, Jan-Willem and {Sugimoto}, Kanako and {Suoranta}, Ville and {Tsutsumi}, Takahiro and {van Bemmel}, Ilse M. and {Verkouter}, Marjolein and {Wells}, Akeem and {Xiong}, Wei and {Szomoru}, Arpad and {Griffith}, Morgan and {Glendenning}, Brian and {Kern}, Jeff},
        title = "{CASA, the Common Astronomy Software Applications for Radio Astronomy}",
      journal = {\pasp},
         year = 2022,
        month = nov,
       volume = {134},
       number = {1041},
          eid = {114501},
        pages = {114501},
          doi = {10.1088/1538-3873/ac9642},
archivePrefix = {arXiv},
       eprint = {2210.02276},
 primaryClass = {astro-ph.IM},
       adsurl = {https://ui.adsabs.harvard.edu/abs/2022PASP..134k4501C}
}

@ARTICLE{Cash1979,
       author = {{Cash}, W.},
        title = "{Parameter estimation in astronomy through application of the likelihood ratio.}",
      journal = {\apj},
         year = 1979,
        month = mar,
       volume = {228},
        pages = {939-947},
          doi = {10.1086/156922},
       adsurl = {https://ui.adsabs.harvard.edu/abs/1979ApJ...228..939C}
}

@ARTICLE{Choi2016,
       author = {{Choi}, Jieun and {Dotter}, Aaron and {Conroy}, Charlie and {Cantiello}, Matteo and {Paxton}, Bill and {Johnson}, Benjamin D.},
        title = "{Mesa Isochrones and Stellar Tracks (MIST). I. Solar-scaled Models}",
      journal = {\apj},
         year = 2016,
        month = jun,
       volume = {823},
       number = {2},
          eid = {102},
        pages = {102},
          doi = {10.3847/0004-637X/823/2/102},
archivePrefix = {arXiv},
       eprint = {1604.08592},
 primaryClass = {astro-ph.SR},
       adsurl = {https://ui.adsabs.harvard.edu/abs/2016ApJ...823..102C}
}

@ARTICLE{Favata2005,
       author = {{Favata}, F. and {Flaccomio}, E. and {Reale}, F. and {Micela}, G. and {Sciortino}, S. and {Shang}, H. and {Stassun}, K.~G. and {Feigelson}, E.~D.},
        title = "{Bright X-Ray Flares in Orion Young Stars from COUP: Evidence for Star-Disk Magnetic Fields?}",
      journal = {\apjs},
         year = 2005,
        month = oct,
       volume = {160},
       number = {2},
        pages = {469-502},
          doi = {10.1086/432542},
archivePrefix = {arXiv},
       eprint = {astro-ph/0506134},
 primaryClass = {astro-ph},
       adsurl = {https://ui.adsabs.harvard.edu/abs/2005ApJS..160..469F}
}

@ARTICLE{Feinstein2024,
       author = {{Feinstein}, Adina D. and {Seligman}, Darryl Z. and {France}, Kevin and {Gagn{\'e}}, Jonathan and {Kowalski}, Adam},
        title = "{Evolution of Flare Activity in GKM Stars Younger Than 300 Myr over Five Years of TESS Observations}",
      journal = {\aj},
         year = 2024,
        month = aug,
       volume = {168},
       number = {2},
          eid = {60},
        pages = {60},
          doi = {10.3847/1538-3881/ad4edf},
archivePrefix = {arXiv},
       eprint = {2405.00850},
 primaryClass = {astro-ph.SR},
       adsurl = {https://ui.adsabs.harvard.edu/abs/2024AJ....168...60F}
}

@ARTICLE{Frost2025,
       author = {{Frost}, A.~J. and {Sana}, H. and {Le Bouquin}, J. -B. and {Perets}, H.~B. and {Bodensteiner}, J. and {Igoshev}, A.~P. and {Banyard}, G. and {Mahy}, L. and {M{\'e}rand}, A. and {Ram{\'\i}rez-Agudelo}, O.~H.},
        title = "{An interferometric study of B star multiplicity}",
      journal = {\aap},
         year = 2025,
        month = sep,
       volume = {701},
          eid = {A171},
        pages = {A171},
          doi = {10.1051/0004-6361/202554344},
archivePrefix = {arXiv},
       eprint = {2505.02300},
 primaryClass = {astro-ph.SR},
       adsurl = {https://ui.adsabs.harvard.edu/abs/2025A&A...701A.171F}
}

@INPROCEEDINGS{Fruscione2006,
       author = {{Fruscione}, Antonella and {McDowell}, Jonathan C. and {Allen}, Glenn E. and {Brickhouse}, Nancy S. and {Burke}, Douglas J. and {Davis}, John E. and {Durham}, Nick and {Elvis}, Martin and {Galle}, Elizabeth C. and {Harris}, Daniel E. and {Huenemoerder}, David P. and {Houck}, John C. and {Ishibashi}, Bish and {Karovska}, Margarita and {Nicastro}, Fabrizio and {Noble}, Michael S. and {Nowak}, Michael A. and {Primini}, Frank A. and {Siemiginowska}, Aneta and {Smith}, Randall K. and {Wise}, Michael},
        title = "{CIAO: Chandra's data analysis system}",
    booktitle = {Observatory Operations: Strategies, Processes, and Systems},
         year = 2006,
       editor = {{Silva}, David R. and {Doxsey}, Rodger E.},
       series = {Society of Photo-Optical Instrumentation Engineers (SPIE) Conference Series},
       volume = {6270},
        month = jun,
          eid = {62701V},
        pages = {62701V},
          doi = {10.1117/12.671760},
       adsurl = {https://ui.adsabs.harvard.edu/abs/2006SPIE.6270E..1VF}
}

@INPROCEEDINGS{Gabriel2004,
       author = {{Gabriel}, C. and {Denby}, M. and {Fyfe}, D.~J. and {Hoar}, J. and {Ibarra}, A. and {Ojero}, E. and {Osborne}, J. and {Saxton}, R.~D. and {Lammers}, U. and {Vacanti}, G.},
        title = "{The XMM-Newton SAS - Distributed Development and Maintenance of a Large Science Analysis System: A Critical Analysis}",
    booktitle = {Astronomical Data Analysis Software and Systems (ADASS) XIII},
         year = 2004,
       editor = {{Ochsenbein}, Francois and {Allen}, Mark G. and {Egret}, Daniel},
       series = {Astronomical Society of the Pacific Conference Series},
       volume = {314},
        month = jul,
        pages = {759},
       adsurl = {https://ui.adsabs.harvard.edu/abs/2004ASPC..314..759G}
}

@ARTICLE{Gagne1997,
       author = {{Gagn{\'e}}, Marc and {Caillault}, Jean-Pierre and {Stauffer}, John R. and {Linsky}, Jeffrey L.},
        title = "{Periodic X-Ray Emission from the O7 V Star {\ensuremath{\theta}}$^{1}$ Orionis C}",
      journal = {\apjl},
         year = 1997,
        month = apr,
       volume = {478},
       number = {2},
        pages = {L87-L90},
          doi = {10.1086/310558},
archivePrefix = {arXiv},
       eprint = {astro-ph/9701145},
 primaryClass = {astro-ph},
       adsurl = {https://ui.adsabs.harvard.edu/abs/1997ApJ...478L..87G}
}

@ARTICLE{Golay:2023,
       author = {{Golay}, Walter W. and {Mutel}, Robert L. and {Lipman}, Dani and {G{\"u}del}, Manuel},
        title = "{A search for thermal gyro-synchrotron emission from hot stellar coronae}",
      journal = {\mnras},
         year = 2023,
        month = jun,
       volume = {522},
       number = {1},
        pages = {1394-1410},
          doi = {10.1093/mnras/stad980},
archivePrefix = {arXiv},
       eprint = {2210.11440},
 primaryClass = {astro-ph.SR},
       adsurl = {https://ui.adsabs.harvard.edu/abs/2023MNRAS.522.1394G}
}

@ARTICLE{Gunderson2025b,
       author = {{Gunderson}, Sean J. and {Codd}, Jackson and {Golay}, Walter W. and {Huenemoerder}, David P. and {Cannon}, John M. and {Fluegel}, J. Alex and {Griffin}, Philip E. and {Haurberg}, Nathalie C. and {Ignace}, Richard and {Moreno}, Alexandrea and {Pradhan}, Pragati and {Riggs}, Alexis and {Wetzel}, James and {Canizares}, Claude R. and {The Macro Consortium}},
        title = "{A Multiwavelength View of {\ensuremath{\rho}} Oph. I. Resolving the X-Ray Source between A and B}",
      journal = {\apj},
         year = 2025,
        month = dec,
       volume = {995},
       number = {1},
          eid = {13},
        pages = {13},
          doi = {10.3847/1538-4357/ae0d7e},
archivePrefix = {arXiv},
       eprint = {2509.26268},
 primaryClass = {astro-ph.SR},
       adsurl = {https://ui.adsabs.harvard.edu/abs/2025ApJ...995...13G}
}

@ARTICLE{Gunderson2025c,
       author = {{Gunderson}, Sean J. and {Huenemoerder}, David P. and {Torrej{\'o}n}, Jos{\'e} M. and {Swarm}, Dustin K. and {Nichols}, Joy S. and {Pradhan}, Pragati and {Ignace}, Richard and {Guenther}, Hans Moritz and {Pollock}, A.~M.~T. and {Schulz}, Norbert S.},
        title = "{A Time-dependent Spectral Analysis of {\ensuremath{\gamma}} Cassiopeiae}",
      journal = {\apj},
         year = 2025,
        month = jan,
       volume = {978},
       number = {1},
          eid = {105},
        pages = {105},
          doi = {10.3847/1538-4357/ad944e},
archivePrefix = {arXiv},
       eprint = {2411.11825},
 primaryClass = {astro-ph.HE},
       adsurl = {https://ui.adsabs.harvard.edu/abs/2025ApJ...978..105G}
}

@ARTICLE{Gudel1993,
       author = {{Guedel}, Manuel and {Benz}, Arnold O.},
        title = "{X-Ray/Microwave Relation of Different Types of Active Stars}",
      journal = {\apjl},
         year = 1993,
        month = mar,
       volume = {405},
        pages = {L63},
          doi = {10.1086/186766},
       adsurl = {https://ui.adsabs.harvard.edu/abs/1993ApJ...405L..63G}
}

@ARTICLE{Gudel2004,
       author = {{G{\"u}del}, Manuel},
        title = "{X-ray astronomy of stellar coronae}",
      journal = {\aapr},
         year = 2004,
        month = sep,
       volume = {12},
       number = {2-3},
        pages = {71-237},
          doi = {10.1007/s00159-004-0023-2},
archivePrefix = {arXiv},
       eprint = {astro-ph/0406661},
 primaryClass = {astro-ph},
       adsurl = {https://ui.adsabs.harvard.edu/abs/2004A&ARv..12...71G}
}

@ARTICLE{Gullikson2016,
       author = {{Gullikson}, Kevin and {Kraus}, Adam and {Dodson-Robinson}, Sarah},
        title = "{The Close Companion Mass-ratio Distribution of Intermediate-mass Stars}",
      journal = {\aj},
         year = 2016,
        month = aug,
       volume = {152},
       number = {2},
          eid = {40},
        pages = {40},
          doi = {10.3847/0004-6256/152/2/40},
archivePrefix = {arXiv},
       eprint = {1604.06456},
 primaryClass = {astro-ph.SR},
       adsurl = {https://ui.adsabs.harvard.edu/abs/2016AJ....152...40G}
}

@INPROCEEDINGS{Houck2000,
       author = {{Houck}, J.~C. and {Denicola}, L.~A.},
        title = "{ISIS: An Interactive Spectral Interpretation System for High Resolution X-Ray Spectroscopy}",
    booktitle = {Astronomical Data Analysis Software and Systems IX},
         year = 2000,
       editor = {{Manset}, Nadine and {Veillet}, Christian and {Crabtree}, Dennis},
       series = {Astronomical Society of the Pacific Conference Series},
       volume = {216},
        month = jan,
        pages = {591},
       adsurl = {https://ui.adsabs.harvard.edu/abs/2000ASPC..216..591H}
}

@ARTICLE{Huenemoerder2010,
       author = {{Huenemoerder}, David P. and {Schulz}, Norbert S. and {Testa}, Paola and {Drake}, Jeremy J. and {Osten}, Rachel A. and {Reale}, Fabio},
        title = "{X-ray Flares of EV Lac: Statistics, Spectra, and Diagnostics}",
      journal = {\apj},
         year = 2010,
        month = nov,
       volume = {723},
       number = {2},
        pages = {1558-1567},
          doi = {10.1088/0004-637X/723/2/1558},
archivePrefix = {arXiv},
       eprint = {1006.2558},
 primaryClass = {astro-ph.HE},
       adsurl = {https://ui.adsabs.harvard.edu/abs/2010ApJ...723.1558H}
}

@ARTICLE{Huenemoerder2013,
       author = {{Huenemoerder}, David P. and {Phillips}, Kenneth J.~H. and {Sylwester}, Janusz and {Sylwester}, Barbara},
        title = "{Stellar Coronae, Solar Flares: A Detailed Comparison of {\ensuremath{\sigma}} GEM, HR 1099, and the Sun in High-resolution X-Rays}",
      journal = {\apj},
         year = 2013,
        month = may,
       volume = {768},
       number = {2},
          eid = {135},
        pages = {135},
          doi = {10.1088/0004-637X/768/2/135},
archivePrefix = {arXiv},
       eprint = {1304.0408},
 primaryClass = {astro-ph.SR},
       adsurl = {https://ui.adsabs.harvard.edu/abs/2013ApJ...768..135H}
}

@ARTICLE{Huenemoerder2012,
       author = {{Huenemoerder}, David P. and {Oskinova}, Lidia M. and {Ignace}, Richard and {Waldron}, Wayne L. and {Todt}, Helge and {Hamaguchi}, Kenji and {Kitamoto}, Shunji},
        title = "{On the Weak-wind Problem in Massive Stars: X-Ray Spectra Reveal a Massive Hot Wind in {\ensuremath{\mu}} Columbae}",
      journal = {\apjl},
         year = 2012,
        month = sep,
       volume = {756},
       number = {2},
          eid = {L34},
        pages = {L34},
          doi = {10.1088/2041-8205/756/2/L34},
archivePrefix = {arXiv},
       eprint = {1208.0820},
 primaryClass = {astro-ph.SR},
       adsurl = {https://ui.adsabs.harvard.edu/abs/2012ApJ...756L..34H}
}

@ARTICLE{Huenemoerder2020,
       author = {{Huenemoerder}, David P. and {Ignace}, Richard and {Miller}, Nathan A. and {Gayley}, Kenneth G. and {Hamann}, Wolf-Rainer and {Lauer}, Jennifer and {Moffat}, Anthony F.~J. and {Naz{\'e}}, Ya{\"e}l and {Nichols}, Joy S. and {Oskinova}, Lidia and {Richardson}, Noel D. and {Waldron}, Wayne},
        title = "{A Deep Exposure in High Resolution X-Rays Reveals the Hottest Plasma in the {\ensuremath{\zeta}} Puppis Wind}",
      journal = {\apj},
         year = 2020,
        month = apr,
       volume = {893},
       number = {1},
          eid = {52},
        pages = {52},
          doi = {10.3847/1538-4357/ab8005},
archivePrefix = {arXiv},
       eprint = {2003.06889},
 primaryClass = {astro-ph.SR},
       adsurl = {https://ui.adsabs.harvard.edu/abs/2020ApJ...893...52H}
}

@ARTICLE{Jilinski2006,
       author = {{Jilinski}, E. and {Daflon}, S. and {Cunha}, K. and {de La Reza}, R.},
        title = "{Radial velocity measurements of B stars in the Scorpius-Centaurus association}",
      journal = {\aap},
         year = 2006,
        month = mar,
       volume = {448},
       number = {3},
        pages = {1001-1006},
          doi = {10.1051/0004-6361:20041614},
archivePrefix = {arXiv},
       eprint = {astro-ph/0601643},
 primaryClass = {astro-ph},
       adsurl = {https://ui.adsabs.harvard.edu/abs/2006A&A...448.1001J}
}

@ARTICLE{Lestrade1993,
       author = {{Lestrade}, Jean-Francois and {Phillips}, Robert B. and {Hodges}, Mark W. and {Preston}, Robert A.},
        title = "{VLBI Astrometric Identification of the Radio-emitting Region in Algol and Determination of the Orientation of the Close Binary}",
      journal = {\apj},
         year = 1993,
        month = jun,
       volume = {410},
        pages = {808},
          doi = {10.1086/172798},
       adsurl = {https://ui.adsabs.harvard.edu/abs/1993ApJ...410..808L}
}

@INPROCEEDINGS{Marshall2004,
       author = {{Marshall}, Herman L. and {Tennant}, Allyn and {Grant}, Catherine E. and {Hitchcock}, Adam P. and {O'Dell}, Stephen L. and {Plucinsky}, Paul P.},
        title = "{Composition of the Chandra ACIS contaminant}",
    booktitle = {X-Ray and Gamma-Ray Instrumentation for Astronomy XIII},
         year = 2004,
       editor = {{Flanagan}, Kathryn A. and {Siegmund}, Oswald H.~W.},
       series = {Society of Photo-Optical Instrumentation Engineers (SPIE) Conference Series},
       volume = {5165},
        month = feb,
        pages = {497-508},
          doi = {10.1117/12.508310},
archivePrefix = {arXiv},
       eprint = {astro-ph/0308332},
 primaryClass = {astro-ph},
       adsurl = {https://ui.adsabs.harvard.edu/abs/2004SPIE.5165..497M}
}

@dataset{Mason2025,
       author = {{Mason}, B.~D. and {Wycoff}, G.~L. and {Hartkopf}, W.~I. and {Douglass}, G.~G. and {Worley}, C.~E.},
        title = "{VizieR Online Data Catalog: The Washington Visual Double Star Catalog (Mason+ 2001-2020)}",
 howpublished = {VizieR On-line Data Catalog: B/wds.  Originally published in: 2001AJ....122.3466M},
         year = 2025,
        month = nov,
          eid = {B/wds},
       adsurl = {https://ui.adsabs.harvard.edu/abs/2025yCat....102026M}
}

@ARTICLE{Moe2025,
       author = {{Moe}, Maxwell and {Oey}, M.~S. and {Vargas-Salazar}, Irene and {Kratter}, Kaitlin M.},
        title = "{The Close Binary Properties of Massive Stars across Different Environments within the LMC}",
      journal = {\apj},
         year = 2025,
        month = oct,
       volume = {991},
       number = {2},
          eid = {182},
        pages = {182},
          doi = {10.3847/1538-4357/ae0034},
archivePrefix = {arXiv},
       eprint = {2508.20319},
 primaryClass = {astro-ph.SR},
       adsurl = {https://ui.adsabs.harvard.edu/abs/2025ApJ...991..182M}
}

@ARTICLE{Naze2026,
       author = {{Naz{\'e}}, Ya{\"e}l and {Tsujimoto}, Masahiro and {Rauw}, Gregor and {Gunderson}, Sean J.},
        title = "{Orbital motion detected in {\ensuremath{\gamma}} Cas Fe K emission lines}",
      journal = {\aap},
         year = 2026,
        month = mar,
       volume = {707},
          eid = {A334},
        pages = {A334},
          doi = {10.1051/0004-6361/202558284},
archivePrefix = {arXiv},
       eprint = {2603.22938},
 primaryClass = {astro-ph.SR},
       adsurl = {https://ui.adsabs.harvard.edu/abs/2026A&A...707A.334N}
}

@ARTICLE{Ndugu2024,
       author = {{Ndugu}, N. and {Bitsch}, B. and {Lienert}, J.~L.},
        title = "{How external photoevaporation changes the chemical composition of the inner disc}",
      journal = {\aap},
         year = 2024,
        month = nov,
       volume = {691},
          eid = {A32},
        pages = {A32},
          doi = {10.1051/0004-6361/202451633},
archivePrefix = {arXiv},
       eprint = {2409.07596},
 primaryClass = {astro-ph.EP},
       adsurl = {https://ui.adsabs.harvard.edu/abs/2024A&A...691A..32N}
}

@ARTICLE{Novakovic2007,
       author = {{Novakovi{\'c}}, B.},
        title = "{Orbits of Five Visual Binary Stars}",
      journal = {Baltic Astronomy},
         year = 2007,
        month = jan,
       volume = {16},
        pages = {435-442},
          doi = {10.48550/arXiv.0712.4242},
archivePrefix = {arXiv},
       eprint = {0712.4242},
 primaryClass = {astro-ph},
       adsurl = {https://ui.adsabs.harvard.edu/abs/2007BaltA..16..435N}
}

@ARTICLE{Nordon2007,
       author = {{Nordon}, R. and {Behar}, E.},
        title = "{Six large coronal X-ray flares observed with Chandra}",
      journal = {\aap},
         year = 2007,
        month = mar,
       volume = {464},
       number = {1},
        pages = {309-321},
          doi = {10.1051/0004-6361:20066449},
archivePrefix = {arXiv},
       eprint = {astro-ph/0611386},
 primaryClass = {astro-ph},
       adsurl = {https://ui.adsabs.harvard.edu/abs/2007A&A...464..309N}
}

@INPROCEEDINGS{Odell2017,
       author = {{O'Dell}, Stephen L. and {Swartz}, Douglas A. and {Tice}, Neil W. and {Plucinsky}, Paul P. and {Marshall}, Herman L. and {Bogdan}, Akos and {Grant}, Catherine E. and {Tennant}, Allyn F. and {Dahmer}, Matthew},
        title = "{Modeling contamination migration on the Chandra X-ray Observatory IV}",
    booktitle = {Society of Photo-Optical Instrumentation Engineers (SPIE) Conference Series},
         year = 2017,
       editor = {{Siegmund}, Oswald H.},
       series = {Society of Photo-Optical Instrumentation Engineers (SPIE) Conference Series},
       volume = {10397},
        month = aug,
          eid = {103970C},
        pages = {103970C},
          doi = {10.1117/12.2274818},
       adsurl = {https://ui.adsabs.harvard.edu/abs/2017SPIE10397E..0CO}
}

@ARTICLE{Pelisoli2020,
       author = {{Pelisoli}, Ingrid and {Vos}, Joris and {Geier}, Stephan and {Schaffenroth}, Veronika and {Baran}, Andrzej S.},
        title = "{Alone but not lonely: Observational evidence that binary interaction is always required to form hot subdwarf stars}",
      journal = {\aap},
         year = 2020,
        month = oct,
       volume = {642},
          eid = {A180},
        pages = {A180},
          doi = {10.1051/0004-6361/202038473},
archivePrefix = {arXiv},
       eprint = {2008.07522},
 primaryClass = {astro-ph.SR},
       adsurl = {https://ui.adsabs.harvard.edu/abs/2020A&A...642A.180P}
}

@ARTICLE{Petite2013,
       author = {{Petit}, V. and {Owocki}, S.~P. and {Wade}, G.~A. and {Cohen}, D.~H. and {Sundqvist}, J.~O. and {Gagn{\'e}}, M. and {Ma{\'\i}z Apell{\'a}niz}, J. and {Oksala}, M.~E. and {Bohlender}, D.~A. and {Rivinius}, T. and {Henrichs}, H.~F. and {Alecian}, E. and {Townsend}, R.~H.~D. and {ud-Doula}, A. and {MiMeS Collaboration}},
        title = "{A magnetic confinement versus rotation classification of massive-star magnetospheres}",
      journal = {\mnras},
         year = 2013,
        month = feb,
       volume = {429},
       number = {1},
        pages = {398-422},
          doi = {10.1093/mnras/sts344},
archivePrefix = {arXiv},
       eprint = {1211.0282},
 primaryClass = {astro-ph.SR},
       adsurl = {https://ui.adsabs.harvard.edu/abs/2013MNRAS.429..398P}
}

@ARTICLE{Pradhan2023,
       author = {{Pradhan}, Pragati and {Huenemoerder}, David P. and {Ignace}, Richard and {Nichols}, Joy S. and {Pollock}, A.~M.~T.},
        title = "{Survey of X-Rays from Massive Stars Observed at High Spectral Resolution with Chandra}",
      journal = {\apj},
         year = 2023,
        month = sep,
       volume = {954},
       number = {2},
          eid = {123},
        pages = {123},
          doi = {10.3847/1538-4357/ace9d6},
archivePrefix = {arXiv},
       eprint = {2308.00758},
 primaryClass = {astro-ph.SR},
       adsurl = {https://ui.adsabs.harvard.edu/abs/2023ApJ...954..123P}
}

@ARTICLE{Renzo2019,
       author = {{Renzo}, M. and {Zapartas}, E. and {de Mink}, S.~E. and {G{\"o}tberg}, Y. and {Justham}, S. and {Farmer}, R.~J. and {Izzard}, R.~G. and {Toonen}, S. and {Sana}, H.},
        title = "{Massive runaway and walkaway stars. A study of the kinematical imprints of the physical processes governing the evolution and explosion of their binary progenitors}",
      journal = {\aap},
         year = 2019,
        month = apr,
       volume = {624},
          eid = {A66},
        pages = {A66},
          doi = {10.1051/0004-6361/201833297},
archivePrefix = {arXiv},
       eprint = {1804.09164},
 primaryClass = {astro-ph.SR},
       adsurl = {https://ui.adsabs.harvard.edu/abs/2019A&A...624A..66R}
}

@ARTICLE{Shultz2025,
       author = {{Shultz}, M.~E. and {Berry}, I. and {Bohlender}, D. and {Catanzaro}, G. and {Giarrusso}, M. and {Klement}, R. and {Labadie-Bartz}, J. and {Leone}, F. and {Leto}, P. and {Neiner}, C. and {Owocki}, S.~P. and {Rivinius}, Th. and {ud-Doula}, A. and {Wade}, G.~A.},
        title = "{Discovery of the binary nature of the magnetospheric B-type star {\ensuremath{\rho}} Oph A}",
      journal = {\aap},
         year = 2025,
        month = aug,
       volume = {700},
          eid = {A14},
        pages = {A14},
          doi = {10.1051/0004-6361/202554023},
archivePrefix = {arXiv},
       eprint = {2505.08007},
 primaryClass = {astro-ph.SR},
       adsurl = {https://ui.adsabs.harvard.edu/abs/2025A&A...700A..14S}
}

@ARTICLE{Shulz2024,
       author = {{Schulz}, Norbert S. and {Huenemoerder}, David P. and {Principe}, David A. and {Gagne}, Marc and {G{\"u}nther}, Hans Moritz and {Kastner}, Joel and {Nichols}, Joy and {Pollock}, Andrew and {Preibisch}, Thomas and {Testa}, Paola and {Reale}, Fabio and {Favata}, Fabio and {Canizares}, Claude R.},
        title = "{The Nature of X-Rays from Young Stellar Objects in the Orion Nebula Cluster{\textemdash}A Chandra HETGS Legacy Project}",
      journal = {\apj},
         year = 2024,
        month = aug,
       volume = {970},
       number = {2},
          eid = {190},
        pages = {190},
          doi = {10.3847/1538-4357/ad47c2},
archivePrefix = {arXiv},
       eprint = {2404.19676},
 primaryClass = {astro-ph.SR},
       adsurl = {https://ui.adsabs.harvard.edu/abs/2024ApJ...970..190S}
}

@article{Shultz2026,
    author = {{Shultz}, M.~E. and {Alecian}, E. and {Berry}, I. and {Krticka}, J. and {Labadie-Bartz}, J. and {Leto}, P. and {Owocki}, S.~P. and {ud-Doula}, A. and {Wade}, A.~G.},
    title = {Revisiting the centrifugal breakout scaling relationship for Hα emission in light of ρ Oph C, the coolest B-type star with an Hα-bright centrifugal magnetosphere},
    journal = {\aap},
    year = {in press}
}

@software{specutils2025,
       author = {{Earl}, Nicholas and {Tollerud}, Erik and {O'Steen}, Ricky and {brechmos} and {Kerzendorf}, Wolfgang and {Busko}, Ivo and {shaileshahuja} and {Lim}, P.~L. and {D'Avella}, Dan and {Robitaille}, Thomas and {Ginsburg}, Adam and {Homeier}, Derek and {Sip{\H{o}}cz}, Brigitta and {Averbukh}, Jesse and {Cherinka}, Brian and {Tocknell}, James and {Ogaz}, Sara and {Geda}, Robel and {Davies}, James and {Conroy}, Kyle and {G{\"u}nther}, Hans Moritz and {Barbary}, Kyle and {Cruz}, Kelle and {Foster}, Jonathan and {Droettboom}, Michael and {Nguyen}, Duy and {Bray}, E.~M. and {Casey}, Andy and {Ferguson}, Henry},
        title = "{astropy/specutils: v2.1.0}",
         year = 2025,
        month = jul,
          eid = {10.5281/zenodo.16615456},
          doi = {10.5281/zenodo.16615456},
      version = {v2.1.0},
    publisher = {Zenodo},
       adsurl = {https://ui.adsabs.harvard.edu/abs/2025zndo..16615456E}
}

@ARTICLE{Tovar2022,
       author = {{Tovar Mendoza}, Guadalupe and {Davenport}, James R.~A. and {Agol}, Eric and {Jackman}, James A.~G. and {Hawley}, Suzanne L.},
        title = "{Llamaradas Estelares: Modeling the Morphology of White-light Flares}",
      journal = {\aj},
         year = 2022,
        month = jul,
       volume = {164},
       number = {1},
          eid = {17},
        pages = {17},
          doi = {10.3847/1538-3881/ac6fe6},
archivePrefix = {arXiv},
       eprint = {2205.05706},
 primaryClass = {astro-ph.SR},
       adsurl = {https://ui.adsabs.harvard.edu/abs/2022AJ....164...17T}
}

@ARTICLE{Townsend2005,
       author = {{Townsend}, R.~H.~D. and {Owocki}, S.~P.},
        title = "{A rigidly rotating magnetosphere model for circumstellar emission from magnetic OB stars}",
      journal = {\mnras},
         year = 2005,
        month = feb,
       volume = {357},
       number = {1},
        pages = {251-264},
          doi = {10.1111/j.1365-2966.2005.08642.x},
archivePrefix = {arXiv},
       eprint = {astro-ph/0408565},
 primaryClass = {astro-ph},
       adsurl = {https://ui.adsabs.harvard.edu/abs/2005MNRAS.357..251T}
}

@ARTICLE{Tsujimoto2018,
       author = {{Tsujimoto}, Masahiro and {Morihana}, Kumiko and {Hayashi}, Takayuki and {Kitaguchi}, Takao},
        title = "{Suzaku and NuSTAR X-ray spectroscopy of {\ensuremath{\gamma}} Cassiopeiae and HD 110432}",
      journal = {\pasj},
         year = 2018,
        month = dec,
       volume = {70},
       number = {6},
          eid = {109},
        pages = {109},
          doi = {10.1093/pasj/psy111},
archivePrefix = {arXiv},
       eprint = {1809.01419},
 primaryClass = {astro-ph.HE},
       adsurl = {https://ui.adsabs.harvard.edu/abs/2018PASJ...70..109T}
}

@ARTICLE{Tsujimoto2023,
       author = {{Tsujimoto}, Masahiro and {Hayashi}, Takayuki and {Morihana}, Kumiko and {Moritani}, Yuki},
        title = "{X-ray and optical spectroscopic study of a {\ensuremath{\gamma}} Cassiopeiae analog source {\ensuremath{\pi}} Aquarii}",
      journal = {\pasj},
         year = 2023,
        month = feb,
       volume = {75},
       number = {1},
        pages = {177-186},
          doi = {10.1093/pasj/psac099},
archivePrefix = {arXiv},
       eprint = {2211.10803},
 primaryClass = {astro-ph.HE},
       adsurl = {https://ui.adsabs.harvard.edu/abs/2023PASJ...75..177T}
}

@ARTICLE{Wade2006,
       author = {{Wade}, G.~A. and {Fullerton}, A.~W. and {Donati}, J.-F. and {Landstreet}, J.~D. and {Petit}, P. and {Strasser}, S.},
        title = "{The magnetic field and confined wind of the O star {\ensuremath{\theta}}$^{1}$ Orionis C}",
      journal = {\aap},
         year = 2006,
        month = may,
       volume = {451},
       number = {1},
        pages = {195-206},
          doi = {10.1051/0004-6361:20054380},
archivePrefix = {arXiv},
       eprint = {astro-ph/0601623},
 primaryClass = {astro-ph},
       adsurl = {https://ui.adsabs.harvard.edu/abs/2006A&A...451..195W}
}

@ARTICLE{RomanLopes2018,
       author = {{Roman-Lopes}, A. and {Rom{\'a}n-Z{\'u}{\~n}iga}, C. and {Tapia}, Mauricio and {Chojnowski}, Drew and {G{\'o}mez Maqueo Chew}, Y. and {Garc{\'\i}a-Hern{\'a}ndez}, D.~A. and {Borissova}, Jura and {Minniti}, Dante and {Covey}, Kevin R. and {Longa-Pe{\~n}a}, Pen{\'e}lope and {Fernandez-Trincado}, J.~G. and {Zamora}, Olga and {Nitschelm}, Christian},
        title = "{Massive Stars in the SDSS-IV/APOGEE SURVEY. I. OB Stars}",
      journal = {\apj},
         year = 2018,
        month = mar,
       volume = {855},
       number = {1},
          eid = {68},
        pages = {68},
          doi = {10.3847/1538-4357/aaac27},
archivePrefix = {arXiv},
       eprint = {1802.01724},
 primaryClass = {astro-ph.GA},
       adsurl = {https://ui.adsabs.harvard.edu/abs/2018ApJ...855...68R}
}

@INPROCEEDINGS{Schlawinl14,
       author = {{Schlawin}, E. and {Herter}, T.~L. and {Henderson}, C. and {Wilson}, J.~C. and {Probst}, R. and {Sprayberry}, D. and {Bonati}, M. and {Schurter}, P. and {James}, D. and {Warner}, M. and {Tighe}, R. and {Adams}, J.~D. and {Mart{\'\i}nez}, M.},
        title = "{Design updates and status of the fourth generation TripleSpec spectrograph}",
    booktitle = {Ground-based and Airborne Instrumentation for Astronomy V},
         year = 2014,
       editor = {{Ramsay}, Suzanne K. and {McLean}, Ian S. and {Takami}, Hideki},
       series = {Society of Photo-Optical Instrumentation Engineers (SPIE) Conference Series},
       volume = {9147},
        month = aug,
          eid = {91472H},
        pages = {91472H},
          doi = {10.1117/12.2055233},
       adsurl = {https://ui.adsabs.harvard.edu/abs/2014SPIE.9147E..2HS}
}

@ARTICLE{Cushing2004,
       author = {{Cushing}, Michael C. and {Vacca}, William D. and {Rayner}, John T.},
        title = "{Spextool: A Spectral Extraction Package for SpeX, a 0.8-5.5 Micron Cross-Dispersed Spectrograph}",
      journal = {\pasp},
         year = 2004,
        month = apr,
       volume = {116},
       number = {818},
        pages = {362-376},
          doi = {10.1086/382907},
       adsurl = {https://ui.adsabs.harvard.edu/abs/2004PASP..116..362C}
}

@ARTICLE{Condon:1998,
       author = {{Condon}, J.~J. and {Cotton}, W.~D. and {Greisen}, E.~W. and {Yin}, Q.~F. and {Perley}, R.~A. and {Taylor}, G.~B. and {Broderick}, J.~J.},
        title = "{The NRAO VLA Sky Survey}",
      journal = {\aj},
         year = 1998,
        month = may,
       volume = {115},
       number = {5},
        pages = {1693-1716},
          doi = {10.1086/300337},
       adsurl = {https://ui.adsabs.harvard.edu/abs/1998AJ....115.1693C}
}

@ARTICLE{Lacy:2020,
       author = {{Lacy}, M. and {Baum}, S.~A. and {Chandler}, C.~J. and {Chatterjee}, S. and {Clarke}, T.~E. and {Deustua}, S. and {English}, J. and {Farnes}, J. and {Gaensler}, B.~M. and {Gugliucci}, N. and {Hallinan}, G. and {Kent}, B.~R. and {Kimball}, A. and {Law}, C.~J. and {Lazio}, T.~J.~W. and {Marvil}, J. and {Mao}, S.~A. and {Medlin}, D. and {Mooley}, K. and {Murphy}, E.~J. and {Myers}, S. and {Osten}, R. and {Richards}, G.~T. and {Rosolowsky}, E. and {Rudnick}, L. and {Schinzel}, F. and {Sivakoff}, G.~R. and {Sjouwerman}, L.~O. and {Taylor}, R. and {White}, R.~L. and {Wrobel}, J. and {Andernach}, H. and {Beasley}, A.~J. and {Berger}, E. and {Bhatnager}, S. and {Birkinshaw}, M. and {Bower}, G.~C. and {Brandt}, W.~N. and {Brown}, S. and {Burke-Spolaor}, S. and {Butler}, B.~J. and {Comerford}, J. and {Demorest}, P.~B. and {Fu}, H. and {Giacintucci}, S. and {Golap}, K. and {G{\"u}th}, T. and {Hales}, C.~A. and {Hiriart}, R. and {Hodge}, J. and {Horesh}, A. and {Ivezi{\'c}}, {\v{Z}}. and {Jarvis}, M.~J. and {Kamble}, A. and {Kassim}, N. and {Liu}, X. and {Loinard}, L. and {Lyons}, D.~K. and {Masters}, J. and {Mezcua}, M. and {Moellenbrock}, G.~A. and {Mroczkowski}, T. and {Nyland}, K. and {O'Dea}, C.~P. and {O'Sullivan}, S.~P. and {Peters}, W.~M. and {Radford}, K. and {Rao}, U. and {Robnett}, J. and {Salcido}, J. and {Shen}, Y. and {Sobotka}, A. and {Witz}, S. and {Vaccari}, M. and {van Weeren}, R.~J. and {Vargas}, A. and {Williams}, P.~K.~G. and {Yoon}, I.},
        title = "{The Karl G. Jansky Very Large Array Sky Survey (VLASS). Science Case and Survey Design}",
      journal = {\pasp},
         year = 2020,
        month = mar,
       volume = {132},
       number = {1009},
          eid = {035001},
        pages = {035001},
          doi = {10.1088/1538-3873/ab63eb},
archivePrefix = {arXiv},
       eprint = {1907.01981},
 primaryClass = {astro-ph.IM},
       adsurl = {https://ui.adsabs.harvard.edu/abs/2020PASP..132c5001L}
}

@ARTICLE{Allison:2022,
       author = {{Allison}, James R. and {Sadler}, E.~M. and {Amaral}, A.~D. and {An}, T. and {Curran}, S.~J. and {Darling}, J. and {Edge}, A.~C. and {Ellison}, S.~L. and {Emig}, K.~L. and {Gaensler}, B.~M. and {Garratt-Smithson}, L. and {Glowacki}, M. and {Grasha}, K. and {Koribalski}, B.~S. and {Lagos}, C. del P. and {Lah}, P. and {Mahony}, E.~K. and {Mao}, S.~A. and {Morganti}, R. and {Moss}, V.~A. and {Pettini}, M. and {Pimbblet}, K.~A. and {Power}, C. and {Salas}, P. and {Staveley-Smith}, L. and {Whiting}, M.~T. and {Wong}, O.~I. and {Yoon}, H. and {Zheng}, Z. and {Zwaan}, M.~A.},
        title = "{The First Large Absorption Survey in H I (FLASH): I. Science goals and survey design}",
      journal = {\pasa},
         year = 2022,
        month = jan,
       volume = {39},
          eid = {e010},
        pages = {e010},
          doi = {10.1017/pasa.2022.3},
archivePrefix = {arXiv},
       eprint = {2110.00469},
 primaryClass = {astro-ph.GA},
       adsurl = {https://ui.adsabs.harvard.edu/abs/2022PASA...39...10A}
}

@ARTICLE{Koribalski:2020,
       author = {{Koribalski}, B{\"a}rbel S. and {Staveley-Smith}, L. and {Westmeier}, T. and {Serra}, P. and {Spekkens}, K. and {Wong}, O.~I. and {Lee-Waddell}, K. and {Lagos}, C.~D.~P. and {Obreschkow}, D. and {Ryan-Weber}, E.~V. and {Zwaan}, M. and {Kilborn}, V. and {Bekiaris}, G. and {Bekki}, K. and {Bigiel}, F. and {Boselli}, A. and {Bosma}, A. and {Catinella}, B. and {Chauhan}, G. and {Cluver}, M.~E. and {Colless}, M. and {Courtois}, H.~M. and {Crain}, R.~A. and {de Blok}, W.~J.~G. and {D{\'e}nes}, H. and {Duffy}, A.~R. and {Elagali}, A. and {Fluke}, C.~J. and {For}, B.-Q. and {Heald}, G. and {Henning}, P.~A. and {Hess}, K.~M. and {Holwerda}, B.~W. and {Howlett}, C. and {Jarrett}, T. and {Jones}, D.~H. and {Jones}, M.~G. and {J{\'o}zsa}, G.~I.~G. and {Jurek}, R. and {J{\"u}tte}, E. and {Kamphuis}, P. and {Karachentsev}, I. and {Kerp}, J. and {Kleiner}, D. and {Kraan-Korteweg}, R.~C. and {L{\'o}pez-S{\'a}nchez}, {\'A}. R. and {Madrid}, J. and {Meyer}, M. and {Mould}, J. and {Murugeshan}, C. and {Norris}, R.~P. and {Oh}, S.-H. and {Oosterloo}, T.~A. and {Popping}, A. and {Putman}, M. and {Reynolds}, T.~N. and {Rhee}, J. and {Robotham}, A.~S.~G. and {Ryder}, S. and {Schr{\"o}der}, A.~C. and {Shao}, Li and {Stevens}, A.~R.~H. and {Taylor}, E.~N. and {van{\^A} der Hulst}, J.~M. and {Verdes-Montenegro}, L. and {Wakker}, B.~P. and {Wang}, J. and {Whiting}, M. and {Winkel}, B. and {Wolf}, C.},
        title = "{WALLABY {\textendash} an SKA Pathfinder H I survey}",
      journal = {\apss},
         year = 2020,
        month = jul,
       volume = {365},
       number = {7},
          eid = {118},
        pages = {118},
          doi = {10.1007/s10509-020-03831-4},
archivePrefix = {arXiv},
       eprint = {2002.07311},
 primaryClass = {astro-ph.GA},
       adsurl = {https://ui.adsabs.harvard.edu/abs/2020Ap&SS.365..118K}
}

@ARTICLE{McConnell:2020,
       author = {{McConnell}, D. and {Hale}, C.~L. and {Lenc}, E. and {Banfield}, J.~K. and {Heald}, George and {Hotan}, A.~W. and {Leung}, James K. and {Moss}, Vanessa A. and {Murphy}, Tara and {O'Brien}, Andrew and {Pritchard}, Joshua and {Raja}, Wasim and {Sadler}, Elaine M. and {Stewart}, Adam and {Thomson}, Alec J.~M. and {Whiting}, M. and {Allison}, James R. and {Amy}, S.~W. and {Anderson}, C. and {Ball}, Lewis and {Bannister}, Keith W. and {Bell}, Martin and {Bock}, Douglas C.-J. and {Bolton}, Russ and {Bunton}, J.~D. and {Chippendale}, A.~P. and {Collier}, J.~D. and {Cooray}, F.~R. and {Cornwell}, T.~J. and {Diamond}, P.~J. and {Edwards}, P.~G. and {Gupta}, N. and {Hayman}, Douglas B. and {Heywood}, Ian and {Jackson}, C.~A. and {Koribalski}, B{\"a}rbel S. and {Lee-Waddell}, Karen and {McClure-Griffiths}, N.~M. and {Ng}, Alan and {Norris}, Ray P. and {Phillips}, Chris and {Reynolds}, John E. and {Roxby}, Daniel N. and {Schinckel}, Antony E.~T. and {Shields}, Matt and {Tremblay}, Chenoa and {Tzioumis}, A. and {Voronkov}, M.~A. and {Westmeier}, Tobias},
        title = "{The Rapid ASKAP Continuum Survey I: Design and first results}",
      journal = {\pasa},
         year = 2020,
        month = nov,
       volume = {37},
          eid = {e048},
        pages = {e048},
          doi = {10.1017/pasa.2020.41},
archivePrefix = {arXiv},
       eprint = {2012.00747},
 primaryClass = {astro-ph.IM},
       adsurl = {https://ui.adsabs.harvard.edu/abs/2020PASA...37...48M}
}

@ARTICLE{Leto2020,
       author = {{Leto}, P. and {Trigilio}, C. and {Buemi}, C.~S. and {Leone}, F. and {Pillitteri}, I. and {Fossati}, L. and {Cavallaro}, F. and {Oskinova}, L.~M. and {Ignace}, R. and {Krti{\v{c}}ka}, J. and {Umana}, G. and {Catanzaro}, G. and {Ingallinera}, A. and {Bufano}, F. and {Riggi}, S. and {Cerrigone}, L. and {Loru}, S. and {Schillir{\'o}}, F. and {Agliozzo}, C. and {Phillips}, N.~M. and {Giarrusso}, M. and {Robrade}, J.},
        title = "{The auroral radio emission of the magnetic B-type star {\ensuremath{\rho}}-=OphC}",
      journal = {\mnras},
         year = 2020,
        month = dec,
       volume = {499},
       number = {1},
        pages = {L72-L76},
          doi = {10.1093/mnrasl/slaa157},
archivePrefix = {arXiv},
       eprint = {2009.02363},
 primaryClass = {astro-ph.SR},
       adsurl = {https://ui.adsabs.harvard.edu/abs/2020MNRAS.499L..72L}
}

@ARTICLE{Leto:2020B,
       author = {{Leto}, P. and {Trigilio}, C. and {Leone}, F. and {Pillitteri}, I. and {Buemi}, C.~S. and {Fossati}, L. and {Cavallaro}, F. and {Oskinova}, L.~M. and {Ignace}, R. and {Krti{\v{c}}ka}, J. and {Umana}, G. and {Catanzaro}, G. and {Ingallinera}, A. and {Bufano}, F. and {Agliozzo}, C. and {Phillips}, N.~M. and {Cerrigone}, L. and {Riggi}, S. and {Loru}, S. and {Munari}, M. and {Gangi}, M. and {Giarrusso}, M. and {Robrade}, J.},
        title = "{Evidence for radio and X-ray auroral emissions from the magnetic B-type star {\ensuremath{\rho}} Oph A}",
      journal = {\mnras},
         year = 2020,
        month = apr,
       volume = {493},
       number = {4},
        pages = {4657-4676},
          doi = {10.1093/mnras/staa587},
archivePrefix = {arXiv},
       eprint = {2002.09251},
 primaryClass = {astro-ph.SR},
       adsurl = {https://ui.adsabs.harvard.edu/abs/2020MNRAS.493.4657L}
}

@ARTICLE{Leto:2021,
       author = {{Leto}, P. and {Trigilio}, C. and {Krti{\v{c}}ka}, J. and {Fossati}, L. and {Ignace}, R. and {Shultz}, M.~E. and {Buemi}, C.~S. and {Cerrigone}, L. and {Umana}, G. and {Ingallinera}, A. and {Bordiu}, C. and {Pillitteri}, I. and {Bufano}, F. and {Oskinova}, L.~M. and {Agliozzo}, C. and {Cavallaro}, F. and {Riggi}, S. and {Loru}, S. and {Todt}, H. and {Giarrusso}, M. and {Phillips}, N.~M. and {Robrade}, J. and {Leone}, F.},
        title = "{A scaling relationship for non-thermal radio emission from ordered magnetospheres: from the top of the main sequence to planets}",
      journal = {\mnras},
         year = 2021,
        month = oct,
       volume = {507},
       number = {2},
        pages = {1979-1998},
          doi = {10.1093/mnras/stab2168},
archivePrefix = {arXiv},
       eprint = {2107.11995},
 primaryClass = {astro-ph.SR},
       adsurl = {https://ui.adsabs.harvard.edu/abs/2021MNRAS.507.1979L}
}

@ARTICLE{Stibbs:1950,
       author = {{Stibbs}, D.~W.~N.},
        title = "{A study of the spectrum and magnetic variable star HD 125248}",
      journal = {\mnras},
         year = 1950,
        month = jan,
       volume = {110},
        pages = {395},
          doi = {10.1093/mnras/110.4.395},
       adsurl = {https://ui.adsabs.harvard.edu/abs/1950MNRAS.110..395S}
}

@ARTICLE{Babcock:1949,
       author = {{Babcock}, H.~W.},
        title = "{Stellar magnetic fields and rotation}",
      journal = {The Observatory},
         year = 1949,
        month = oct,
       volume = {69},
        pages = {191-192},
       adsurl = {https://ui.adsabs.harvard.edu/abs/1949Obs....69..191B}
}

@ARTICLE{Leone:1994,
       author = {{Leone}, F. and {Trigilio}, C. and {Umana}, G.},
        title = "{Radio emission from magnetic chemically peculiar stars: results of the 1992 VLA survey.}",
      journal = {\aap},
         year = 1994,
        month = mar,
       volume = {283},
        pages = {908-910},
       adsurl = {https://ui.adsabs.harvard.edu/abs/1994A&A...283..908L}
}

@ARTICLE{Leto:2006,
       author = {{Leto}, P. and {Trigilio}, C. and {Buemi}, C.~S. and {Umana}, G. and {Leone}, F.},
        title = "{Stellar magnetosphere reconstruction from radio data. Multi-frequency VLA observations and 3D-simulations of <ASTROBJ>CU Virginis</ASTROBJ>}",
      journal = {\aap},
         year = 2006,
        month = nov,
       volume = {458},
       number = {3},
        pages = {831-839},
          doi = {10.1051/0004-6361:20054511},
archivePrefix = {arXiv},
       eprint = {astro-ph/0610395},
 primaryClass = {astro-ph},
       adsurl = {https://ui.adsabs.harvard.edu/abs/2006A&A...458..831L}
}

@ARTICLE{Leto:2012,
       author = {{Leto}, P. and {Trigilio}, C. and {Buemi}, C.~S. and {Leone}, F. and {Umana}, G.},
        title = "{Searching for a CU Virginis-type cyclotron maser from {\ensuremath{\sigma}} Orionis E: the role of the magnetic quadrupole component}",
      journal = {\mnras},
         year = 2012,
        month = jun,
       volume = {423},
       number = {2},
        pages = {1766-1774},
          doi = {10.1111/j.1365-2966.2012.20997.x},
archivePrefix = {arXiv},
       eprint = {1203.6475},
 primaryClass = {astro-ph.SR},
       adsurl = {https://ui.adsabs.harvard.edu/abs/2012MNRAS.423.1766L}
}

@ARTICLE{Shultz:2019,
       author = {{Shultz}, M.~E. and {Wade}, G.~A. and {Rivinius}, Th and {Alecian}, E. and {Neiner}, C. and {Petit}, V. and {Owocki}, S. and {ud-Doula}, A. and {Kochukhov}, O. and {Bohlender}, D. and {Keszthelyi}, Z. and {MiMeS Collaboration} and {BinaMIcS Collaboration}},
        title = "{The magnetic early B-type stars - III. A main-sequence magnetic, rotational, and magnetospheric biography}",
      journal = {\mnras},
         year = 2019,
        month = nov,
       volume = {490},
       number = {1},
        pages = {274-295},
          doi = {10.1093/mnras/stz2551},
archivePrefix = {arXiv},
       eprint = {1909.02530},
 primaryClass = {astro-ph.SR},
       adsurl = {https://ui.adsabs.harvard.edu/abs/2019MNRAS.490..274S}
}

@ARTICLE{Drake:1987,
       author = {{Drake}, Stephen A. and {Abbott}, David C. and {Bastian}, T.~S. and {Bieging}, J.~H. and {Churchwell}, E. and {Dulk}, G. and {Linsky}, Jeffrey L.},
        title = "{The Discovery of Nonthermal Radio Emission from Magnetic Bp--Ap Stars}",
      journal = {\apj},
         year = 1987,
        month = nov,
       volume = {322},
        pages = {902},
          doi = {10.1086/165784},
       adsurl = {https://ui.adsabs.harvard.edu/abs/1987ApJ...322..902D}
}

@ARTICLE{Linsky:1992,
       author = {{Linsky}, Jeffrey L. and {Drake}, Stephen A. and {Bastian}, T.~S.},
        title = "{Radio Emission from Chemically Peculiar Stars}",
      journal = {\apj},
         year = 1992,
        month = jul,
       volume = {393},
        pages = {341},
          doi = {10.1086/171509},
       adsurl = {https://ui.adsabs.harvard.edu/abs/1992ApJ...393..341L}
}

@ARTICLE{Trigilio:2004,
       author = {{Trigilio}, C. and {Leto}, P. and {Umana}, G. and {Leone}, F. and {Buemi}, C.~S.},
        title = "{A three-dimensional model for the radio emission of magnetic chemically peculiar stars}",
      journal = {\aap},
         year = 2004,
        month = may,
       volume = {418},
        pages = {593-605},
          doi = {10.1051/0004-6361:20040060},
archivePrefix = {arXiv},
       eprint = {astro-ph/0402432},
 primaryClass = {astro-ph},
       adsurl = {https://ui.adsabs.harvard.edu/abs/2004A&A...418..593T}
}

@ARTICLE{Leto:2026,
       author = {{Leto}, P. and {Owocki}, S. and {Trigilio}, C. and {Cavallaro}, F. and {Das}, B. and {Shultz}, M.~E. and {Buemi}, C.~S. and {Umana}, G. and {Fossati}, L. and {Ignace}, R. and {Krti{\v{c}}ka}, J. and {Oskinova}, L.~M. and {Pillitteri}, I. and {Bordiu}, C. and {Bufano}, F. and {Cerrigone}, L. and {Ingallinera}, A. and {Loru}, S. and {Riggi}, S. and {Ruggeri}, A.~C. and {ud-Doula}, A. and {Leone}, F.},
        title = "{A scaling relationship for nonthermal radio emission from ordered magnetospheres: II. Investigating the efficiency of relativistic electron production in the magnetospheres of BA-type stars}",
      journal = {\aap},
         year = 2026,
        month = feb,
       volume = {706},
          eid = {A241},
        pages = {A241},
          doi = {10.1051/0004-6361/202557214},
archivePrefix = {arXiv},
       eprint = {2511.05378},
 primaryClass = {astro-ph.SR},
       adsurl = {https://ui.adsabs.harvard.edu/abs/2026A&A...706A.241L}
}

@ARTICLE{Leto:2018,
       author = {{Leto}, P. and {Trigilio}, C. and {Oskinova}, L.~M. and {Ignace}, R. and {Buemi}, C.~S. and {Umana}, G. and {Ingallinera}, A. and {Leone}, F. and {Phillips}, N.~M. and {Agliozzo}, C. and {Todt}, H. and {Cerrigone}, L.},
        title = "{A combined multiwavelength VLA/ALMA/Chandra study unveils the complex magnetosphere of the B-type star HR5907}",
      journal = {\mnras},
         year = 2018,
        month = may,
       volume = {476},
       number = {1},
        pages = {562-579},
          doi = {10.1093/mnras/sty244},
archivePrefix = {arXiv},
       eprint = {1801.08738},
 primaryClass = {astro-ph.SR},
       adsurl = {https://ui.adsabs.harvard.edu/abs/2018MNRAS.476..562L}
}

@ARTICLE{Usov:1992,
       author = {{Usov}, V.~V. and {Melrose}, D.~B.},
        title = "{X-Ray Emission from Single Magnetic Early-Type Stars}",
      journal = {\apj},
         year = 1992,
        month = aug,
       volume = {395},
        pages = {575},
          doi = {10.1086/171677},
       adsurl = {https://ui.adsabs.harvard.edu/abs/1992ApJ...395..575U}
}

@ARTICLE{Shultz:2022,
       author = {{Shultz}, M.~E. and {Owocki}, S.~P. and {ud-Doula}, A. and {Biswas}, A. and {Bohlender}, D. and {Chandra}, P. and {Das}, B. and {David-Uraz}, A. and {Khalack}, V. and {Kochukhov}, O. and {Landstreet}, J.~D. and {Leto}, P. and {Monin}, D. and {Neiner}, C. and {Rivinius}, Th and {Wade}, G.~A.},
        title = "{MOBSTER - VI. The crucial influence of rotation on the radio magnetospheres of hot stars}",
      journal = {\mnras},
         year = 2022,
        month = jun,
       volume = {513},
       number = {1},
        pages = {1429-1448},
          doi = {10.1093/mnras/stac136},
archivePrefix = {arXiv},
       eprint = {2201.05512},
 primaryClass = {astro-ph.SR},
       adsurl = {https://ui.adsabs.harvard.edu/abs/2022MNRAS.513.1429S}
}

@ARTICLE{Owocki:2022,
       author = {{Owocki}, S.~P. and {Shultz}, M.~E. and {ud-Doula}, A. and {Chandra}, P. and {Das}, B. and {Leto}, P.},
        title = "{Centrifugal breakout reconnection as the electron acceleration mechanism powering the radio magnetospheres of early-type stars}",
      journal = {\mnras},
         year = 2022,
        month = jun,
       volume = {513},
       number = {1},
        pages = {1449-1458},
          doi = {10.1093/mnras/stac341},
archivePrefix = {arXiv},
       eprint = {2202.05449},
 primaryClass = {astro-ph.SR},
       adsurl = {https://ui.adsabs.harvard.edu/abs/2022MNRAS.513.1449O}
}

@ARTICLE{Das:2025,
       author = {{Das}, Barnali and {Driessen}, Laura Nicole and {Shultz}, Matt E. and {Pritchard}, Joshua and {Rose}, Kovi and {Wang}, Yuanming and {Lee}, Yu Wing Joshua and {Sivakoff}, Gregory and {Zic}, Andrew and {Murphy}, Tara},
        title = "{VAST-MeMeS: Characterising non-thermal radio emission from magnetic massive stars using the Australian SKA Pathfinder}",
      journal = {\pasa},
         year = 2025,
        month = oct,
       volume = {42},
          eid = {e147},
        pages = {e147},
          doi = {10.1017/pasa.2025.10090},
archivePrefix = {arXiv},
       eprint = {2505.09148},
 primaryClass = {astro-ph.SR},
       adsurl = {https://ui.adsabs.harvard.edu/abs/2025PASA...42..147D}
}

@ARTICLE{Das:2021,
       author = {{Das}, Barnali and {Chandra}, Poonam},
        title = "{Ultra-wideband, Multiepoch Radio Study of the First Discovered ``Main-sequence Radio Pulse Emitter'' CU Vir}",
      journal = {\apj},
         year = 2021,
        month = nov,
       volume = {921},
       number = {1},
          eid = {9},
        pages = {9},
          doi = {10.3847/1538-4357/ac1075},
archivePrefix = {arXiv},
       eprint = {2107.00849},
 primaryClass = {astro-ph.SR},
       adsurl = {https://ui.adsabs.harvard.edu/abs/2021ApJ...921....9D}
}

@ARTICLE{Mutel:2008,
       author = {{Mutel}, R.~L. and {Christopher}, I.~W. and {Pickett}, J.~S.},
        title = "{Cluster multispacecraft determination of AKR angular beaming}",
      journal = {\grl},
         year = 2008,
        month = apr,
       volume = {35},
       number = {7},
          eid = {L07104},
        pages = {L07104},
          doi = {10.1029/2008GL033377},
archivePrefix = {arXiv},
       eprint = {0803.0078},
 primaryClass = {astro-ph},
       adsurl = {https://ui.adsabs.harvard.edu/abs/2008GeoRL..35.7104M}
}

@ARTICLE{Melrose:1982,
       author = {{Melrose}, D.~B. and {Dulk}, G.~A.},
        title = "{Electron-cyclotron masers as the source of certain solar and stellar radio bursts.}",
      journal = {\apj},
         year = 1982,
        month = aug,
       volume = {259},
        pages = {844-858},
          doi = {10.1086/160219},
       adsurl = {https://ui.adsabs.harvard.edu/abs/1982ApJ...259..844M}
}

@ARTICLE{Mutel:2010,
       author = {{Mutel}, R.~L. and {Menietti}, J.~D. and {Gurnett}, D.~A. and {Kurth}, W. and {Schippers}, P. and {Lynch}, C. and {Lamy}, L. and {Arridge}, C. and {Cecconi}, B.},
        title = "{CMI growth rates for Saturnian kilometric radiation}",
      journal = {\grl},
         year = 2010,
        month = oct,
       volume = {37},
       number = {19},
          eid = {L19105},
        pages = {L19105},
          doi = {10.1029/2010GL044940},
       adsurl = {https://ui.adsabs.harvard.edu/abs/2010GeoRL..3719105M}
}

@ARTICLE{Lamy:2011,
       author = {{Lamy}, L. and {Cecconi}, B. and {Zarka}, P. and {Canu}, P. and {Schippers}, P. and {Kurth}, W.~S. and {Mutel}, R.~L. and {Gurnett}, D.~A. and {Menietti}, D. and {Louarn}, P.},
        title = "{Emission and propagation of Saturn kilometric radiation: Magnetoionic modes, beaming pattern, and polarization state}",
      journal = {Journal of Geophysical Research (Space Physics)},
         year = 2011,
        month = apr,
       volume = {116},
       number = {A4},
          eid = {A04212},
        pages = {A04212},
          doi = {10.1029/2010JA016195},
archivePrefix = {arXiv},
       eprint = {1101.3666},
 primaryClass = {astro-ph.EP},
       adsurl = {https://ui.adsabs.harvard.edu/abs/2011JGRA..116.4212L}
}

@ARTICLE{Lee:2013,
       author = {{Lee}, Sang-Yun and {Yi}, Sibaek and {Lim}, Dayeh and {Kim}, Hee-Eun and {Seough}, Jungjoon and {Yoon}, Peter H.},
        title = "{Loss cone-driven cyclotron maser instability}",
      journal = {Journal of Geophysical Research (Space Physics)},
         year = 2013,
        month = nov,
       volume = {118},
       number = {11},
        pages = {7036-7044},
          doi = {10.1002/2013JA019298},
       adsurl = {https://ui.adsabs.harvard.edu/abs/2013JGRA..118.7036L}
}

@ARTICLE{Trigilio:2011,
       author = {{Trigilio}, Corrado and {Leto}, Paolo and {Umana}, Grazia and {Buemi}, Carla S. and {Leone}, Francesco},
        title = "{Auroral Radio Emission from Stars: The Case of CU Virginis}",
      journal = {\apjl},
         year = 2011,
        month = sep,
       volume = {739},
       number = {1},
          eid = {L10},
        pages = {L10},
          doi = {10.1088/2041-8205/739/1/L10},
archivePrefix = {arXiv},
       eprint = {1104.3268},
 primaryClass = {astro-ph.SR},
       adsurl = {https://ui.adsabs.harvard.edu/abs/2011ApJ...739L..10T}
}

@ARTICLE{Treumann:2006,
       author = {{Treumann}, Rudolf A.},
        title = "{The electron-cyclotron maser for astrophysical application}",
      journal = {\aapr},
         year = 2006,
        month = aug,
       volume = {13},
       number = {4},
        pages = {229-315},
          doi = {10.1007/s00159-006-0001-y},
       adsurl = {https://ui.adsabs.harvard.edu/abs/2006A&ARv..13..229T}
}

@ARTICLE{Leto:2016,
       author = {{Leto}, P. and {Trigilio}, C. and {Buemi}, C.~S. and {Umana}, G. and {Ingallinera}, A. and {Cerrigone}, L.},
        title = "{3D modelling of stellar auroral radio emission}",
      journal = {\mnras},
         year = 2016,
        month = jun,
       volume = {459},
       number = {2},
        pages = {1159-1169},
          doi = {10.1093/mnras/stw639},
archivePrefix = {arXiv},
       eprint = {1603.02423},
 primaryClass = {astro-ph.SR},
       adsurl = {https://ui.adsabs.harvard.edu/abs/2016MNRAS.459.1159L}
}

@ARTICLE{Biswas:2025,
       author = {{Biswas}, Ayan and {Das}, Barnali and {Barron}, James A. and {Wade}, Gregg A. and {Holgado}, Gonzalo},
        title = "{A Nonstop Aurora? The Intriguing Radio Emission from the Rapidly Rotating Magnetic Massive Star HR 5907}",
      journal = {\apj},
         year = 2025,
        month = feb,
       volume = {980},
       number = {2},
          eid = {260},
        pages = {260},
          doi = {10.3847/1538-4357/adae02},
archivePrefix = {arXiv},
       eprint = {2501.10813},
 primaryClass = {astro-ph.SR},
       adsurl = {https://ui.adsabs.harvard.edu/abs/2025ApJ...980..260B}
}

@ARTICLE{Kochukhov:2019,
       author = {{Kochukhov}, O. and {Shultz}, M. and {Neiner}, C.},
        title = "{Magnetic field topologies of the bright, weak-field Ap stars {\ensuremath{\theta}} Aurigae and ɛ Ursae Majoris}",
      journal = {\aap},
         year = 2019,
        month = jan,
       volume = {621},
          eid = {A47},
        pages = {A47},
          doi = {10.1051/0004-6361/201834279},
archivePrefix = {arXiv},
       eprint = {1811.04928},
 primaryClass = {astro-ph.SR},
       adsurl = {https://ui.adsabs.harvard.edu/abs/2019A&A...621A..47K}
}

@ARTICLE{Braithwaite:2004,
       author = {{Braithwaite}, Jonathan and {Spruit}, Hendrik C.},
        title = "{A fossil origin for the magnetic field in A stars and white dwarfs}",
      journal = {\nat},
         year = 2004,
        month = oct,
       volume = {431},
       number = {7010},
        pages = {819-821},
          doi = {10.1038/nature02934},
archivePrefix = {arXiv},
       eprint = {astro-ph/0502043},
 primaryClass = {astro-ph},
       adsurl = {https://ui.adsabs.harvard.edu/abs/2004Natur.431..819B}
}

@ARTICLE{Wade:2016,
       author = {{Wade}, G.~A. and {Neiner}, C. and {Alecian}, E. and {Grunhut}, J.~H. and {Petit}, V. and {Batz}, B. de and {Bohlender}, D.~A. and {Cohen}, D.~H. and {Henrichs}, H.~F. and {Kochukhov}, O. and {Landstreet}, J.~D. and {Manset}, N. and {Martins}, F. and {Mathis}, S. and {Oksala}, M.~E. and {Owocki}, S.~P. and {Rivinius}, Th. and {Shultz}, M.~E. and {Sundqvist}, J.~O. and {Townsend}, R.~H.~D. and {ud-Doula}, A. and {Bouret}, J.-C. and {Braithwaite}, J. and {Briquet}, M. and {Carciofi}, A.~C. and {David-Uraz}, A. and {Folsom}, C.~P. and {Fullerton}, A.~W. and {Leroy}, B. and {Marcolino}, W.~L.~F. and {Moffat}, A.~F.~J. and {Naz{\'e}}, Y. and {Louis}, N. St and {Auri{\`e}re}, M. and {Bagnulo}, S. and {Bailey}, J.~D. and {Barb{\'a}}, R.~H. and {Blaz{\`e}re}, A. and {B{\"o}hm}, T. and {Catala}, C. and {Donati}, J.-F. and {Ferrario}, L. and {Harrington}, D. and {Howarth}, I.~D. and {Ignace}, R. and {Kaper}, L. and {L{\"u}ftinger}, T. and {Prinja}, R. and {Vink}, J.~S. and {Weiss}, W.~W. and {Yakunin}, I.},
        title = "{The MiMeS survey of magnetism in massive stars: introduction and overview}",
      journal = {\mnras},
         year = 2016,
        month = feb,
       volume = {456},
       number = {1},
        pages = {2-22},
          doi = {10.1093/mnras/stv2568},
archivePrefix = {arXiv},
       eprint = {1511.08425},
 primaryClass = {astro-ph.SR},
       adsurl = {https://ui.adsabs.harvard.edu/abs/2016MNRAS.456....2W}
}

@ARTICLE{Shultz:2018,
       author = {{Shultz}, M.~E. and {Wade}, G.~A. and {Rivinius}, Th and {Neiner}, C. and {Alecian}, E. and {Bohlender}, D. and {Monin}, D. and {Sikora}, J. and {MiMeS Collaboration} and {BinaMIcS Collaboration}},
        title = "{The magnetic early B-type stars I: magnetometry and rotation}",
      journal = {\mnras},
         year = 2018,
        month = jan,
       volume = {475},
       number = {4},
        pages = {5144-5178},
          doi = {10.1093/mnras/sty103},
archivePrefix = {arXiv},
       eprint = {1801.02924},
 primaryClass = {astro-ph.SR},
       adsurl = {https://ui.adsabs.harvard.edu/abs/2018MNRAS.475.5144S}
}

@ARTICLE{Landstreet:2007,
       author = {{Landstreet}, J.~D. and {Bagnulo}, S. and {Andretta}, V. and {Fossati}, L. and {Mason}, E. and {Silaj}, J. and {Wade}, G.~A.},
        title = "{Searching for links between magnetic fields and stellar evolution: II. The evolution of magnetic fields as revealed by observations of Ap stars in open clusters and associations}",
      journal = {\aap},
         year = 2007,
        month = aug,
       volume = {470},
       number = {2},
        pages = {685-698},
          doi = {10.1051/0004-6361:20077343},
archivePrefix = {arXiv},
       eprint = {0706.0330},
 primaryClass = {astro-ph},
       adsurl = {https://ui.adsabs.harvard.edu/abs/2007A&A...470..685L}
}

@ARTICLE{Landstreet:2008,
       author = {{Landstreet}, J.~D. and {Silaj}, J. and {Andretta}, V. and {Bagnulo}, S. and {Berdyugina}, S.~V. and {Donati}, J.-F. and {Fossati}, L. and {Petit}, P. and {Silvester}, J. and {Wade}, G.~A.},
        title = "{Searching for links between magnetic fields and stellar evolution. III. Measurement of magnetic fields in open cluster Ap stars with ESPaDOnS}",
      journal = {\aap},
         year = 2008,
        month = apr,
       volume = {481},
       number = {2},
        pages = {465-480},
          doi = {10.1051/0004-6361:20078884},
archivePrefix = {arXiv},
       eprint = {0803.0877},
 primaryClass = {astro-ph},
       adsurl = {https://ui.adsabs.harvard.edu/abs/2008A&A...481..465L}
}

@ARTICLE{Sikora:2019,
       author = {{Sikora}, J. and {Wade}, G.~A. and {Power}, J. and {Neiner}, C.},
        title = "{A volume-limited survey of mCP stars within 100 pc II: rotational and magnetic properties}",
      journal = {\mnras},
         year = 2019,
        month = mar,
       volume = {483},
       number = {3},
        pages = {3127-3145},
          doi = {10.1093/mnras/sty2895},
archivePrefix = {arXiv},
       eprint = {1811.05635},
 primaryClass = {astro-ph.SR},
       adsurl = {https://ui.adsabs.harvard.edu/abs/2019MNRAS.483.3127S}
}

@ARTICLE{Das:2020,
       author = {{Das}, Barnali and {Mondal}, Surajit and {Chandra}, Poonam},
        title = "{A 3D Framework to Explore the Propagation Effects in Stars Exhibiting Electron Cyclotron Maser Emission}",
      journal = {\apj},
         year = 2020,
        month = sep,
       volume = {900},
       number = {2},
          eid = {156},
        pages = {156},
          doi = {10.3847/1538-4357/aba8fd},
archivePrefix = {arXiv},
       eprint = {2007.06822},
 primaryClass = {astro-ph.SR},
       adsurl = {https://ui.adsabs.harvard.edu/abs/2020ApJ...900..156D}
}

@ARTICLE{Pillitteri:2016,
       author = {{Pillitteri}, I. and {Wolk}, S.~J. and {Chen}, H.~H. and {Goodman}, A.},
        title = "{First stars of the {\ensuremath{\rho}} Ophiuchi dark cloud. XMM-Newton view of {\ensuremath{\rho}} Oph and its neighbors}",
      journal = {\aap},
         year = 2016,
        month = aug,
       volume = {592},
          eid = {A88},
        pages = {A88},
          doi = {10.1051/0004-6361/201628284},
archivePrefix = {arXiv},
       eprint = {1605.07365},
 primaryClass = {astro-ph.SR},
       adsurl = {https://ui.adsabs.harvard.edu/abs/2016A&A...592A..88P}
}

@ARTICLE{Das:2024,
       author = {{Das}, Barnali and {Chandra}, Poonam and {Petit}, V{\'e}ronique},
        title = "{Coherent Radio Emission from ``Main-sequence Radio Pulse Emitters'': A New Stellar Diagnostic to Probe 3D Magnetospheric Structures}",
      journal = {\apj},
         year = 2024,
        month = oct,
       volume = {974},
       number = {2},
          eid = {267},
        pages = {267},
          doi = {10.3847/1538-4357/ad71c5},
archivePrefix = {arXiv},
       eprint = {2408.11242},
 primaryClass = {astro-ph.SR},
       adsurl = {https://ui.adsabs.harvard.edu/abs/2024ApJ...974..267D}
}

@ARTICLE{Das:2022c,
       author = {{Das}, Barnali and {Chandra}, Poonam and {Petit}, V{\'e}ronique},
        title = "{What leads to premature upper cut-off frequencies of auroral radio emission from hot magnetic stars?}",
      journal = {\mnras},
         year = 2022,
        month = sep,
       volume = {515},
       number = {2},
        pages = {2008-2014},
          doi = {10.1093/mnras/stac1894},
archivePrefix = {arXiv},
       eprint = {2207.00470},
 primaryClass = {astro-ph.SR},
       adsurl = {https://ui.adsabs.harvard.edu/abs/2022MNRAS.515.2008D}
}

@ARTICLE{Callingham:2021,
       author = {{Callingham}, J.~R. and {Vedantham}, H.~K. and {Shimwell}, T.~W. and {Pope}, B.~J.~S. and {Davis}, I.~E. and {Best}, P.~N. and {Hardcastle}, M.~J. and {R{\"o}ttgering}, H.~J.~A. and {Sabater}, J. and {Tasse}, C. and {van Weeren}, R.~J. and {Williams}, W.~L. and {Zarka}, P. and {de Gasperin}, F. and {Drabent}, A.},
        title = "{The population of M dwarfs observed at low radio frequencies}",
      journal = {Nature Astronomy},
         year = 2021,
        month = dec,
       volume = {5},
        pages = {1233-1239},
          doi = {10.1038/s41550-021-01483-0},
archivePrefix = {arXiv},
       eprint = {2110.03713},
 primaryClass = {astro-ph.SR},
       adsurl = {https://ui.adsabs.harvard.edu/abs/2021NatAs...5.1233C}
}

@ARTICLE{Pritchard:2021,
       author = {{Pritchard}, Joshua and {Murphy}, Tara and {Zic}, Andrew and {Lynch}, Christene and {Heald}, George and {Kaplan}, David L. and {Anderson}, Craig and {Banfield}, Julie and {Hale}, Catherine and {Hotan}, Aidan and {Lenc}, Emil and {Leung}, James K. and {McConnell}, David and {Moss}, Vanessa A. and {Raja}, Wasim and {Stewart}, Adam J. and {Whiting}, Matthew},
        title = "{A circular polarization survey for radio stars with the Australian SKA Pathfinder}",
      journal = {\mnras},
         year = 2021,
        month = apr,
       volume = {502},
       number = {4},
        pages = {5438-5454},
          doi = {10.1093/mnras/stab299},
archivePrefix = {arXiv},
       eprint = {2102.01801},
 primaryClass = {astro-ph.SR},
       adsurl = {https://ui.adsabs.harvard.edu/abs/2021MNRAS.502.5438P}
}

@ARTICLE{Cesaroni:2023,
       author = {{Cesaroni}, R. and {Moscadelli}, L. and {Caratti o Garatti}, A. and {Eisl{\"o}ffel}, J. and {Fedriani}, R. and {Neri}, R. and {Ray}, T. and {Sanna}, A. and {Stecklum}, B.},
        title = "{Radio outburst from a massive (proto)star. II. A portrait in space and time of the expanding radio jet from S255IR NIRS 3}",
      journal = {\aap},
         year = 2023,
        month = dec,
       volume = {680},
          eid = {A110},
        pages = {A110},
          doi = {10.1051/0004-6361/202347468},
archivePrefix = {arXiv},
       eprint = {2310.18002},
 primaryClass = {astro-ph.GA},
       adsurl = {https://ui.adsabs.harvard.edu/abs/2023A&A...680A.110C}
}

@ARTICLE{Driessen:2024,
       author = {{Driessen}, Laura Nicole and {Pritchard}, Joshua and {Murphy}, Tara and {Heald}, George and {Robrade}, Jan and {Das}, Barnali and {Duchesne}, Stefan William and {Kaplan}, David L. and {Lenc}, Emil and {Lynch}, Christene R. and {Mitchell-Bolton}, Jackson and {Pope}, Benjamin J.~S. and {Rose}, Kovi and {Stelzer}, Beate and {Wang}, Yuanming and {Zic}, Andrew},
        title = "{The Sydney Radio Star Catalogue: Properties of radio stars at megahertz to gigahertz frequencies}",
      journal = {\pasa},
         year = 2024,
        month = nov,
       volume = {41},
          eid = {e084},
        pages = {e084},
          doi = {10.1017/pasa.2024.72},
archivePrefix = {arXiv},
       eprint = {2404.07418},
 primaryClass = {astro-ph.SR},
       adsurl = {https://ui.adsabs.harvard.edu/abs/2024PASA...41...84D}
}

@ARTICLE{Caldwell2020,
       author = {{Caldwell}, Douglas A. and {Tenenbaum}, Peter and {Twicken}, Joseph D. and {Jenkins}, Jon M. and {Ting}, Eric and {Smith}, Jeffrey C. and {Hedges}, Christina and {Fausnaugh}, Michael M. and {Rose}, Mark and {Burke}, Christopher},
        title = "{TESS Science Processing Operations Center FFI Target List Products}",
      journal = {Research Notes of the American Astronomical Society},
         year = 2020,
        month = nov,
       volume = {4},
       number = {11},
          eid = {201},
        pages = {201},
          doi = {10.3847/2515-5172/abc9b3},
archivePrefix = {arXiv},
       eprint = {2011.05495},
 primaryClass = {astro-ph.EP},
       adsurl = {https://ui.adsabs.harvard.edu/abs/2020RNAAS...4..201C}
}

@ARTICLE{Munoz:2020,
       author = {{Munoz}, M.~S. and {Wade}, G.~A. and {Naz{\'e}}, Y. and {Puls}, J. and {Bagnulo}, S. and {Szyma{\'n}ski}, M.~K.},
        title = "{Modelling the photometric variability of magnetic massive stars with the Analytical Dynamical Magnetosphere model}",
      journal = {\mnras},
         year = 2020,
        month = feb,
       volume = {492},
       number = {1},
        pages = {1199-1213},
          doi = {10.1093/mnras/stz2904},
archivePrefix = {arXiv},
       eprint = {1910.05792},
 primaryClass = {astro-ph.SR},
       adsurl = {https://ui.adsabs.harvard.edu/abs/2020MNRAS.492.1199M}
}

@ARTICLE{Alecian:2014,
       author = {{Alecian}, E. and {Kochukhov}, O. and {Petit}, V. and {Grunhut}, J. and {Landstreet}, J. and {Oksala}, M.~E. and {Wade}, G.~A. and {Hussain}, G. and {Neiner}, C. and {Bohlender}, D. and {MiMeS Collaboration}},
        title = "{Discovery of new magnetic early-B stars within the MiMeS HARPSpol survey}",
      journal = {\aap},
         year = 2014,
        month = jul,
       volume = {567},
          eid = {A28},
        pages = {A28},
          doi = {10.1051/0004-6361/201323286},
archivePrefix = {arXiv},
       eprint = {1404.5508},
 primaryClass = {astro-ph.SR},
       adsurl = {https://ui.adsabs.harvard.edu/abs/2014A&A...567A..28A}
}

@ARTICLE{Das2025,
       author = {{Das}, B. and {Owocki}, S.~P.},
        title = "{The effect of collisional cooling of energetic electrons on radio emission from the centrifugal magnetospheres of magnetic hot stars}",
      journal = {\pasa},
         year = 2025,
        month = oct,
       volume = {42},
          eid = {e138},
        pages = {e138},
          doi = {10.1017/pasa.2025.10095},
archivePrefix = {arXiv},
       eprint = {2509.14561},
 primaryClass = {astro-ph.SR},
       adsurl = {https://ui.adsabs.harvard.edu/abs/2025PASA...42..138D}
}

@article{Naze2014,
    doi = {10.1088/0067-0049/215/1/10},
    url = {https://doi.org/10.1088/0067-0049/215/1/10},
    year = {2014},
    month = {oct},
    publisher = {The American Astronomical Society},
    volume = {215},
    number = {1},
    pages = {10},
    author = {Nazé, Yaël and Petit, Véronique and Rinbrand, Melanie and Cohen, David and Owocki, Stan and ud-Doula, Asif and Wade, Gregg A.},
    title = {X-RAY EMISSION FROM MAGNETIC MASSIVE STARS*},
    journal = {The Astrophysical Journal Supplement Series},
}

@ARTICLE{Strom1969,
       author = {{Strom}, S.~E. and {Strom}, K.~M.},
        title = "{Effect of Silicon Opacity on b- and A-Star Atmospheres}",
      journal = {\apj},
         year = 1969,
        month = jan,
       volume = {155},
        pages = {17},
          doi = {10.1086/149843},
       adsurl = {https://ui.adsabs.harvard.edu/abs/1969ApJ...155...17S}
}

@ARTICLE{Kochukhov2009,
       author = {{Kochukhov}, O. and {Shulyak}, D. and {Ryabchikova}, T.},
        title = "{A self-consistent empirical model atmosphere, abundance and stratification analysis of the benchmark roAp star {\ensuremath{\alpha}} Circini}",
      journal = {\aap},
         year = 2009,
        month = jun,
       volume = {499},
       number = {3},
        pages = {851-863},
          doi = {10.1051/0004-6361/200911653},
archivePrefix = {arXiv},
       eprint = {0903.3512},
 primaryClass = {astro-ph.SR},
       adsurl = {https://ui.adsabs.harvard.edu/abs/2009A&A...499..851K}
}

@ARTICLE{Kochukhov2005,
       author = {{Kochukhov}, O. and {Khan}, S. and {Shulyak}, D.},
        title = "{Stellar model atmospheres with magnetic line blanketing}",
      journal = {\aap},
         year = 2005,
        month = apr,
       volume = {433},
       number = {2},
        pages = {671-682},
          doi = {10.1051/0004-6361:20042300},
archivePrefix = {arXiv},
       eprint = {astro-ph/0412262},
 primaryClass = {astro-ph},
       adsurl = {https://ui.adsabs.harvard.edu/abs/2005A&A...433..671K}
}

@ARTICLE{He2019,
       author = {{He}, Lin and {Wang}, Song and {Liu}, Jifeng and {Soria}, Roberto and {Bai}, Zhongrui and {Yang}, Huiqin and {Bai}, Yu and {Guo}, Jincheng},
        title = "{A Combined Chandra and LAMOST Study of Stellar Activity}",
      journal = {\apj},
         year = 2019,
        month = feb,
       volume = {871},
       number = {2},
          eid = {193},
        pages = {193},
          doi = {10.3847/1538-4357/aaf8b7},
archivePrefix = {arXiv},
       eprint = {1812.05763},
 primaryClass = {astro-ph.SR},
       adsurl = {https://ui.adsabs.harvard.edu/abs/2019ApJ...871..193H}
}

@ARTICLE{Ducati2003,
       author = {{Ducati}, Jorge R. and {Ribeiro}, Daiana and {Rembold}, Sandro B.},
        title = "{A Method for Simultaneous Determination of A$_{V}$ and R and Applications}",
      journal = {\apj},
         year = 2003,
        month = may,
       volume = {588},
       number = {1},
        pages = {344-352},
          doi = {10.1086/368376},
       adsurl = {https://ui.adsabs.harvard.edu/abs/2003ApJ...588..344D}
}
\bibliographystyle{aasjournalv7}



\appendix
\section{Radio Fluxes}
\restartappendixnumbering

\begin{deluxetable*}{l l l c c c c c}
\centering
\tablecaption{Radio fluxes of \rOC\ reported in this manuscript \label{table:radio-fluxes}}
\tablehead{
\colhead{Telescope} & \colhead{Project Code} & \colhead{\Change{Start--}Mid\Change{--Stop}} & \colhead{\Change{Phase Range}} & \colhead{Frequency} & \colhead{Bandwidth} & \colhead{Stokes $I$} & \colhead{Stokes $V$} \\
\colhead{} & \colhead{} & \colhead{[UTC]} & \colhead{\Change{$\phi_\mathrm{start}$--$\phi_\mathrm{mid}$--$\phi_\mathrm{stop}$}} & \colhead{[MHz]} & \colhead{[MHz]} & \colhead{[mJy]} & \colhead{[mJy]}}
\startdata
VLA  & NVSS            & 1997-09-27 \Change{22:56:09--}22:57:34\Change{--22:58:59} & \Change{0.949--0.950--0.951}                  & 1400   &   50 & $9.7\pm0.6$    & --            \\
--   & VLASS1.2        & 2019-06-29 \Change{03:56:09--}03:56:24\Change{--03:56:39} & \Change{0.699--0.699--0.699}                  & 3000   & 2048 & $13.89\pm0.16$ & --            \\
--   & VLA/19A-446     & 2019-08-26 \Change{00:44:05--}00:45:39\Change{--00:47:12} & \Change{0.682--0.683--0.684}                  & 1500   & 1024 & $20.72\pm0.04$ & $3.85\pm0.04$ \\
--   & VLASS2.2        & 2022-02-08 \Change{12:46:43--}12:46:58\Change{--12:47:13} & \Change{0.577--0.577--0.577}                  & 3000   & 2048 & $24.93\pm0.13$ & --            \\
--   & VLASS3.2        & 2024-09-24 \Change{23:23:58--}23:24:13\Change{--23:24:28} & \Change{0.172--0.172--0.172}                  & 3000   & 2048 & $18.91\pm0.17$ & --            \\
\hline
ASKAP & RACS (AS110)\Change{\tablenotemark{a}}   & 2019-04-25 \Change{16:15:37--}16:23:07\Change{--16:30:37} & \Change{0.053--0.059--0.065}                    & 887.5  & 288  & $6.4\pm0.4$    & --            \\
--    & RACS (AS110)\Change{\tablenotemark{a}}   & 2021-01-18 \Change{01:16:30--}01:24:00\Change{--01:31:30} & \Change{0.211--0.217--0.223}                    & 1367.5 & 288  & $12.39\pm0.14$ & $<0.38$       \\
--    & RACS (AS110)\Change{\tablenotemark{a}}   & 2022-01-20 \Change{01:26:33--}01:34:03\Change{--01:41:33} & \Change{0.037--0.043--0.049}                    & 1655.5 & 288  & $12.5\pm0.2$   & $1.22\pm0.19$ \\
--    & RACS (AS110)   & 2022-04-11 \Change{18:12:38--}18:20:08\Change{--18:27:38} & \Change{0.607--0.613--0.619}                    & 887.5  & 288  & $9.2\pm0.2$    & $1.62\pm0.18$ \\
--    & WALLABY (AS202)\Change{\tablenotemark{a}}& 2023-09-07 \Change{05:02:45--}09:02:45\Change{--13:02:45} & \Change{0.948--1.141--1.334\tablenotemark{b,c}} & 1367.5 & 288  & $13.81\pm0.04$ & $0.48\pm0.03$ \\
--    & RACS (AS110)\Change{\tablenotemark{a}}   & 2024-01-18 \Change{01:35:09--}01:42:39\Change{--01:50:09} & \Change{0.734--0.740--0.746}                    & 943.5  & 288  & $17.35\pm0.16$ & $<0.39$       \\
--    & RACS (AS110)   & 2024-11-16 \Change{05:40:28--}05:47:58\Change{--05:55:28} & \Change{0.666--0.672--0.678}                    & 1367.5 & 288  & $8.65\pm0.16$  & $<0.47$       \\
--    & WALLABY (AS202)& 2025-04-24 \Change{14:17:43--}18:17:43\Change{--22:17:43} & \Change{0.131--0.324--0.517\tablenotemark{c}}                    & 1367.5 & 288  & $12.32\pm0.04$ & $4.59\pm0.04$ \\
--    & WALLABY (AS202)& 2025-05-11 \Change{14:05:56--}18:05:56\Change{--22:05:56} & \Change{0.800--0.993--1.186\tablenotemark{b,c}} & 1367.5 & 288  & $12.93\pm0.04$ & $<0.11$       \\
--    & FLASH (AS209)  & 2025-10-08 \Change{06:39:55--}07:39:55\Change{--08:39:55} & \Change{0.073--0.121--0.169}   & 855.5  & 288  & $7.17\pm0.08$  & $1.80\pm0.07$
\enddata
\tablenotetext{a}{\Change{Stokes $I$ also reported in \citet{Das:2025}.}}
\tablenotetext{b}{\Change{$\phi>1$ indicates a crossing through $\phi=0$.}}
\tablenotetext{c}{\Change{These images are generated from $\approx8\,$hr integrations and are not shown in the middle panels of \autoref{fig:radio-lcs}. Manual imaging at higher time resolution would provide new light curves spanning $\sim1/3$ of a rotational phase coverage each.}}
\tablecomments{\Change{The three times give the start, midpoint, and stop of each observation in UTC; the midpoint and stop share the calendar date of the start. Rotational phases are computed using the period $P = 0.8639$\,d and reference epoch $T_0 = \mathrm{HJD}\,2458555.073$ \citep{Leto2020}.} We list the background r.m.s. of the image to indicate detection significance and do not include systematic flux uncertainties. Upper limits are listed as $3\sigma$.}
\end{deluxetable*}

\end{document}